\documentclass[a4paper,11pt]{article}
\pdfoutput=1
\usepackage{jheppub}

\usepackage{graphicx}
\usepackage{placeins}
\usepackage{url}
\usepackage{multirow}
\usepackage{paralist}
\usepackage{amssymb}
\usepackage{amsmath}
\usepackage{xspace}
\usepackage{booktabs}
\usepackage[symbol]{footmisc}

\begin{document}
\begin{flushright}
\texttt{MPP-2022-35}
\end{flushright}

\title{Optimizing Observables with Machine Learning for Better Unfolding}

\affiliation[a]{Department of Physics and Astronomy, University of California, Riverside, CA 92521, USA}
\affiliation[b]{Thomas Jefferson National Accelerator Facility, 12000 Jefferson Ave., Newport News, VA 23606, United States}
\affiliation[c]{Max-Planck-Institut f{\"u}r Physik, F{\"o}hringer Ring 6, 80805 M{\"u}nchen, Germany}

\affiliation[d]{Physics Division, Lawrence Berkeley National Laboratory, Berkeley, CA 94720, USA}
\affiliation[e]{Berkeley Institute for Data Science, University of California, Berkeley, CA 94720, USA}

\author[a,b]{Miguel Arratia,}
\author[c]{Daniel Britzger,}
\author[a]{Owen Long,}
\author[d,e]{and Benjamin Nachman}

\abstract{
Most measurements in particle and nuclear physics use matrix-based unfolding algorithms to correct for detector effects. 
In nearly all cases, the observable is defined analogously at the particle and detector level. 
We point out that while the particle-level observable needs to be physically motivated to link with theory, the detector-level need not be and can be optimized.  
We show that using deep learning to define detector-level observables has the capability to improve the measurement when combined with standard unfolding methods. }

\maketitle

\section{Introduction}
\label{sec:intro}
The measurement of cross sections in particle and nuclear physics experiments usually proceeds as a two step process. First, an observable is constructed and secondly the data are corrected for detector effects (see e.g.,~\cite{Behnke:1517556}). The latter step is the so-called \textit{unfolding} and is based on Monte Carlo event simulations to correct the data for limited resolution, acceptance effects or mis-tagging. 
The process begins by defining the observable of
interest
$\mathcal{O}_p:\mathbb{R}^{m_p} \rightarrow \mathbb{R}^{n_p}$,
where the particle-level input
$x_p\in\mathbb{R}^{m_p}$ can be a list of four-vectors, and usually
$n_p=1$.  For example, if $\mathcal{O}_p$ is the $p_T$ of the leading
jet, then $\mathcal{O}_p$ acts on all the particles in the event, runs
jet clustering, and outputs the $p_T$ of the hardest jet.
Then, an observable
$\mathcal{O}_d:\mathbb{R}^{m_d} \rightarrow \mathbb{R}^{n_d}$
is defined analogously at detector-level, e.g.\ using energy-flow
objects.
The definition of the observable at detector level should yield a reasonable resolution, corresponding to a strong correlation between the detector-level and particle-level observables.
Next, a histogram is filled at detector level from the event counts.  An event simulation is then used to populate a matrix quantifying detector distortions by linking the event counts between particle-level and detector-level bins. Unfolding is then the art of regularized matrix
inversion (see
Ref.~\cite{Cowan:2002in,Blobel:2203257,doi:10.1002/9783527653416.ch6,Balasubramanian:2019itp}
for recent reviews). 

In most cases, $\mathcal{O}_d$ is defined using detector-level
quantities as input to the equivalent definition of the observable at
particle-level:  $\mathcal{O}_d(x_d)\equiv\mathcal{O}_p(x_d)$.
However, although this often yields sufficient resolution, a drawback of this
approach is that further quantities are commonly not included in the
unfolding, nor in the reconstruction, which then can degrade performance.  For example, if the jet
energy response worsens at high pseudorapdity, $\eta$,
then the measurement of jet $p_T$ would not be optimal,
since $\eta$ is not used in the calculation of $\mathcal{O}$ and it is
also excluded in the unfolding\footnote{These effects are suppressed using dedicated energy
  calibrations that depend on many features to improve their
  resolution~\cite{CMS:2011shu,CMS:2016lmd,
    ATLAS:2017bje,ATLAS:2014hvo,ATLAS:2019oxp}. However, most
  observables lack dedicated calibrations; for example,
  ATLAS~\cite{ATLAS:2019kwg} and CMS~\cite{CMS:2018ypj} do not have a
  dedicated calibration for the
  $n$-subjettiness~\cite{Thaler:2010tr,Thaler:2011gf}.}. In this
situation, it may be advantageous to consider $\eta$ when defining
$\mathcal{O}_d$ or to simultaneously unfold $p_T$ and $\eta$ (as e.g.\ done in Refs.~\cite{H1:2014cbm,H1:2016goa}). The same reasoning would apply
to any variable driving jet performance. However, once more variables
are involved, this leads to a dimensionality problem, or to limited
statistics of data and/or the simulations.

Matrix-based unfolding methods do not scale well to many dimensions.
This is because the migration matrices become large when many quantities are considered~\cite{HERMES:2012kpt,H1:2014cbm,H1:2016goa} or it is technically challenging to consider migrations in further quantities~\cite{H1:2010pfw,H1:2013okq,H1:2020lzc}.
Thus a variety of multidimensional machine learning (ML) techniques have been proposed~\cite{Andreassen:2019cjw,Bellagente:2020piv,Vandegar:2020yvw,Arratia:2021otl}, and first ML-based unfolding techniques were already applied to collider data~\cite{2108.12376}. However, these pose computational challenges, or are often
difficult to implement in existing analysis workflows, and thus low-dimensional unfolding might still continue to be the most prevalent approach in the near future.  Additionally, ML approaches still focus on the case $\mathcal{O}_d(x_d)=\mathcal{O}_p(x_d)$, even if $\mathcal{O}$ is multi-dimensional.

We note that there is freedom to define the
detector-level quantity $\mathcal{O}_d$, and it need not be a simple analog of $\mathcal{O}_p$.
In fact, $\mathcal{O}_d$ could even have a different dimensionality than $\mathcal{O}_p$.  The only core requirement is that the particle-level observable $\mathcal{O}_p$
is linked to theory, and is a useful physics quantity.

We propose to use deep learning to define an improved definition of
$\mathcal{O}_d$, given the aim of measuring $\mathcal{O}_p$ and using a large number of inputs to better account for detector effects. We use a regression model to learn $\mathcal{O}_p$ given $x_d$, with     $\mathcal{O}_p$ being the observable for a cross section measurement\footnote{Many proposals do this for observables used in other tasks~\cite{CMS:2020uim,Kieseler:2021jxc,Belayneh:2019vyx,ATL-PHYS-PUB-2020-018,Akchurin:2021afn,Akchurin:2021ahx,Polson:2021kvr,Pata:2021oez,ATL-PHYS-PUB-2018-013,ATL-PHYS-PUB-2020-001,CMS:2019uxx,Haake:2018hqn,Haake:2019pqd,Baldi:2020hjm,Komiske:2017ubm,ATL-PHYS-PUB-2019-028,Maier:2021ymx,Kasieczka:2020vlh,ArjonaMartinez:2018eah,Diefenthaler:2021rdj,Arratia:2021tsq,Liu:2020pzv,EXO:2018bpx,Baldi:2018qhe,Abbasi:2021ryj,IceCube:2020yct,Carloni:2021zbc} (see also Ref.~\cite{Feickert:2021ajf}).}. Using standard loss functions, direct regression is prior-dependent so we use standard unfolding techniques to mitigate it\footnote{There is a close connection between reconstruction and unfolding.  When the reconstruction has no noise, then unfolding is unnecesary.  In general, reconstruction moves around the features while unfolding moves around the cross section (e.g. $x$-axis versus $y$-axis of a histogram).  We thank J. Thaler for many useful discussions about this connection. }. Our approach to define $\mathcal{O}_d$ is compatible with both binned and unbinned~\cite{Lindemann:1995ut,Aslan:2003vu,DEMBINSKI2013410,Gagunashvili:2010zw,Glazov:2017vni,Datta:2018mwd,Andreassen:2019cjw,Bellagente:2020piv,Bellagente:2019uyp,Andreassen:2021zzk,bunse2018unification,Ruhe2019MiningFS, Howard:2021pos,Vandegar:2020yvw,Arratia:2021otl} unfolding methods.
\begin{figure}[thbp!]
    \centering
    \includegraphics[width=0.95\textwidth]{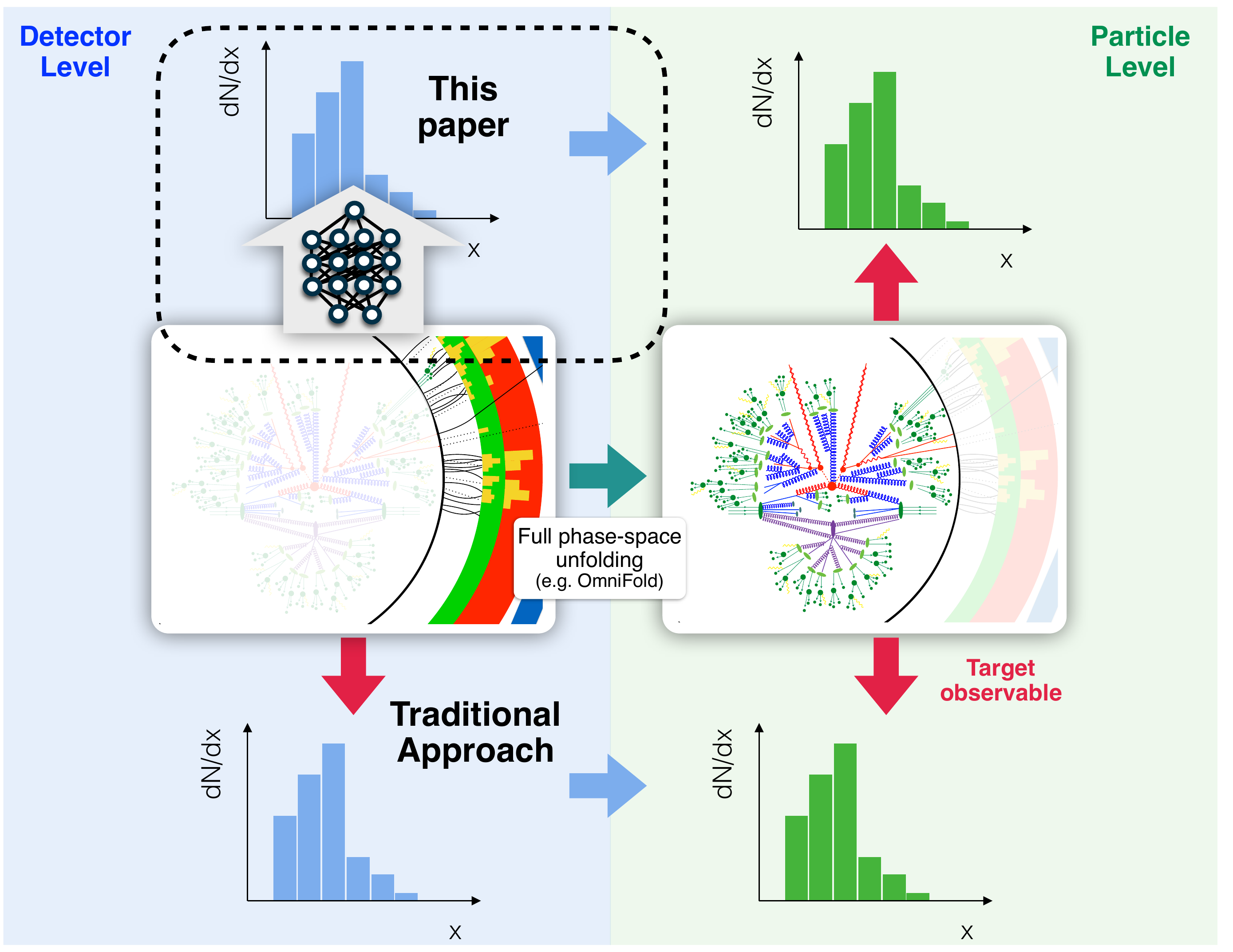}
    \caption{A schematic diagram of unfolding methods, including the one proposed in this paper.  The left-hand side images represent detector-level events while the right-hand side images correspond to particle-level events.  Horizontal arrows correspond to cross section corrections (e.g., via regularized matrix inversion) while vertical arrows correspond to observable definitions.  The traditional path uses the same observable definition at particle-level and detector-level while we propose to use machine learning to define the detector-level observable.  The middle horizontal arrow correspond to a full phase space unfolding where observables (and bins) can be constructed after the unfolding (see Ref.~\cite{Andreassen:2019cjw}).  Such an approach is complementary to the one proposed in this paper.}
    \label{fig:method}
\end{figure}
The workflow is schematically displayed in Figure~\ref{fig:method} and compared to other reconstruction and unfolding procedures.

While we are not aware of other proposals to use ML for observable reconstruction, a variety of related ideas have been studied.  For example, the Simplified Template Cross Section (STXS) protocol for Higgs studies~\cite{LHCHiggsCrossSectionWorkingGroup:2016ypw,Berger:2019wnu} identifies bins of $\mathcal{O}_p$, but does not specify $\mathcal{O}_d$.  Recent ATLAS~\cite{ATLAS-CONF-2020-026} and CMS~\cite{CMS:2021kom} measurements have used Boosted Decision Trees to construct $\mathcal{O}_d$ as a classification problem with the STXS bins as classes. A related idea was presented in Ref.~\cite{Glazov:2017vni}, with an iterative, binned unfolding based on ML classifiers.  The results  Ref.~\cite{Glazov:2017vni} are one-dimensional, but the author mentions adding additional features.

The remainder of this paper is organized as follows.  Section~\ref{sec:methods} provides the technical details of our approach.  We then provide explicit examples using a full detector simulation in Sec.~\ref{sec:results}.  The paper ends with conclusions and outlook in Sec.~\ref{sec:conclusion}.

\section{Machine Learning-Assisted Reconstruction for better Unfolding}
\label{sec:methods}
\subsection{Technical setup}

The focus will be on the combination of an ML-based reconstruction of an observable and the subsequent application of an unfolding algorithm using summary statistics.
It was already demonstrated in Ref.~\cite{Arratia:2021tsq} that ML-reconstructed observables have the capability to improve the resolution and reduce the bias when compared to classically calculated observables.
Hence, we build our study upon this previous ML-setup, and extend it to further observables, and study the unfolding properties of such observables.

The most widely-used regularized unfolding methods are \textsc{TUnfold}~\cite{Schmitt:2012kp}, Iterative Bayesian (also known as Lucy-Richardson)~\cite{1974AJ.....79..745L,Richardson:72,DAgostini:1994fjx}, and SVD~\cite{Hocker:1995kb}. We use \textsc{TUnfold}, which is based on a least-squares fit that includes Tikhonov regularization~\cite{Blobel:1984ku,Tikhonov} to control statistical fluctuations. We use version 17.6 of the \textsc{TUnfold} package in the 6.24 distribution of \textsc{Root}~\cite{Brun:1997pa}.
The response matrix is constructed from 2D histograms with 10 bins in the particle-level value and 20 bins of the detector-level value.
The input to the unfolding is a pseudo-dataset from the MC sample, and the migration matrix is defined by a statistically independent sample, or, alternatively, using an independent MC generator.
The scan for the optimal value of the regularization parameter $\tau$  is performed with the {\tt ScanLcurve} method with 30 points~\cite{Schmitt:2012kp}.
All other settings of \textsc{TUnfold} are set to their default values.

All deep neural network (DNN) models are implemented in \textsc{Keras}~\cite{chollet2015keras} and \textsc{TensorFlow}~\cite{tensorflow} and optimized using \textsc{Adam}~\cite{adam}
with the \textsc{Huber}~\cite{10.1214/aoms/1177703732} loss function as described in our earlier work~\cite{Arratia:2021tsq}.  The network models use 7 sequential hidden layers with between 64 and 1024 nodes per layer.

The metrics that we will use to compare the standard reconstruction and the ML-based reconstruction are
the statistical uncertainties and global correlation coefficients~\cite{Schmitt:2012kp} for the unfolded distribution.
We also compare the response matrices for the methods and the matrix of correlation coefficients for the unfolded results.

\subsection{Simulated Datasets}

We study simulated data of electron-proton collisions at a center-of-mass energy of $\sqrt{s}=319\,$GeV from the H1 experiment at HERA~\cite{H1:1996jzy,H1:1996prr}.
These simulated data are well understood and all aspects of the data are in general in good agreement with data from the real experiment, and the
H1 detector simulation includes all sub-detectors and run-dependent effects.
The simulated data are the same as were used in Ref.~\cite{Arratia:2021tsq} and are briefly described in the following.

Two simulated samples of deep-inelastic scattering (DIS) were created by the H1 Collaboration, each with a different event generator: \textsc{Rapgap}~3.1~\cite{Jung:1993gf}
or
\textsc{Djangoh}~1.4~\cite{Charchula:1994kf}, with beam energies $E_e=27.6$ GeV and $E_p=920$ GeV, for the lepton and proton, respectively. Both generators use the
\textsc{Heracles} routines~\cite{Spiesberger:237380,Kwiatkowski:1990cx,Kwiatkowski:1990es}
for Quantum Electrodynamic (QED) radiation. The simulated events are reconstructed in the same way as data.  An energy-flow
algorithm~\cite{energyflowthesis,energyflowthesis2,energyflowthesis3}
is used to define objects whose sum yields the Hadronic Final State (HFS) four-vector.  The scattered electron candidate is defined using the standard H1 approach~\cite{H1:2012qti,H1:2014cbm,H1:2021wkz}.  
Standard selections~\cite{H1:2012qti,H1:2014cbm} are applied to suppress backgrounds and mis-measured events. The simulated events are processed by H1's computing
environment~\cite{Britzger:2021xcx}.  Altogether, $\mathcal{O}(10^8)$
events were simulated for each generator.
The kinematic region is defined through $Q^2>220$~GeV$^2$ and for the purpose of a reduction of QED initial-state radiation, the reconstructed events must fulfill $45$~GeV~$<E-P_z<62$~GeV, where $E$ and $P_z$ are calculated from the sum of the 4-vectors of the scattered electron and the HFS.
This requirement, as well as detector specific acceptance losses, like those from non-instrumented regions (cracks), and cuts to reduce backgrounds, will be taken into account as acceptance corrections.

\section{Numerical Results}
\label{sec:results}
In Ref.~\cite{Arratia:2021tsq}, we introduced a ML approach to reconstruct observables, and applied it to reconstruct the event kinematics in neutral-current DIS events, which are the momentum transfer squared $Q^2$, the longitudinal momentum fraction $x$, and the inelasticity $y$.
Since the benefit was most distinct in $x$ and $y$, and the ML-reconstructed observables promised a possible extension of the measurement phase space, we will focus on these two DIS kinematic observables in the following. 
Subsequently we will then study the properties of our method when applied
to the 1-jettiness event shape observable $\tau_1^b$.

\subsection{Neutral-current DIS kinematic observables}

Inputs to the regression DNN are the 4-vectors of the scattered electron, hadronic final state (HFS), and photons identified as QED radiation candidates.
The HFS is defined as the sum of all particle candidates that are not associated with the scattered electron. Outputs of the regression DNN are the DIS kinematic variables $Q^2$, $y$, and $x$.
In the presence of QED radiation, the exact definition of the observables at particle level is described and discussed in Ref.~\cite{Arratia:2021tsq}.
Likewise as in Ref.~\cite{Arratia:2021tsq}, the DNN method yields better resolution and smaller biases than any other classical reconstruction methods and the DNN reconstructed observables benefit from the usage of more measured quantities for the calculation of the observables in comparison to classical measurements, and from the inherent classification of QED radiative effects.

\begin{figure}
   \begin{center}
\includegraphics[width=0.32\linewidth]{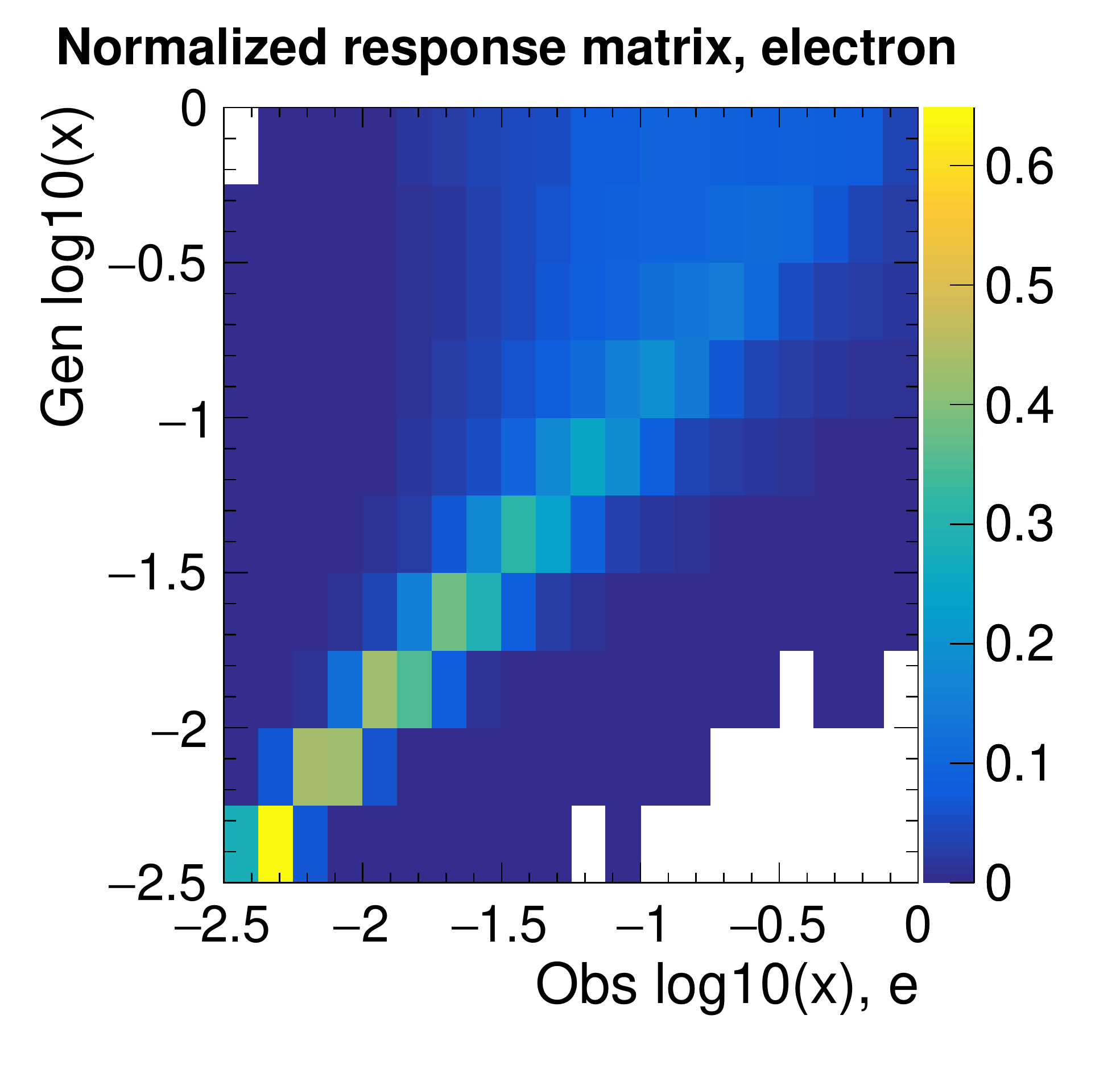}
\includegraphics[width=0.32\linewidth]{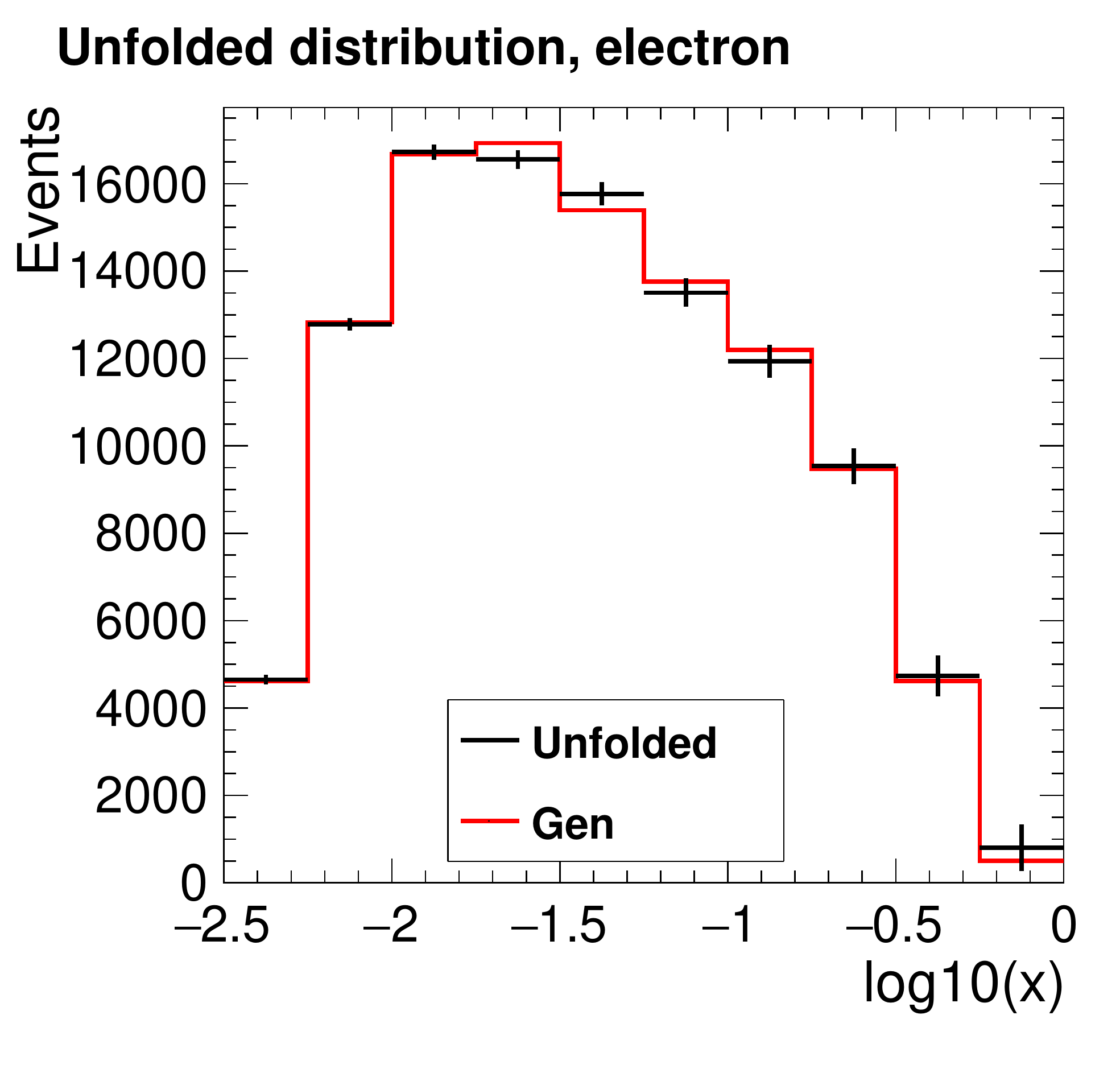}
\includegraphics[width=0.32\linewidth]{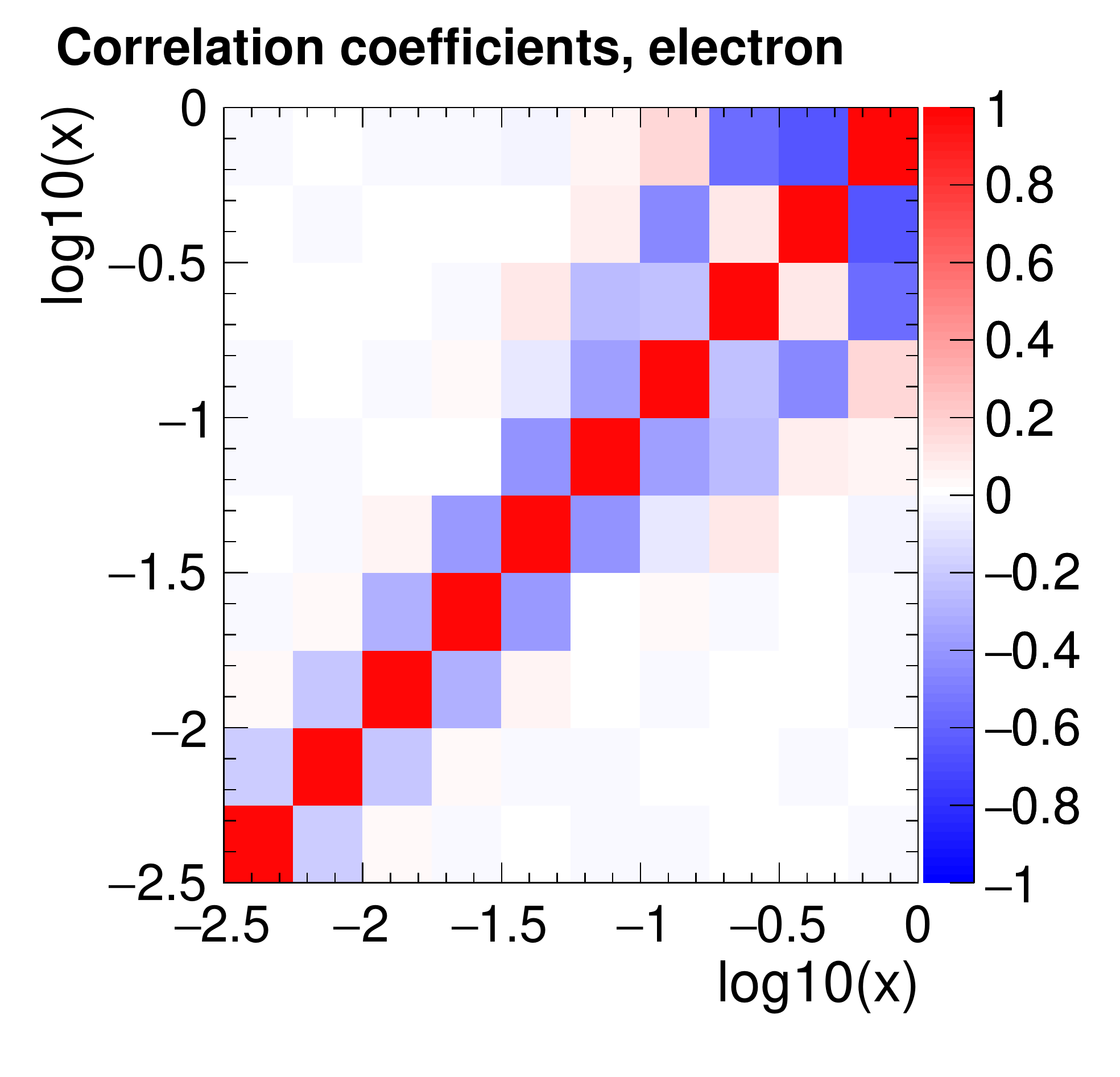} \\
\includegraphics[width=0.32\linewidth]{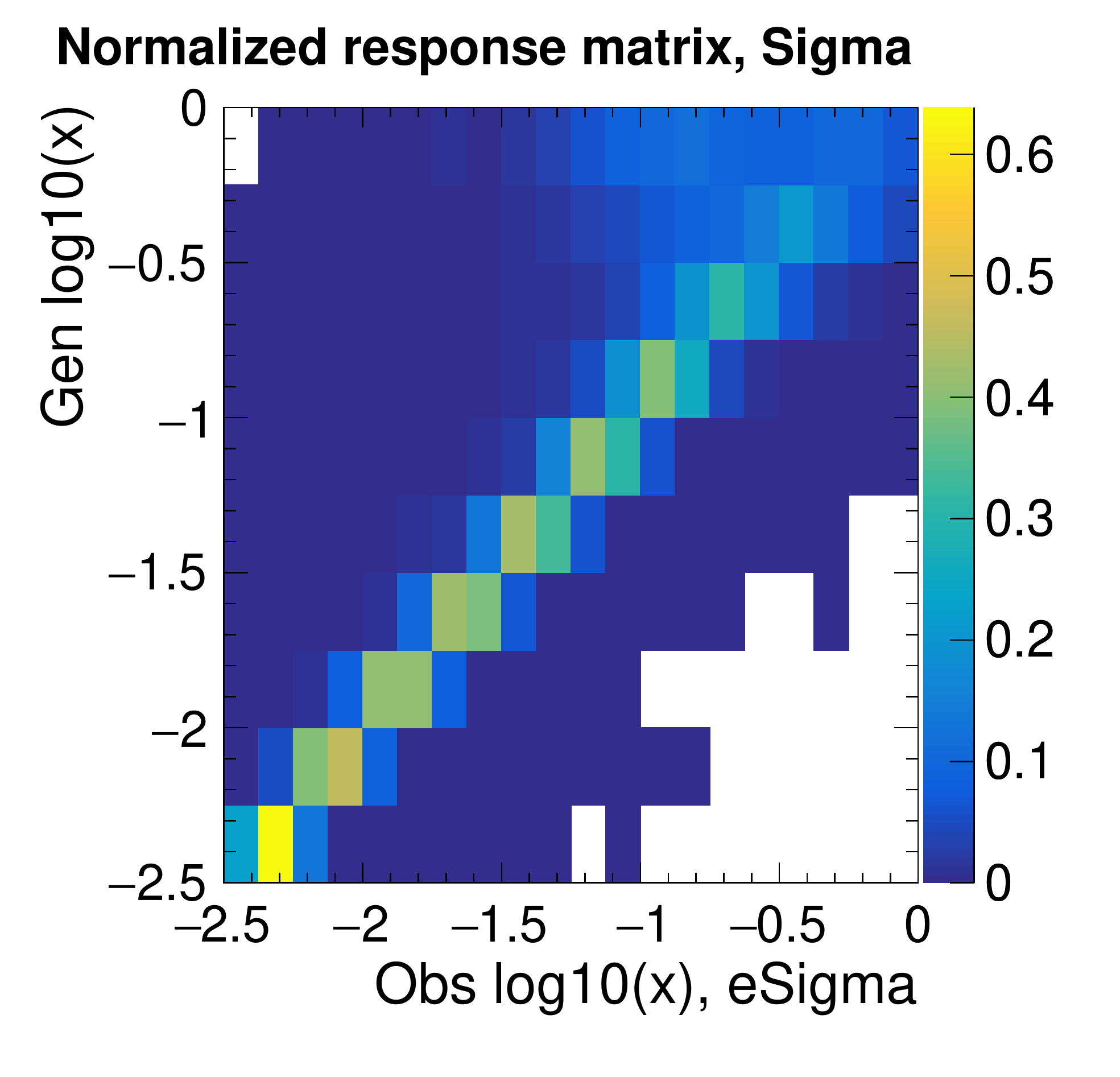}
\includegraphics[width=0.32\linewidth]{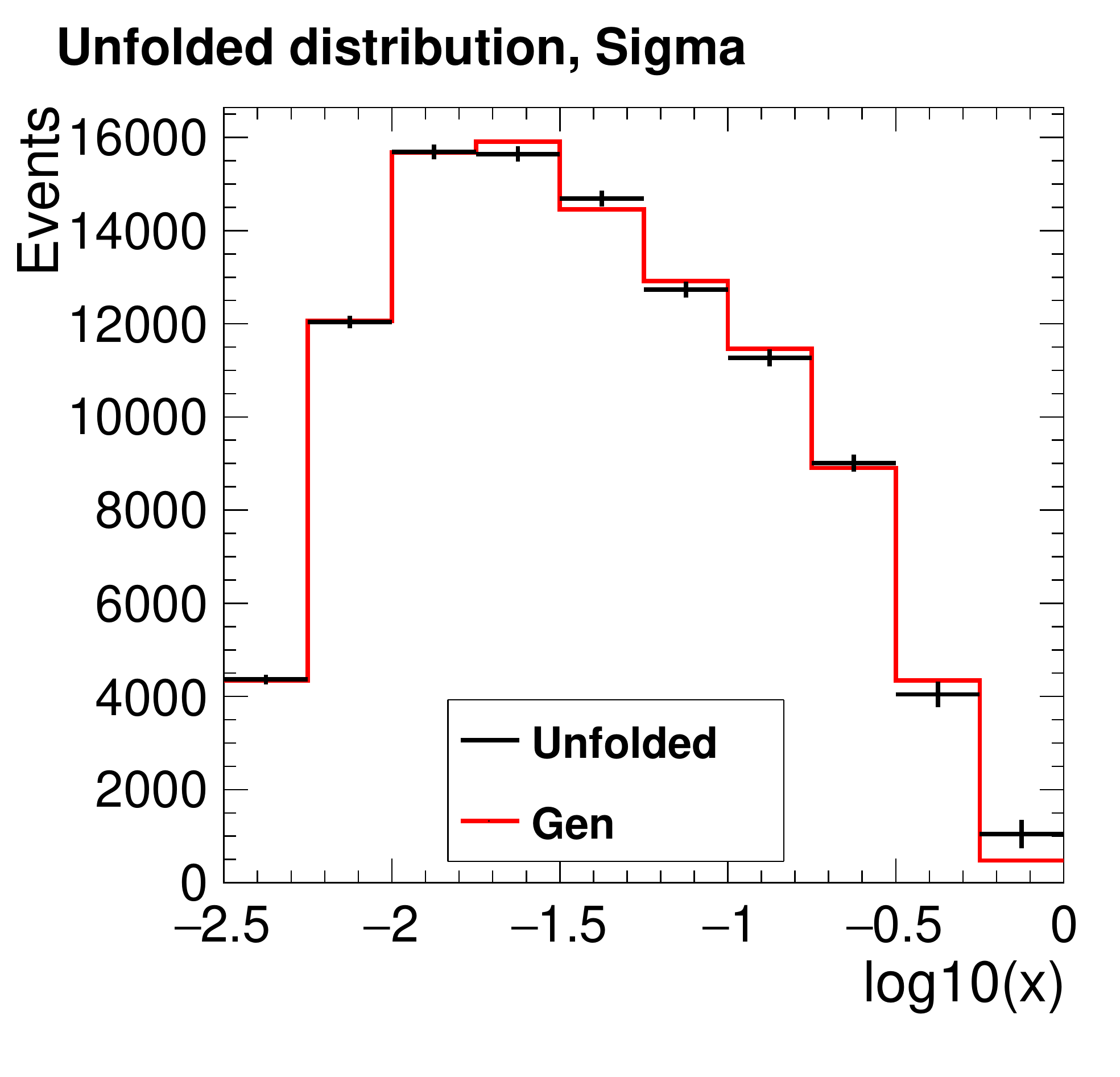}
\includegraphics[width=0.32\linewidth]{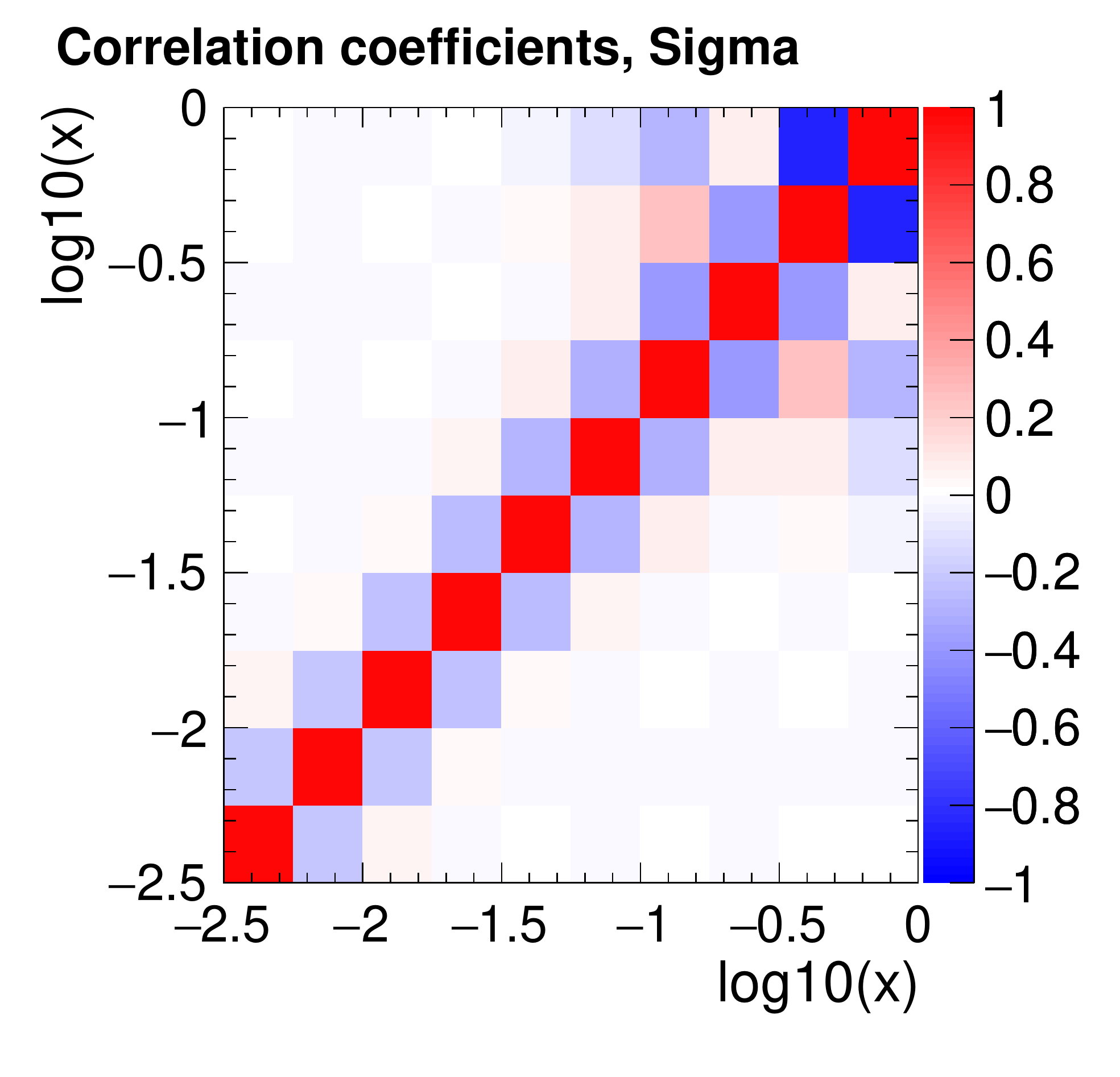} \\
\includegraphics[width=0.32\linewidth]{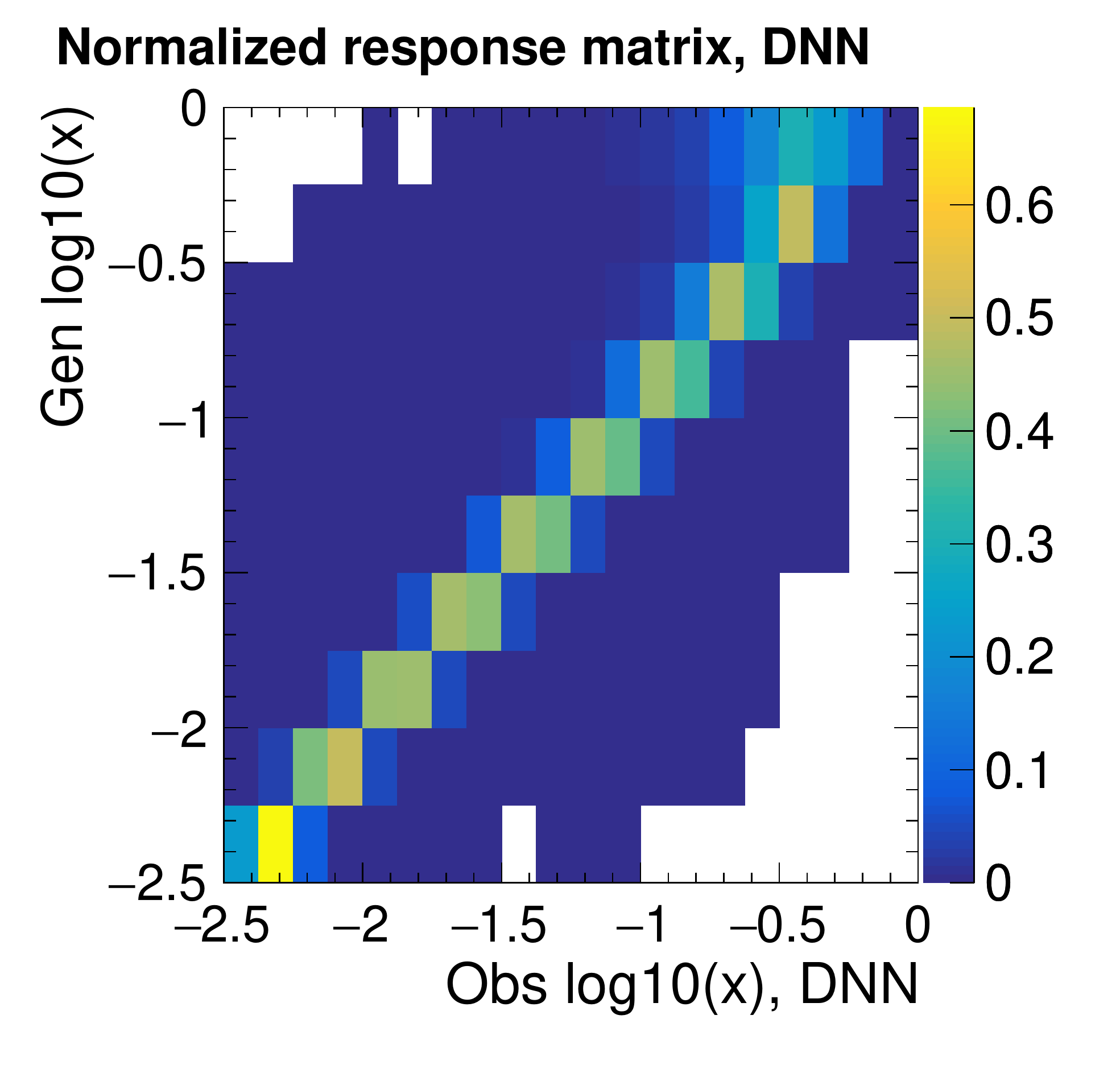}
\includegraphics[width=0.32\linewidth]{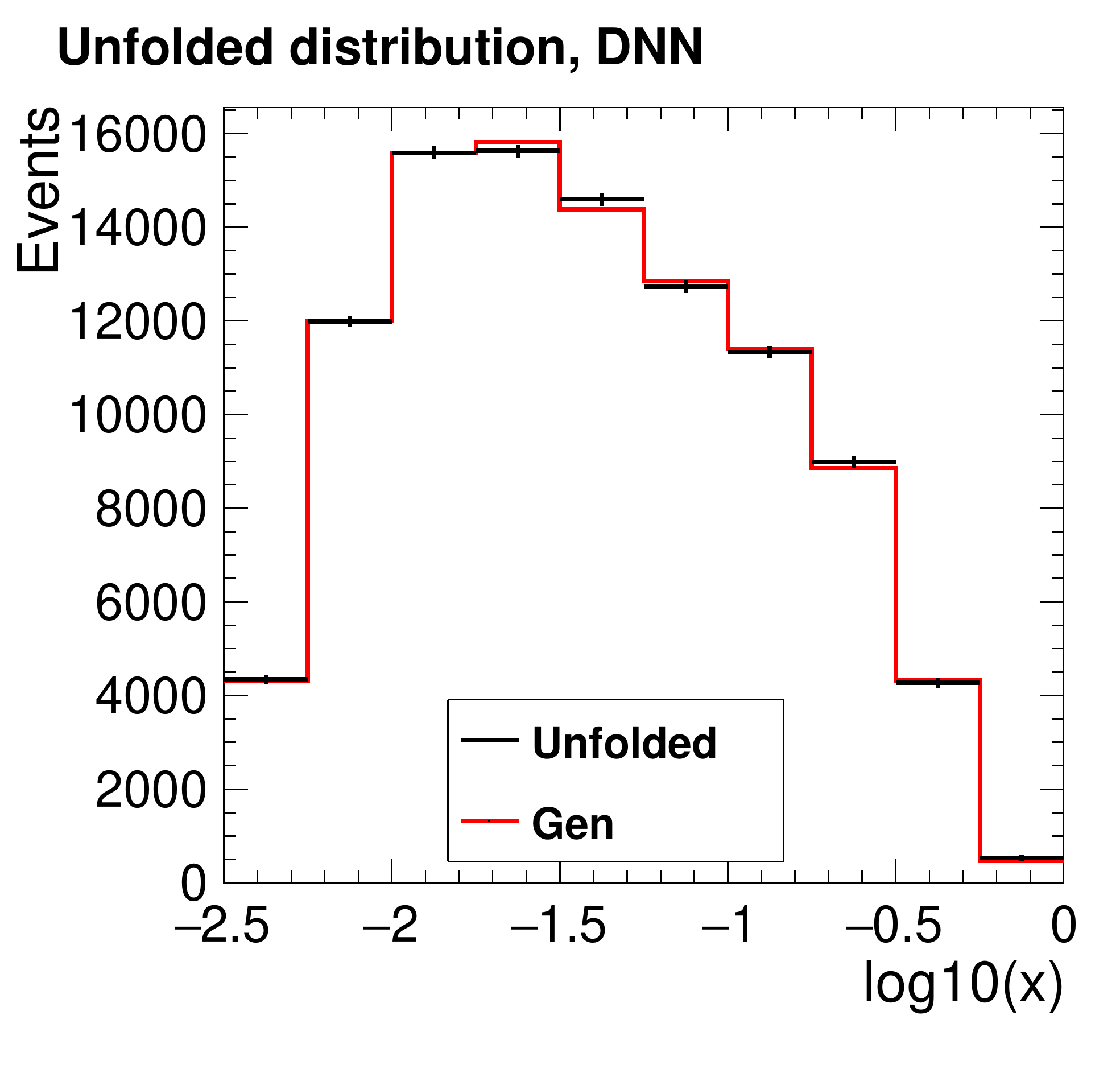}
\includegraphics[width=0.32\linewidth]{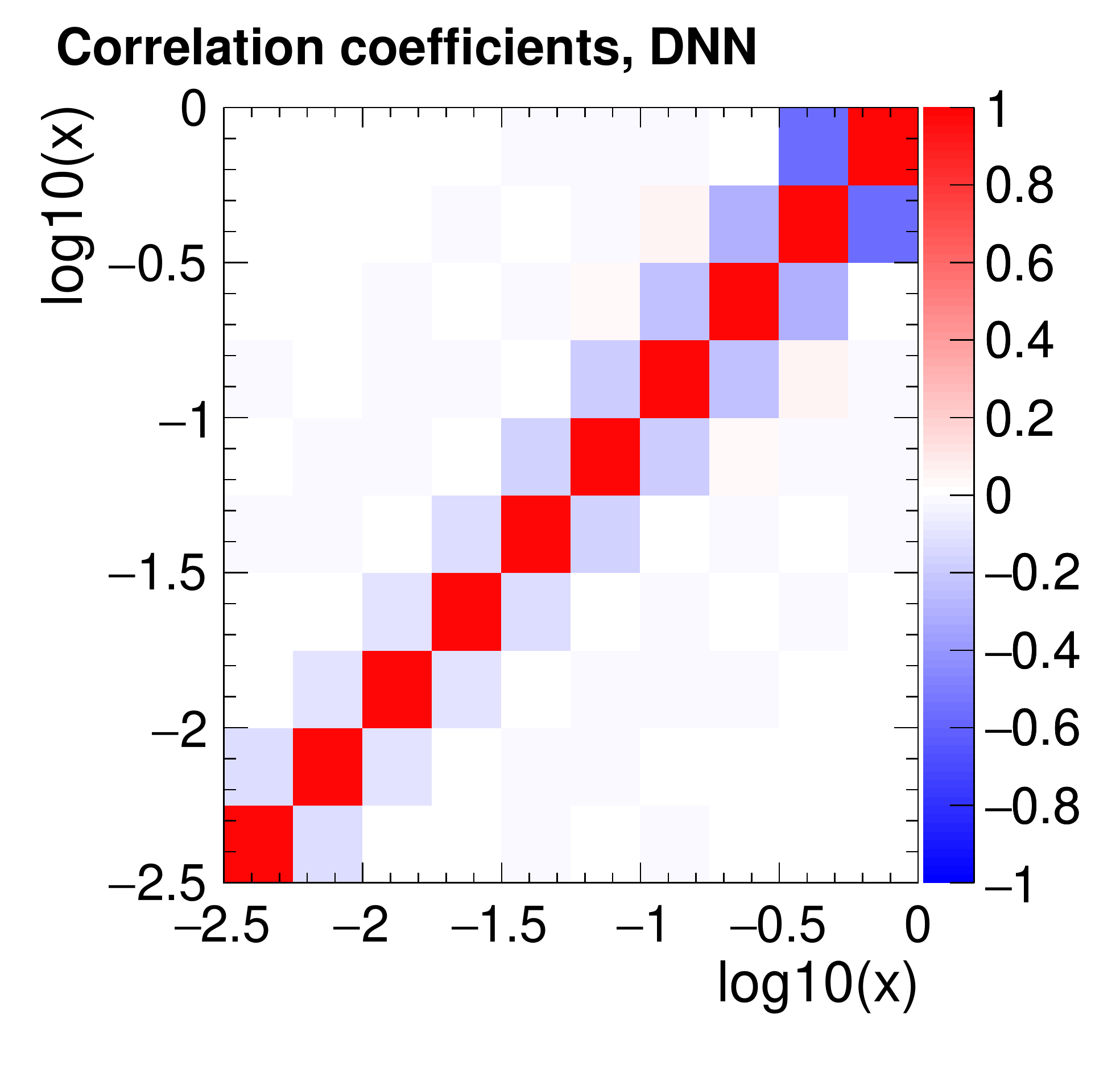} \\
   \caption{
      Examples of unfolding $\log_{10}(x)$ for samples of $10^5$ events.
      The response matrix (left), unfolded and gen distributions (middle), and unfolding
      correlation matrix (right) are shown for the
      electron (top), Sigma (middle), and DNN (bottom) methods.
   }
   \label{fig:dis-unfold-x}
   \end{center}
\end{figure}

\begin{figure}
   \begin{center}
\includegraphics[width=0.32\linewidth]{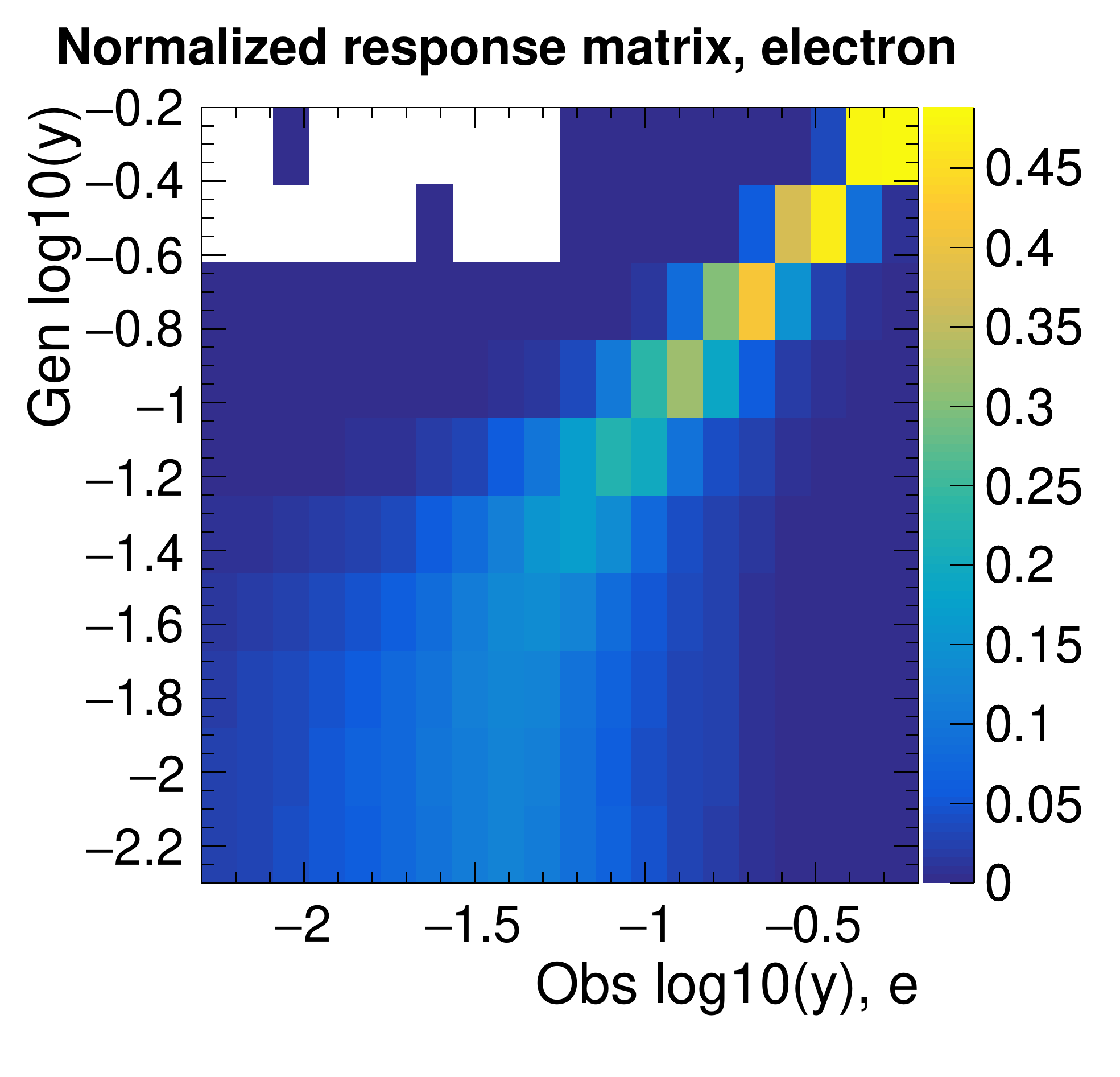}
\includegraphics[width=0.32\linewidth]{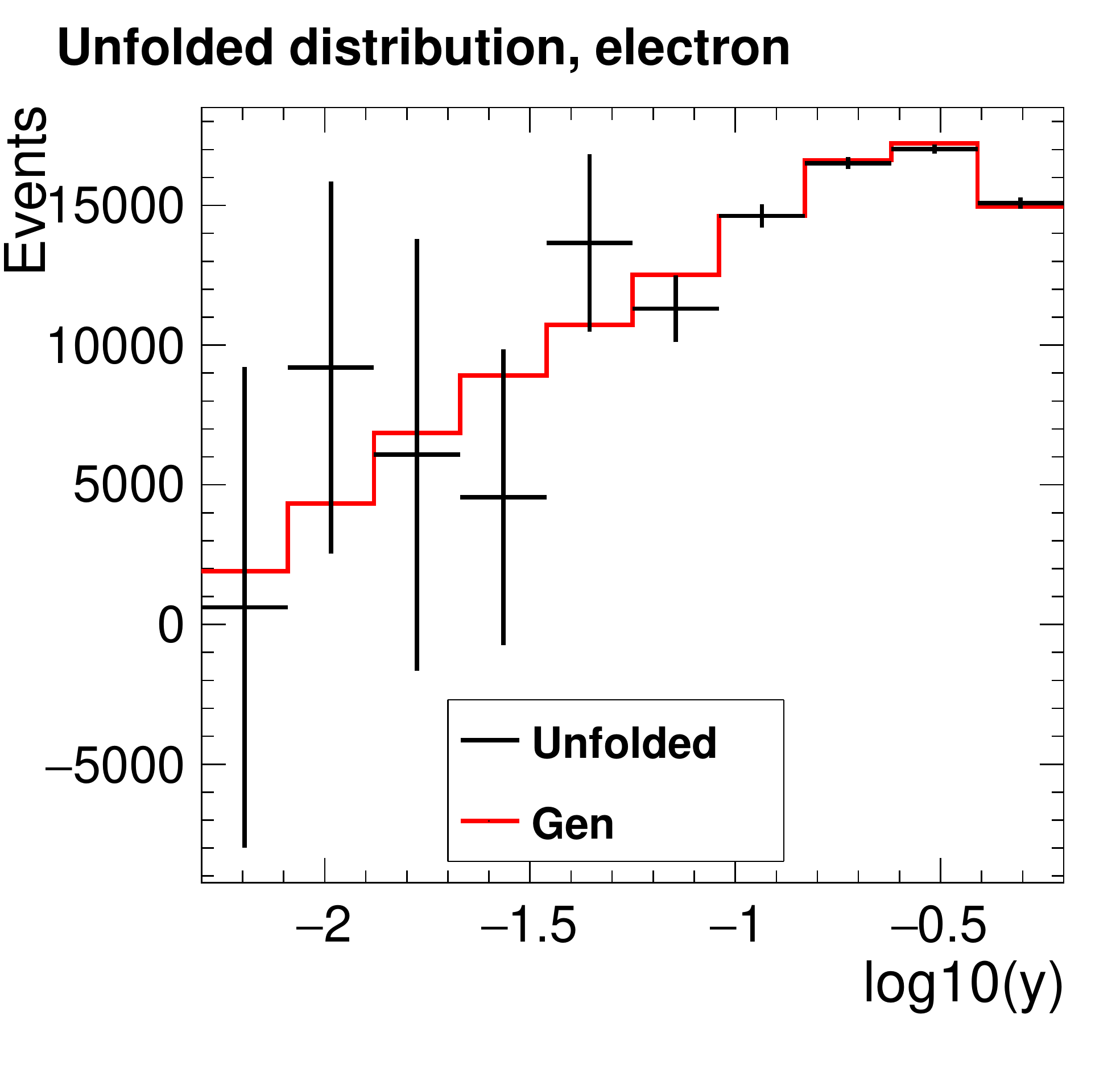}
\includegraphics[width=0.32\linewidth]{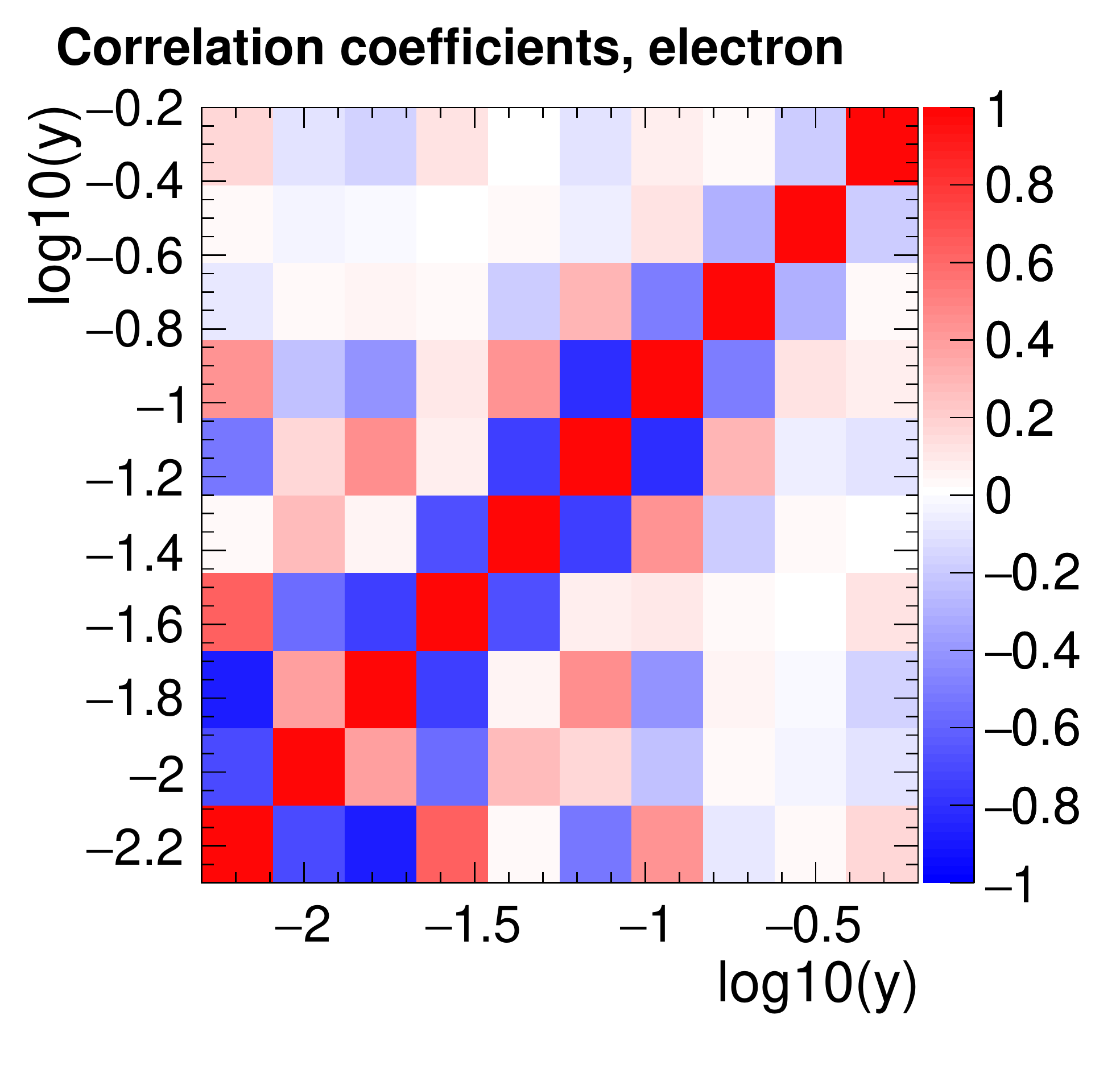} \\
\includegraphics[width=0.32\linewidth]{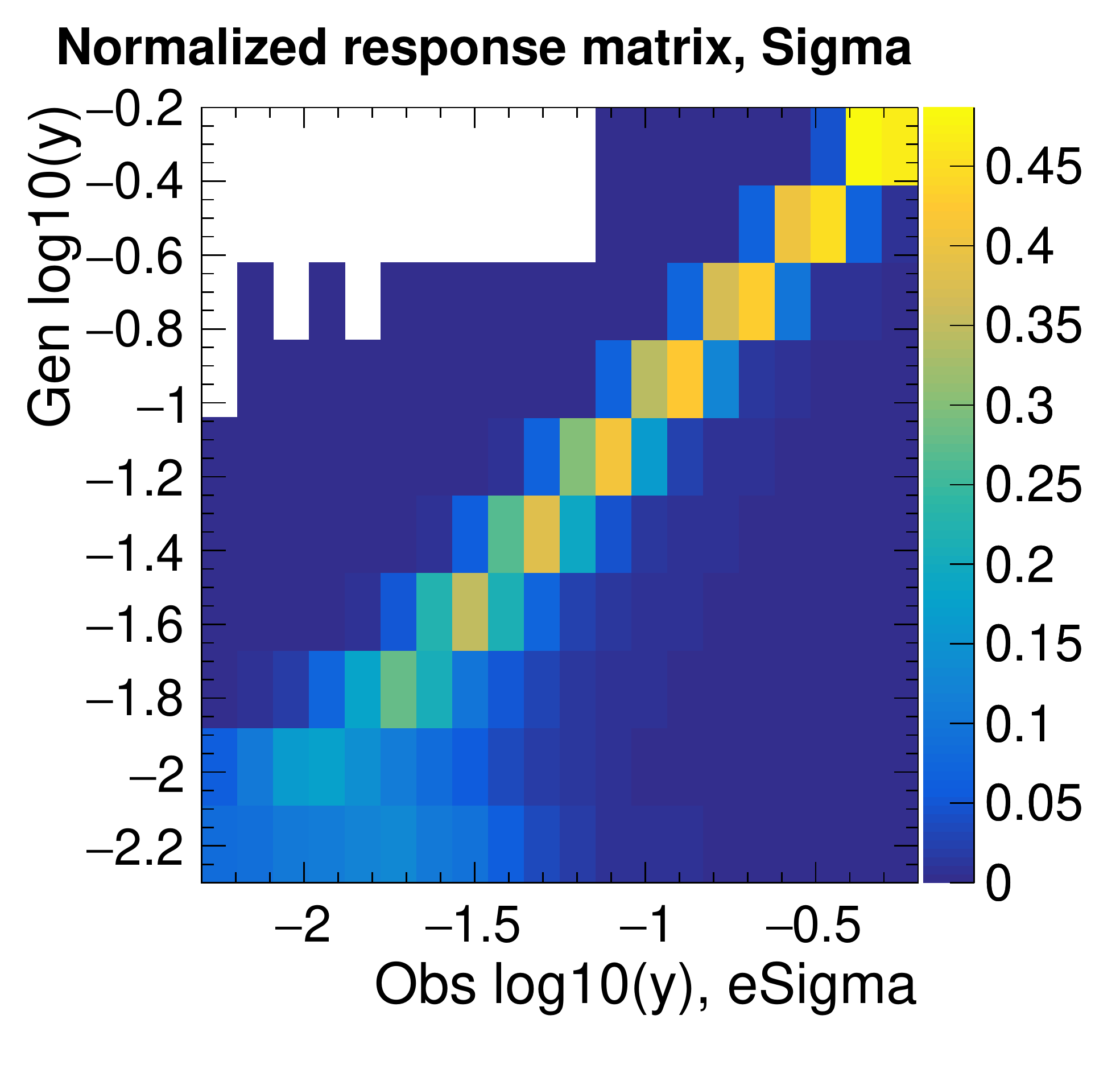}
\includegraphics[width=0.32\linewidth]{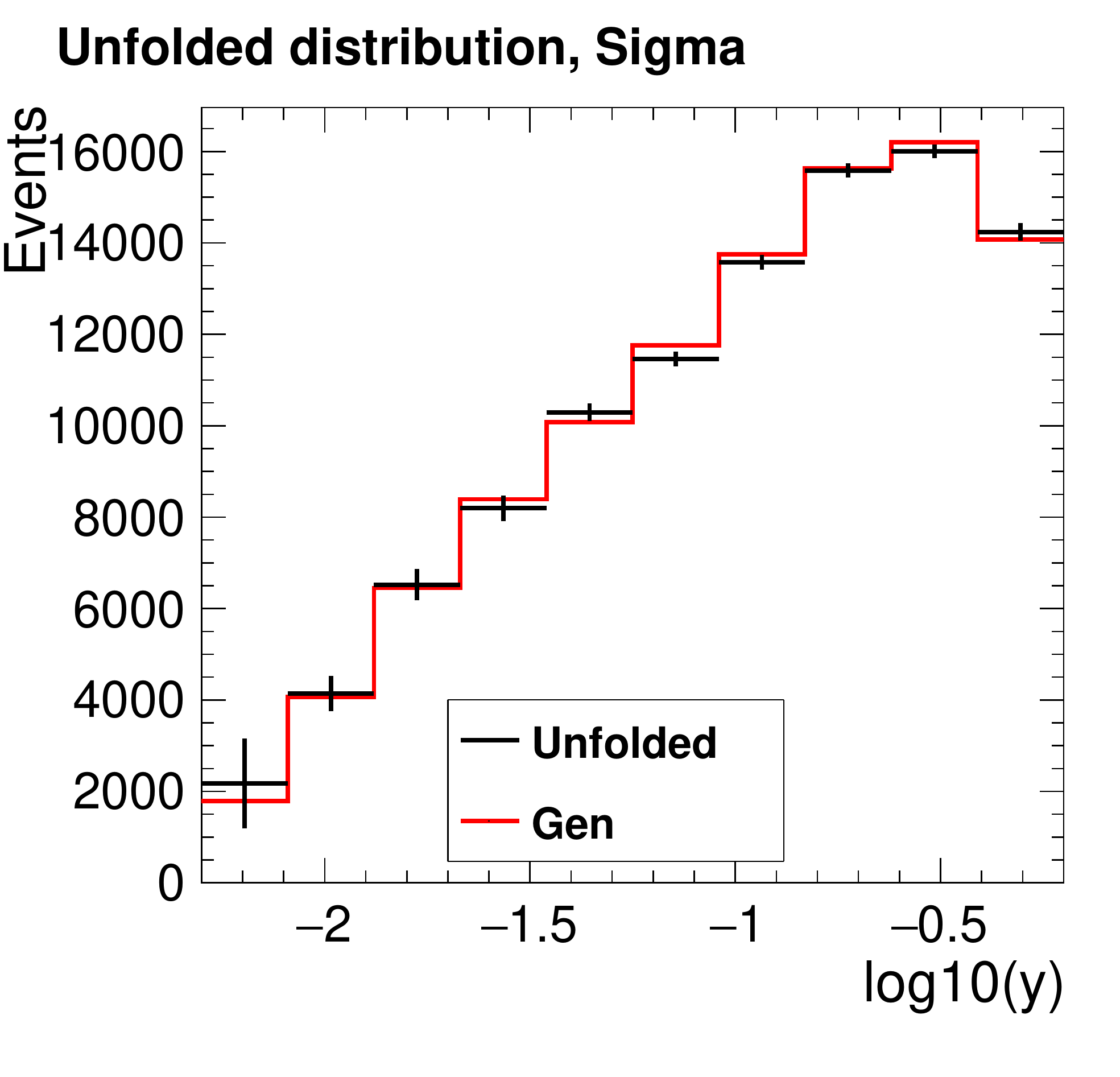}
\includegraphics[width=0.32\linewidth]{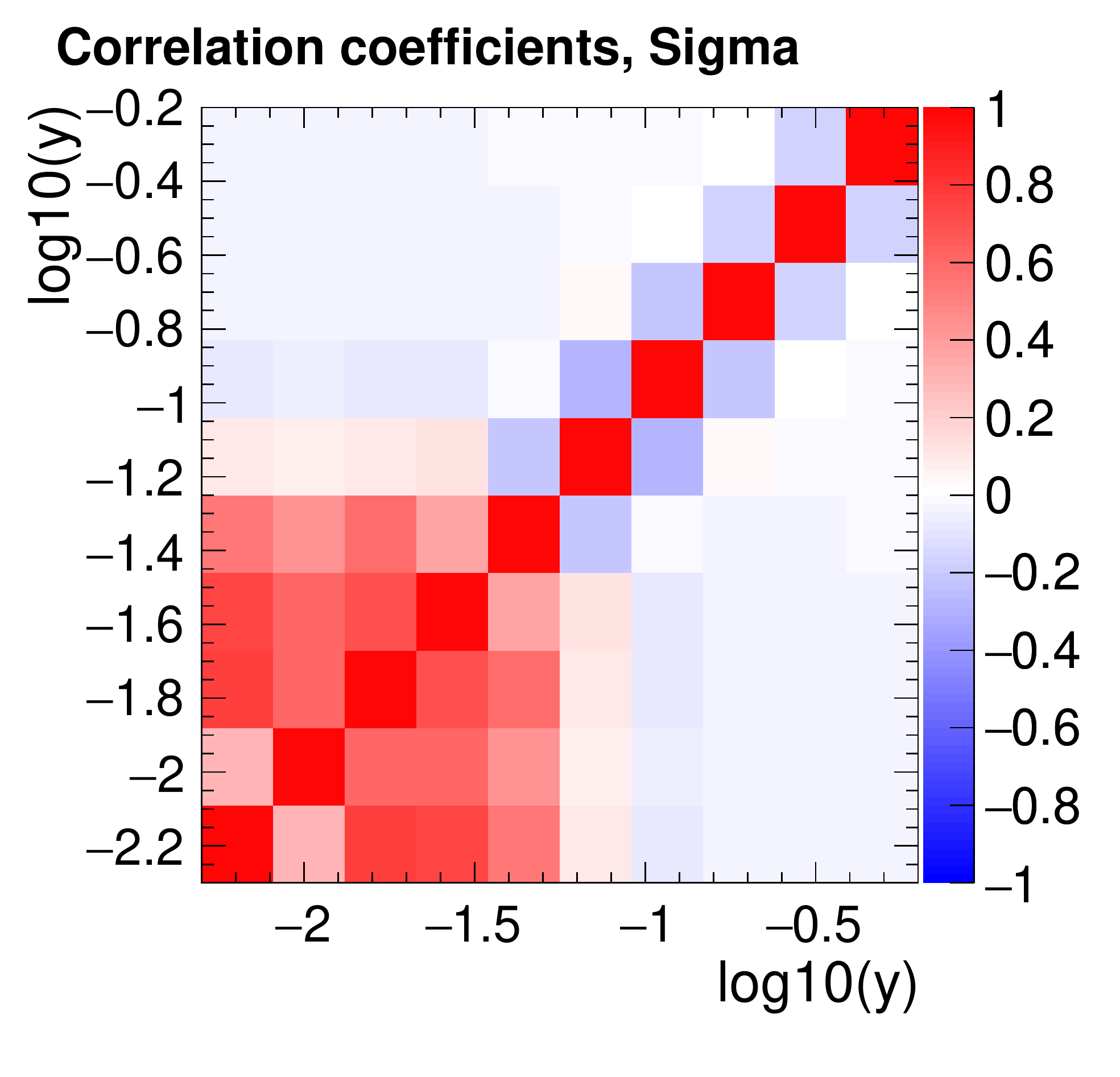} \\
\includegraphics[width=0.32\linewidth]{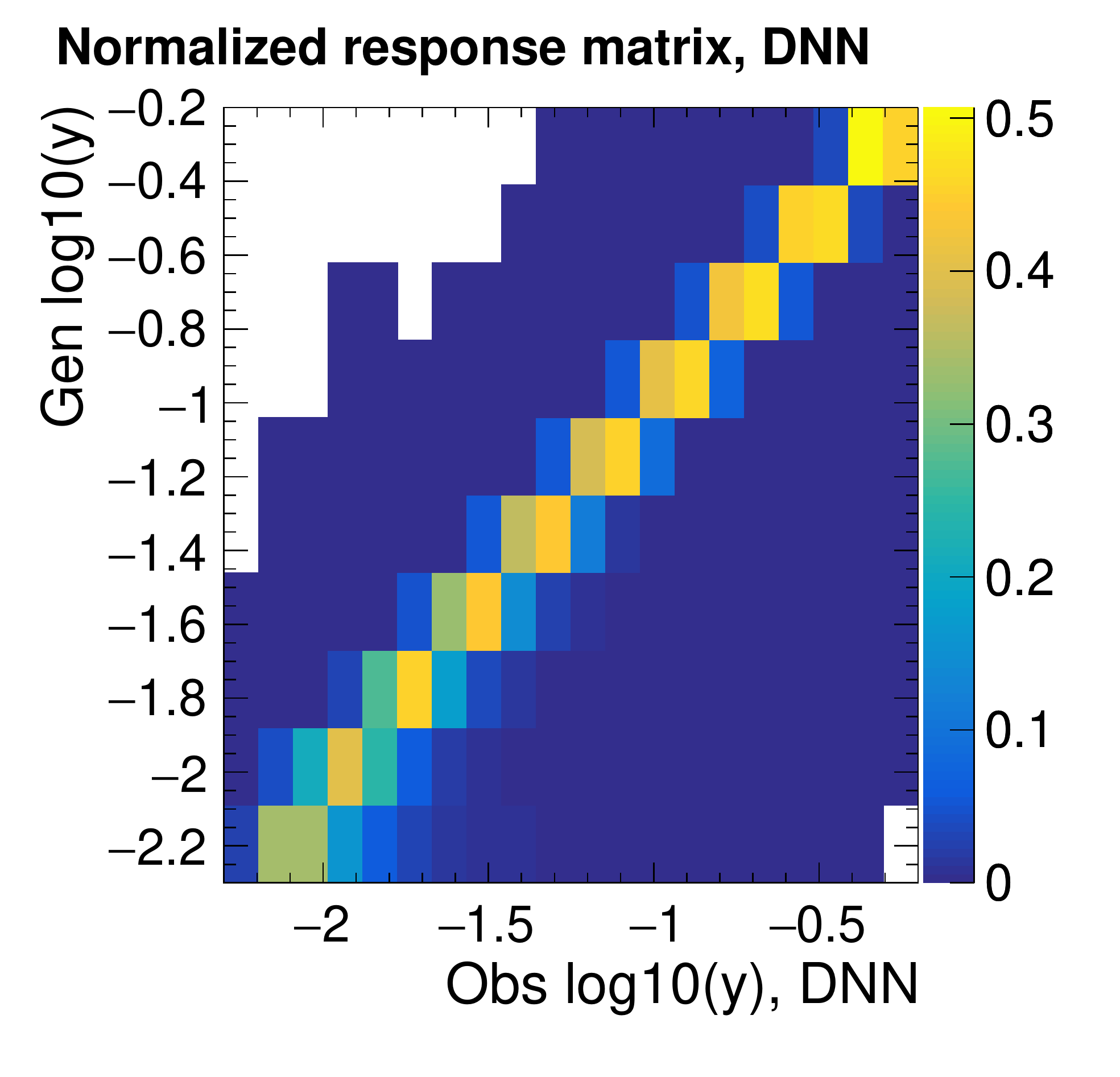}
\includegraphics[width=0.32\linewidth]{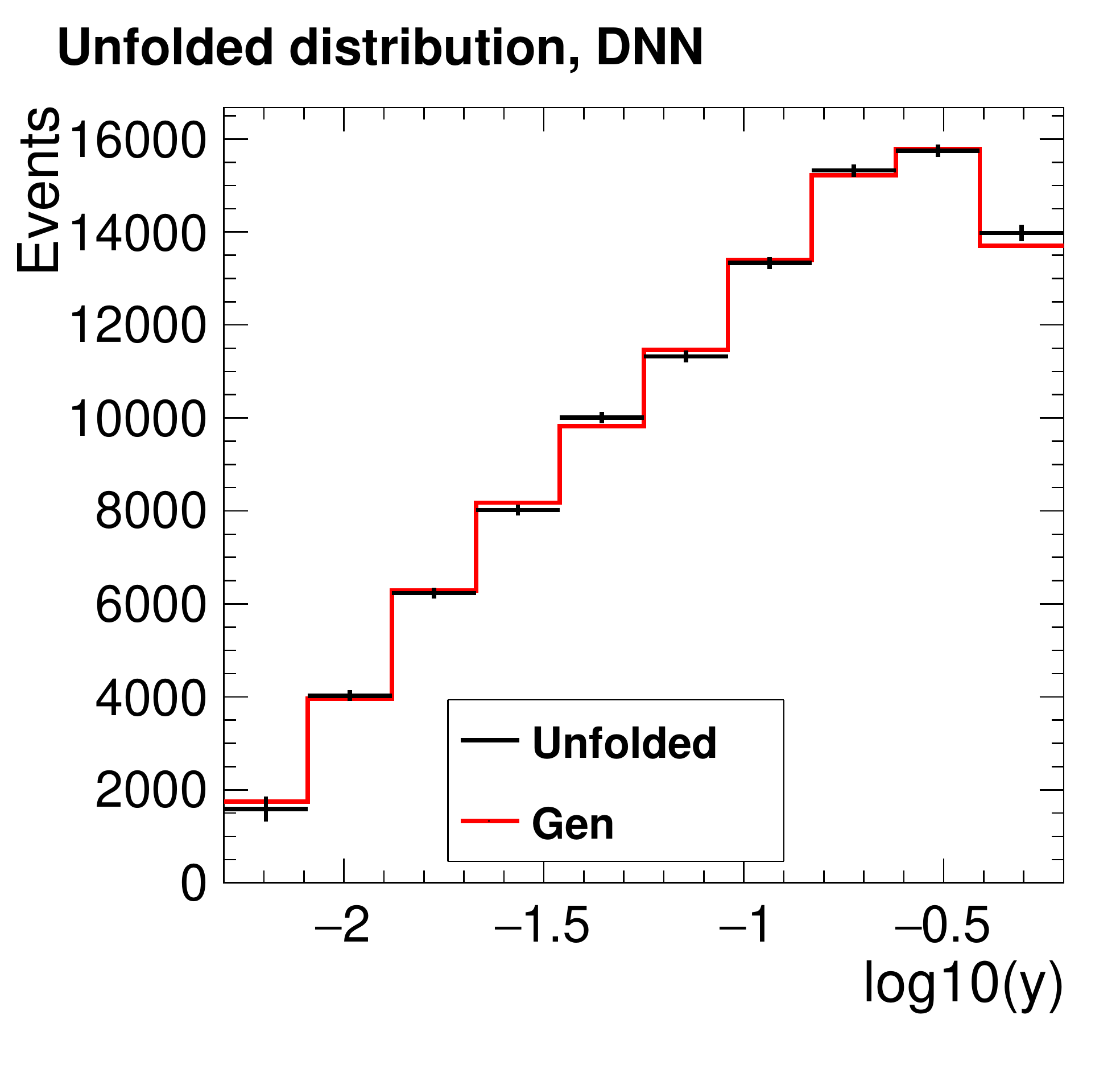}
\includegraphics[width=0.32\linewidth]{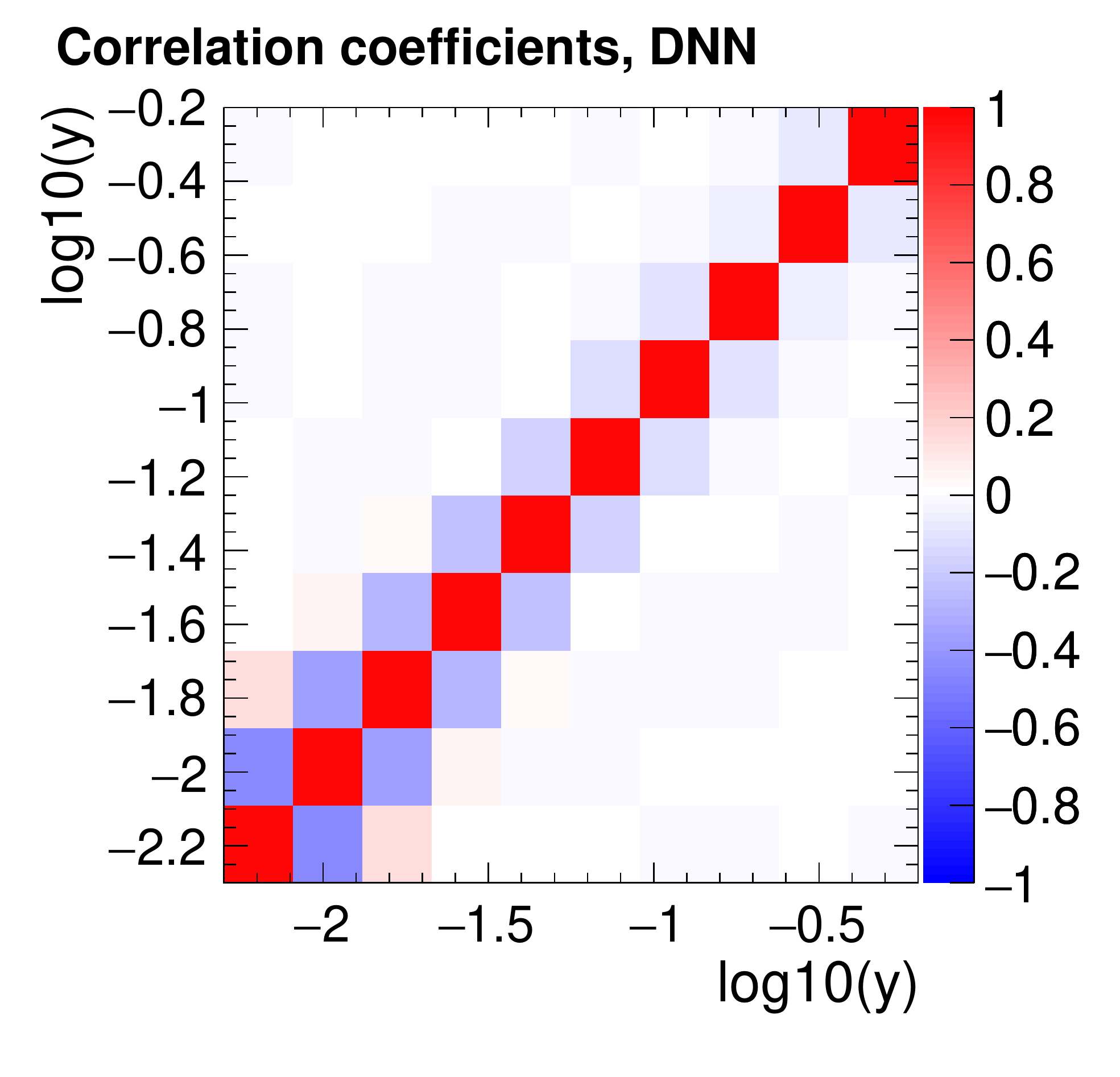} \\
   \caption{
      Examples of unfolding $\log_{10}(y)$ for samples of $10^5$ events.
      The response matrix (left), unfolded and gen distributions (middle), and unfolding
      correlation matrix (right) are shown for the
      electron (top), Sigma (middle), and DNN (bottom) methods.
      }
         \label{fig:dis-unfold-y}
   \end{center}
\end{figure}

Next, we study examples of unfolding in one dimension and quantify the properties of such ML-assisted observables.
The $10\times20$ migration matrices are defined for $x$ in the range $-2.5<\log_{10}(x)<1$, and for $y$ in the range $-2.3<\log_{10}(y)<-0.2$, and the regularization parameter is obtained from the $L$-curve scan.
Figures~\ref{fig:dis-unfold-x} and~\ref{fig:dis-unfold-y} show the normalized response matrices and the results of the unfolding for $\log_{10}(x)$ and $\log_{10}(y)$, respectively.
%
%
%
%
%
%
Each figure compares the DNN-based reconstruction method with two standard methods: the electron method, and Sigma method~\cite{Bassler:1994uq}.
The electron method only uses the scattered electron to compute the DIS kinematic quantities, and the Sigma method is built from HFS inputs. At particle level, all three reconstruction methods use the equivalent particle-level definition for $x$ and $y$~\cite{Arratia:2021tsq}. Hence, only the definition of $x_d$ and $\mathcal{O}_d(x_d)$ changes, but not $\mathcal{O}_p(x_p)$.

The response matrix for the DNN method is the most diagonal, as expected, since the training of the DNN was performed against the particle-level observables. In contrast, the two classical reconstruction methods have not only larger off-diagonal elements, but also some asymmetries.
All three methods have reasonably good resolution at low $x$, and at high $y$, while the electron and Sigma methods suffer from poor resolution at high $x$ and low $y$.
The so-called \emph{influence matrix} or \emph{posterior response matrix} can be considered as an alternative representation of the migration matrix and it includes the effect of the regularization~\cite{Schmelling:2022rci}. These matrices, and some properties, are discussed in Appendix~\ref{sec:PostMigMa}.

The unfolded distribution for each method agrees well with the generated (Gen) distribution within uncertainties, demonstrating good closure of the unfolding.
The classical methods have sizable statistical uncertainties at high $x$ or low $y$, while those from the DNN method are visibly smaller.
When regarding the related correlation matrices, the electron and Sigma methods show significant correlations beyond neighboring bins in areas where the resolution is poor, and even some correlations between distant bins are observed. In contrast, the DNN method has the most diagonal matrix, where the non-zero off-diagonal elements are mostly small correlations between neighboring bins, and no correlations between far apart bins is observed.
The average global correlation coefficients $\rho_\text{avg}$~\cite{Schmitt:2012kp} of these matrices are presented in Table~\ref{tab:rhoavg}, and a clear reduction is observed when using the DNN method.
\begin{table}[h]
  \centering
  \small
  \begin{tabular}{cccc}
    \toprule
    Observable & \multicolumn{3}{c}{Reconstruction method} \\
               & electron & Sigma & DNN \\
    \midrule
    $x$ & 0.692 & 0.611 & 0.400 \\
    $y$ & 0.837 & 0.707 & 0.442 \\
    \bottomrule
  \end{tabular}
  \caption{Value of the average global correlation coefficient $\rho_\text{avg}$ for the unfolding of $x$ and $y$ when using the electron, Sigma or DNN reconstruction method.}
  \label{tab:rhoavg}
\end{table}
In summary, the DNN method directly improves the resolution, which is seen from reduced correlations, and improves mis-reconstruction (e.g.\ in the presence of QED radiation), which is seen from the absence of distant correlations.

\begin{figure}
   \begin{center}
\includegraphics[width=0.32\linewidth]{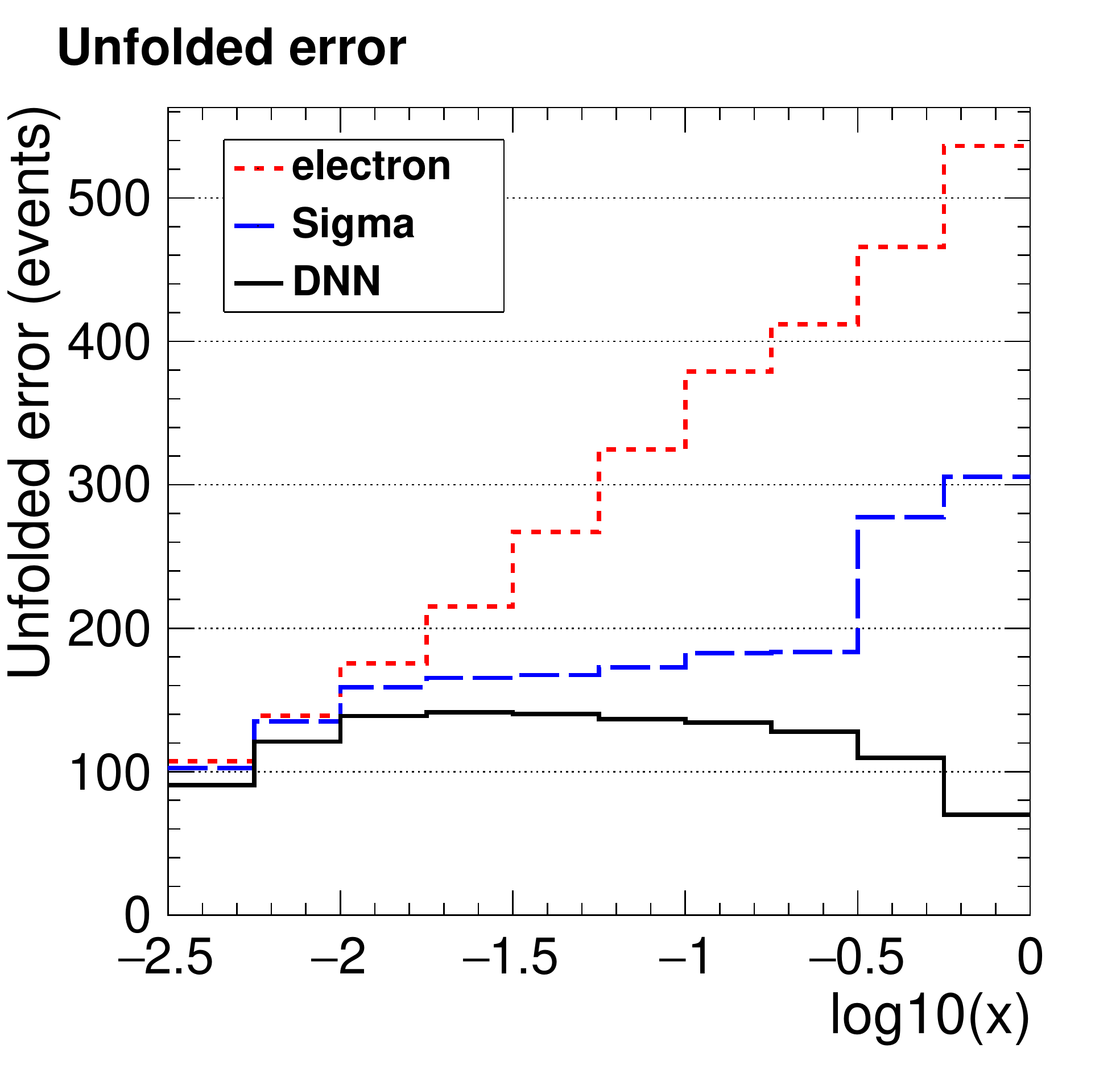}
\includegraphics[width=0.32\linewidth]{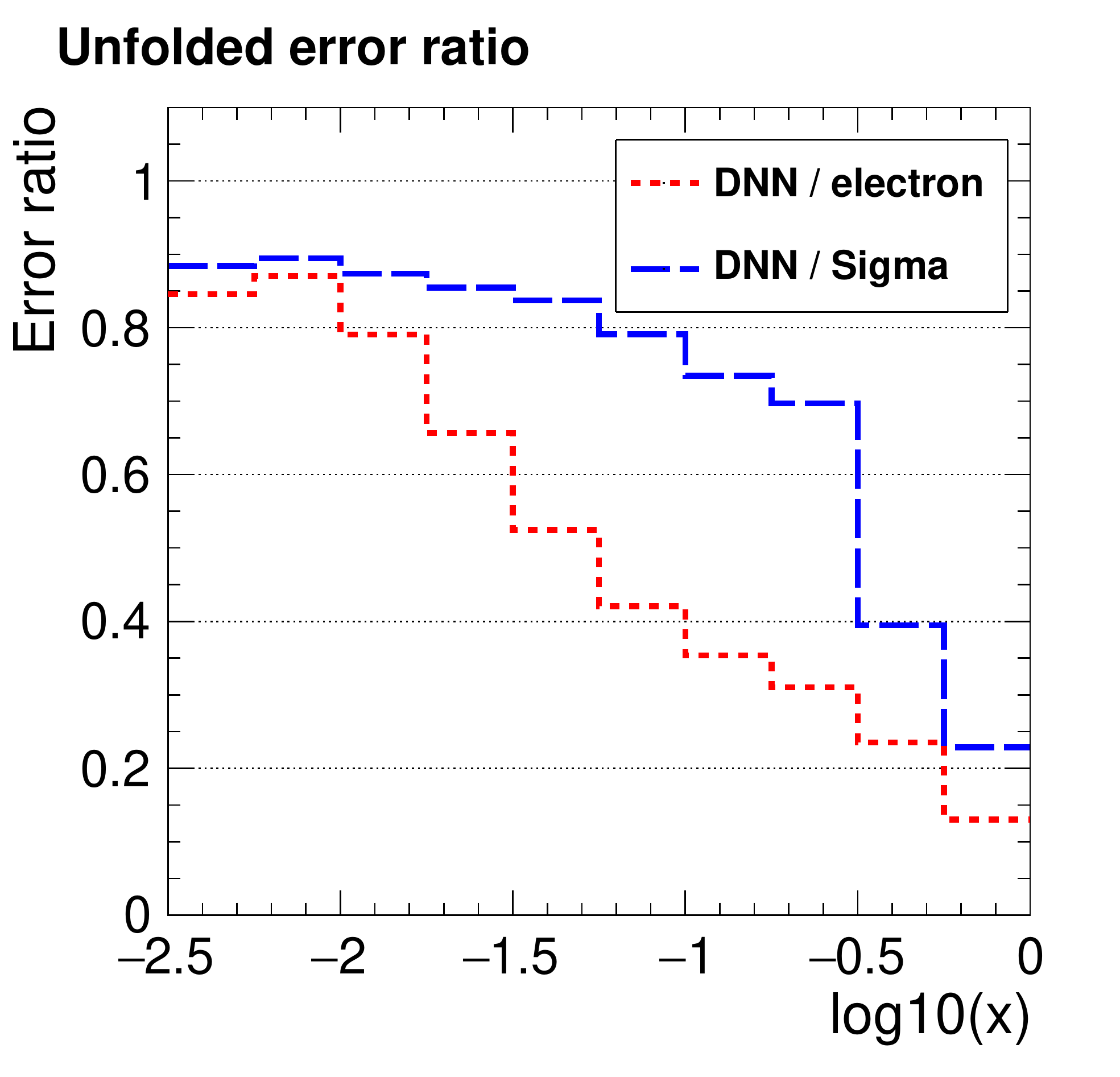}
\includegraphics[width=0.32\linewidth]{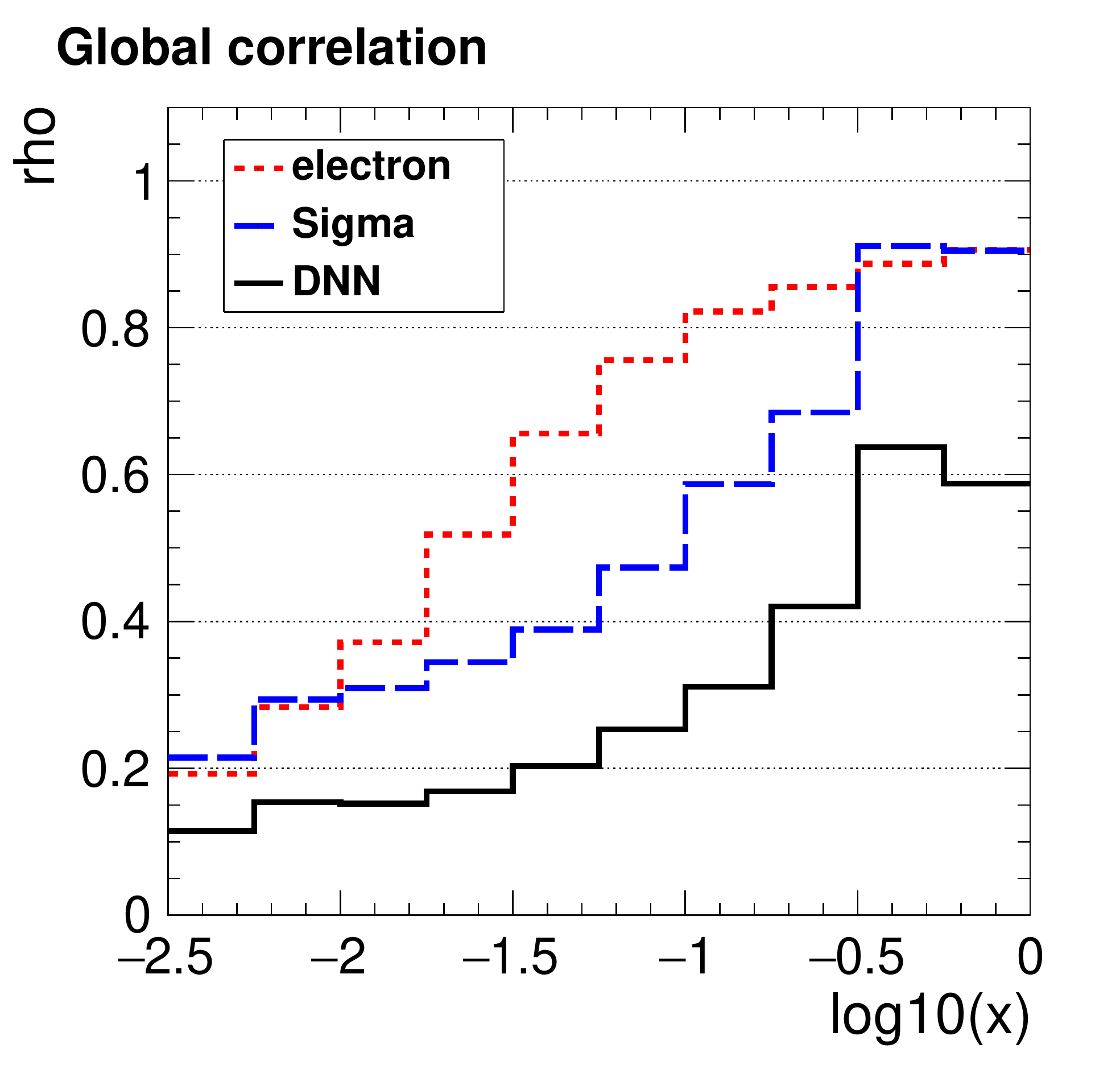}
\includegraphics[width=0.32\linewidth]{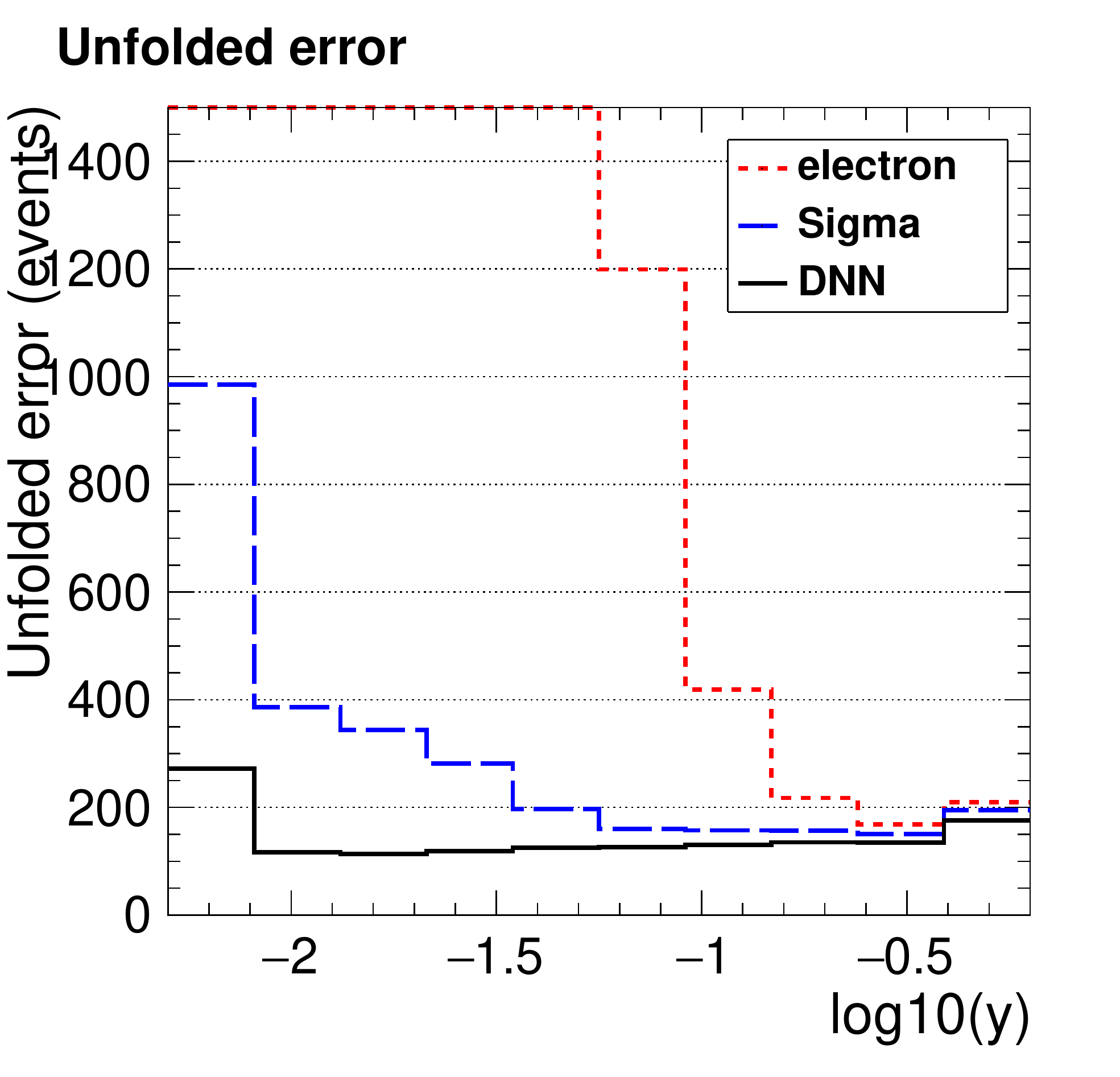}
\includegraphics[width=0.32\linewidth]{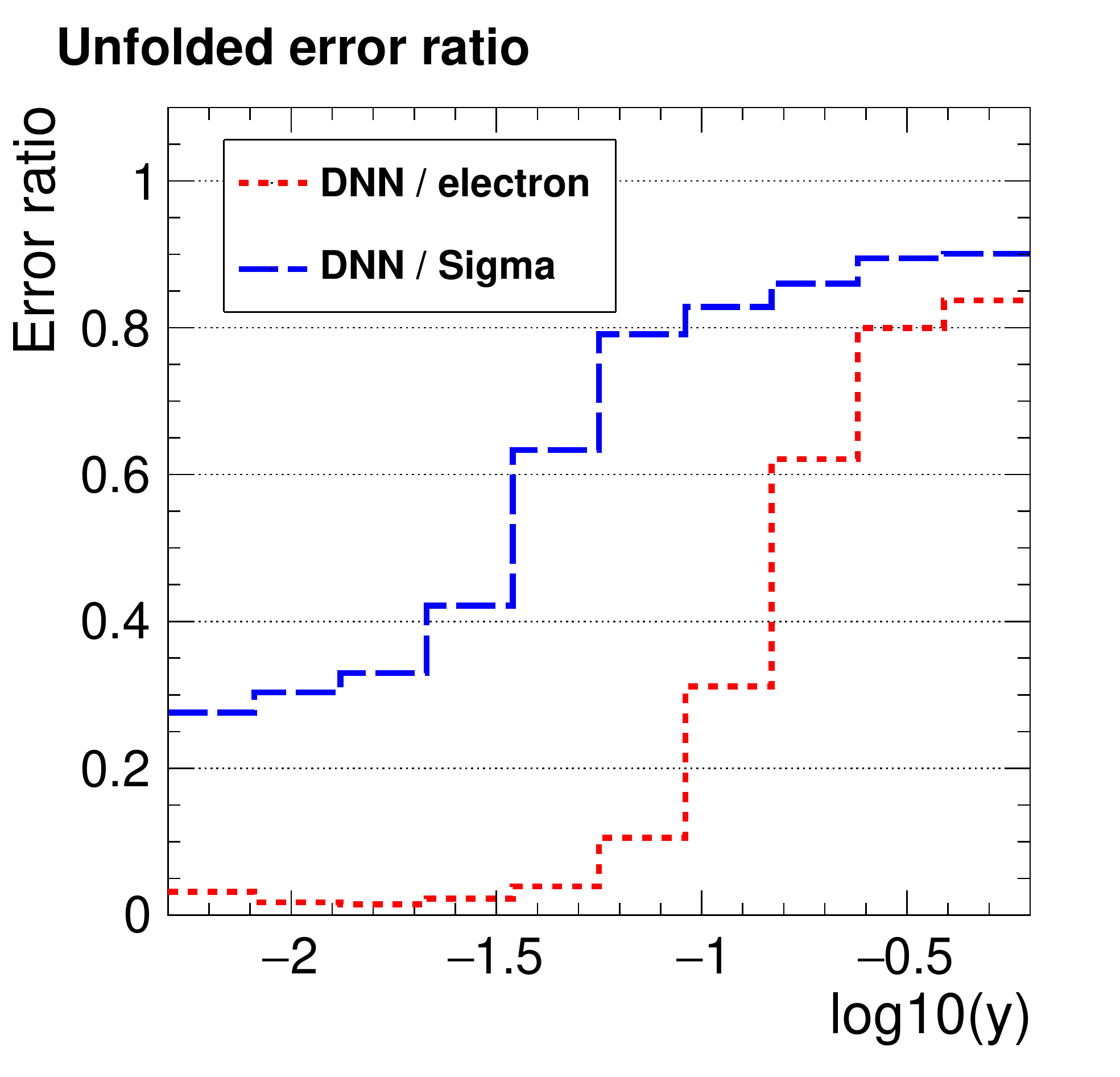}
\includegraphics[width=0.32\linewidth]{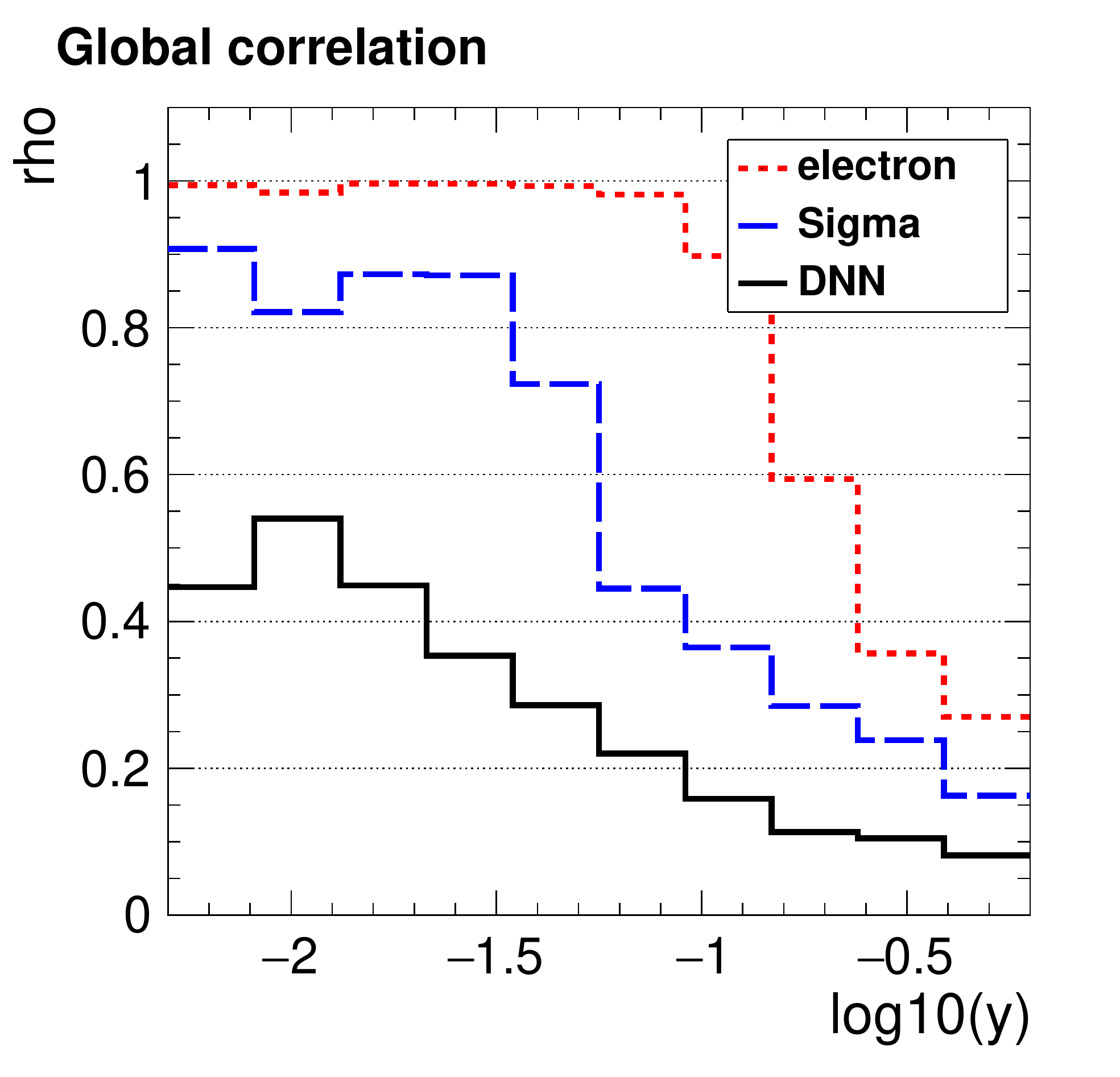}
\caption{
   Comparison of unfolding results for $\log_{10}(x)$ (top) and $\log_{10}(y)$ (bottom) using the electron, Sigma, and DNN methods.
   Distributions for the statistical error on the unfolded result (left), the ratio of errors for the methods (middle), and the global correlation coefficient (right) are shown.
}
\label{fig:dis-unfold-errors-and-correlations-comparison}
   \end{center}
\end{figure}

Figure~\ref{fig:dis-unfold-errors-and-correlations-comparison} examines the statistical errors and the global correlation coefficients~\cite{Schmitt:2012kp} of the unfolding results for the electron, Sigma, and DNN methods in greater detail.
At low $x$ and high $y$, all three methods have good resolution, which results in similar size statistical errors. However, the uncertainties of the DNN methods are still about 10\,\% smaller at lowest $x$ or highest $y$, and the global correlation coefficient for the DNN method is found to be smaller by about a factor of 2 or more.
At high $x$ and low $y$, the DNN method shows significantly lower statistical errors and smaller global correlations in all bins.
The uncertainties of the electron method become large, due to the particularly large global correlation coefficients at low $y$.
Furthermore, with higher $x$ or lower $y$ both classical methods have a successively increasing uncertainty, whereas the DNN shows continuously reduced uncertainties towards these kinematic regions (except the bin at lowest $y$).

Lastly, we study the properties of the unfolding when the migration matrix is obtained from a different Monte Carlo event generator as the (pseudo-)data sample.
Such tests are of great importance in real data analyses, since a possible bias of the unfolding method due the selection of a certain physics generator is assessed.
The differences of the two Monte Carlo models, when unfolded with the simulation from the other generator, is commonly considered as an uncertainty in the data analysis and is denoted as \emph{(generator) model systematic uncertainty}.
In order to reveal generator model systematic uncertainties at a statistically significant level, the size of the simulated event sample for these studies is $10^7$, which is 100 times larger than those used in Figures~\ref{fig:dis-unfold-x} through~\ref{fig:dis-unfold-errors-and-correlations-comparison}.
The technical closure plots when using the \textsc{Rapgap} event generator are displayed in Appendix~\ref{sec:closureRapgap}, and an excellent closure of the unfolded result is observed for all three reconstruction methods.
\begin{figure}
   \begin{center}
\includegraphics[width=0.32\linewidth]{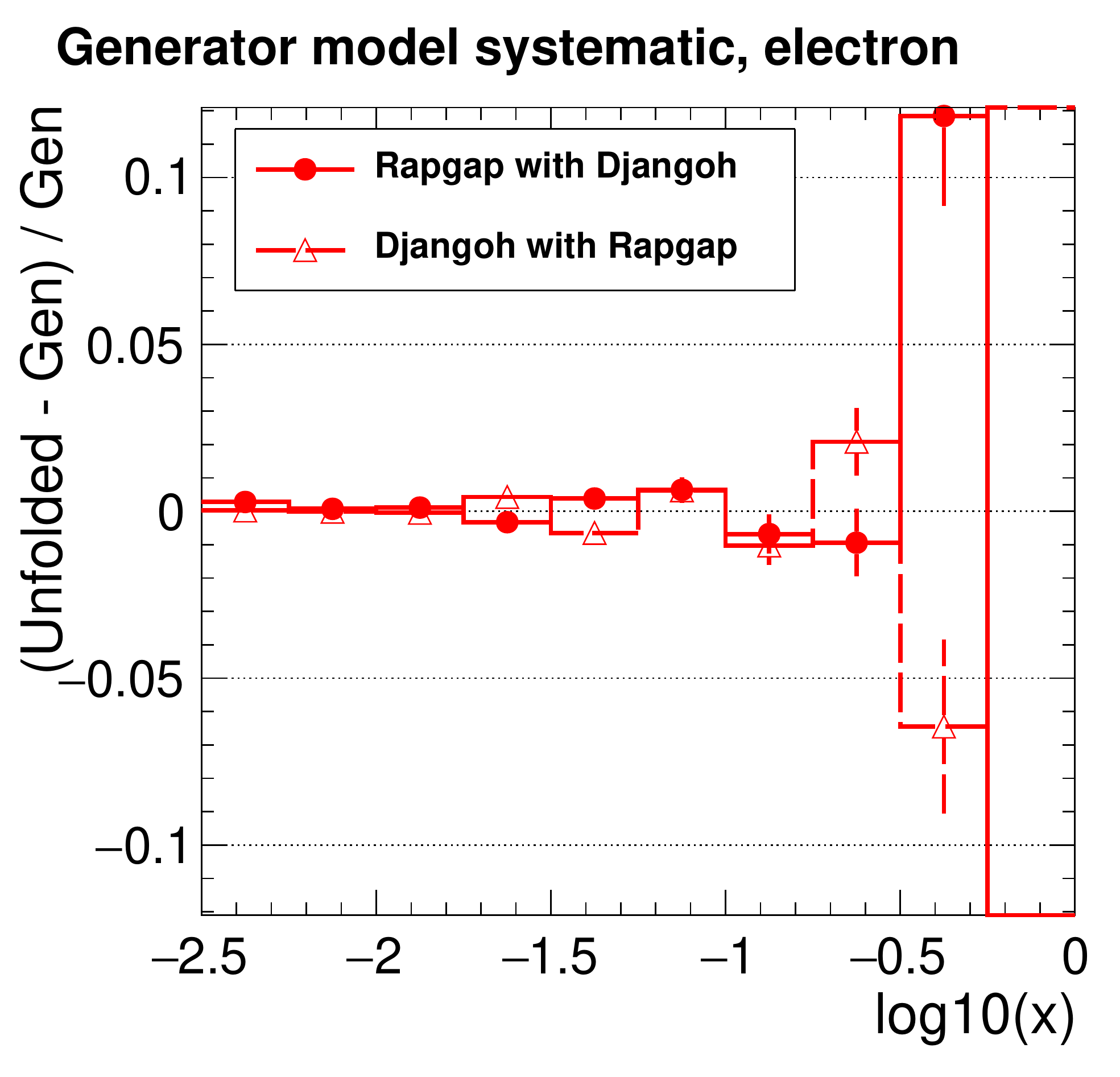}
\includegraphics[width=0.32\linewidth]{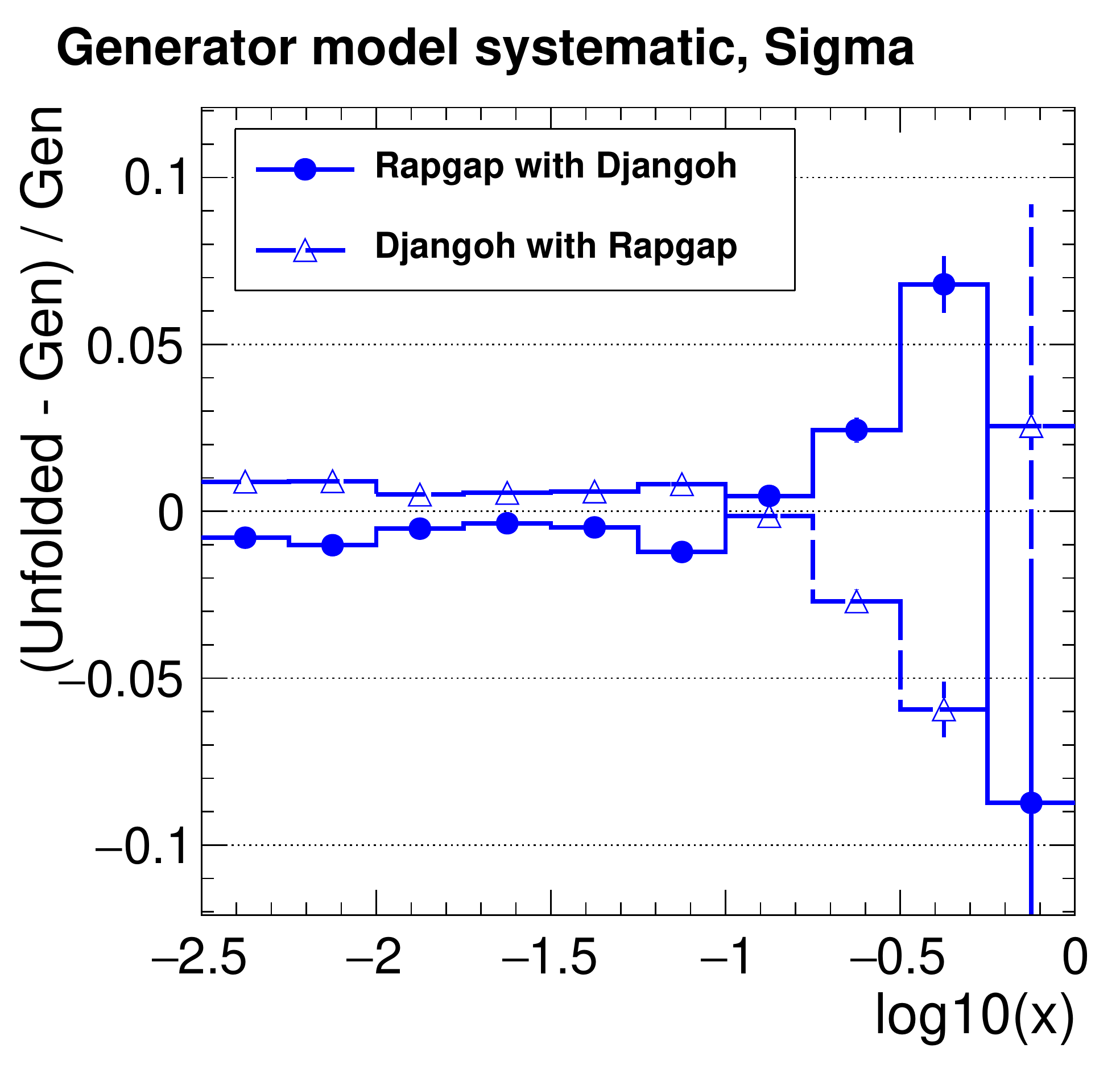}
\includegraphics[width=0.32\linewidth]{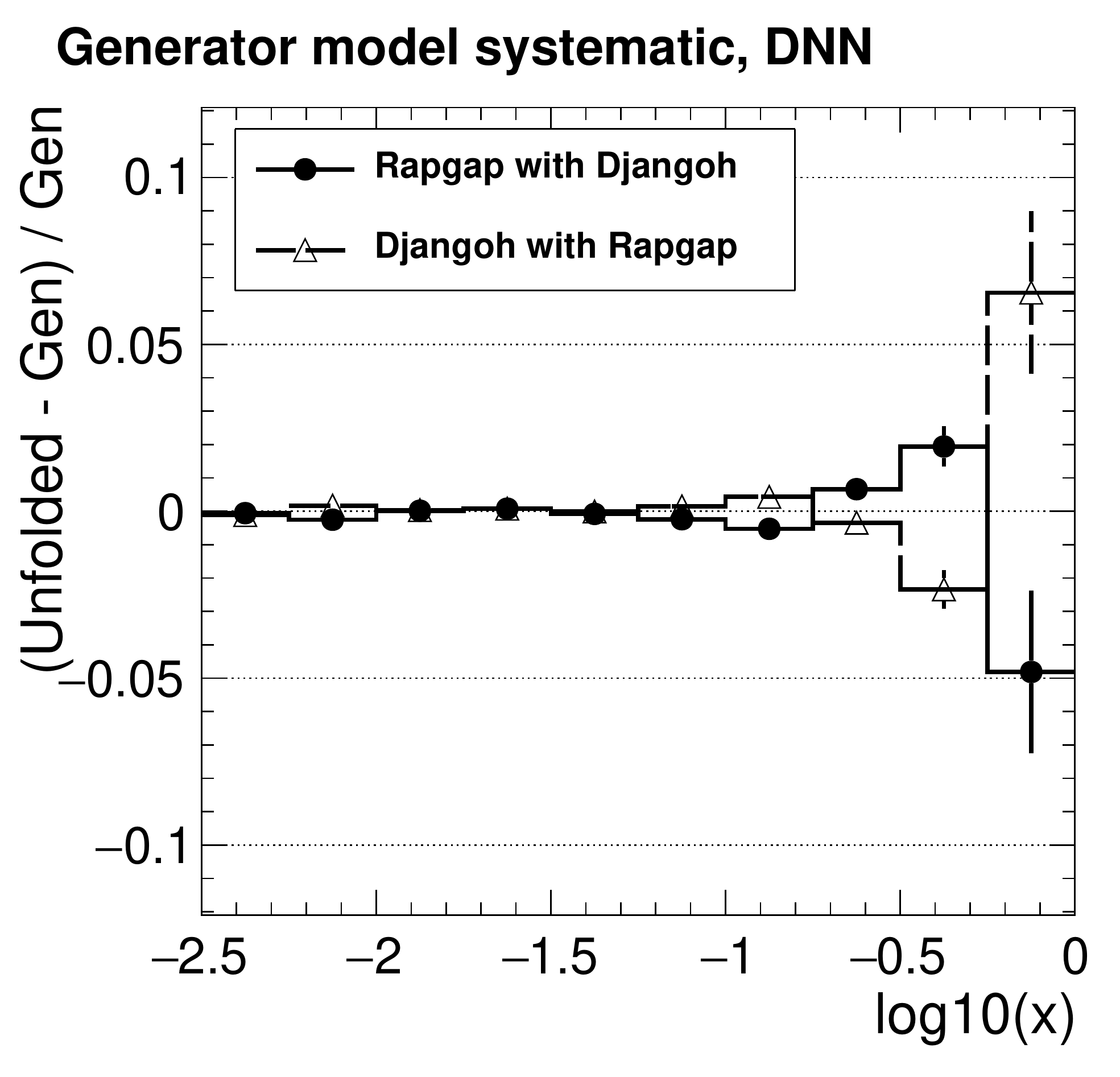} \\
\includegraphics[width=0.32\linewidth]{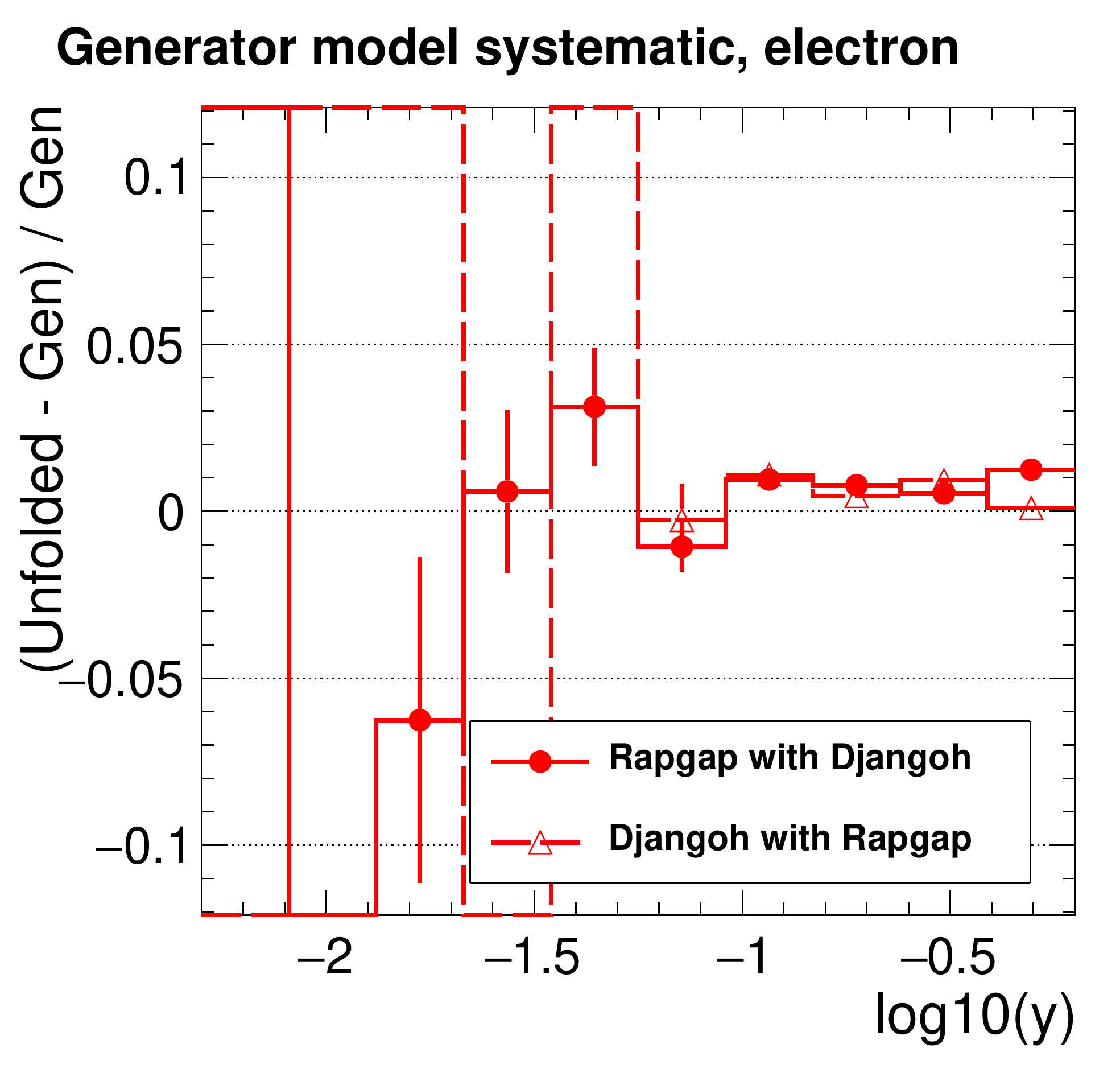} 
\includegraphics[width=0.32\linewidth]{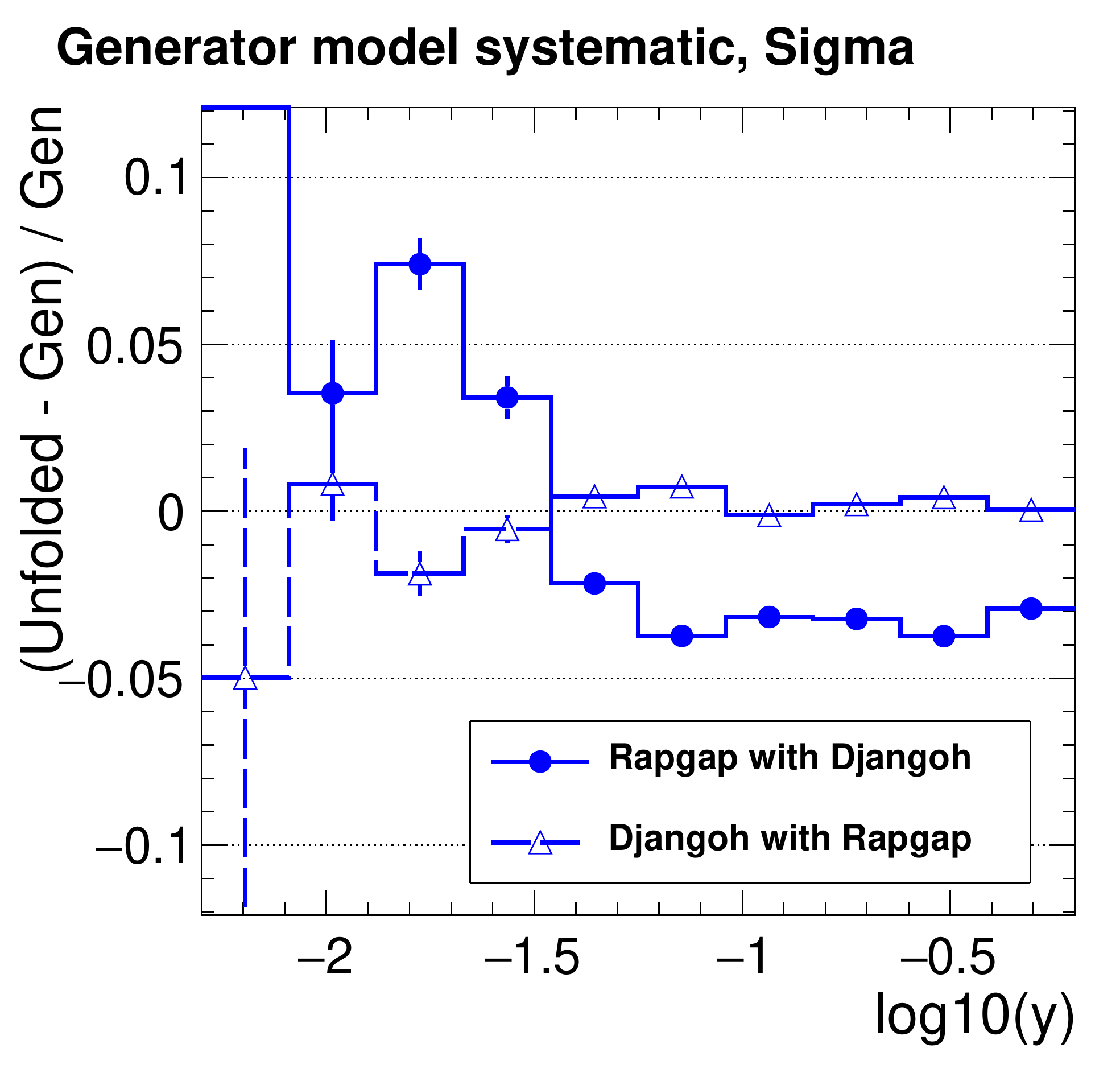} 
\includegraphics[width=0.32\linewidth]{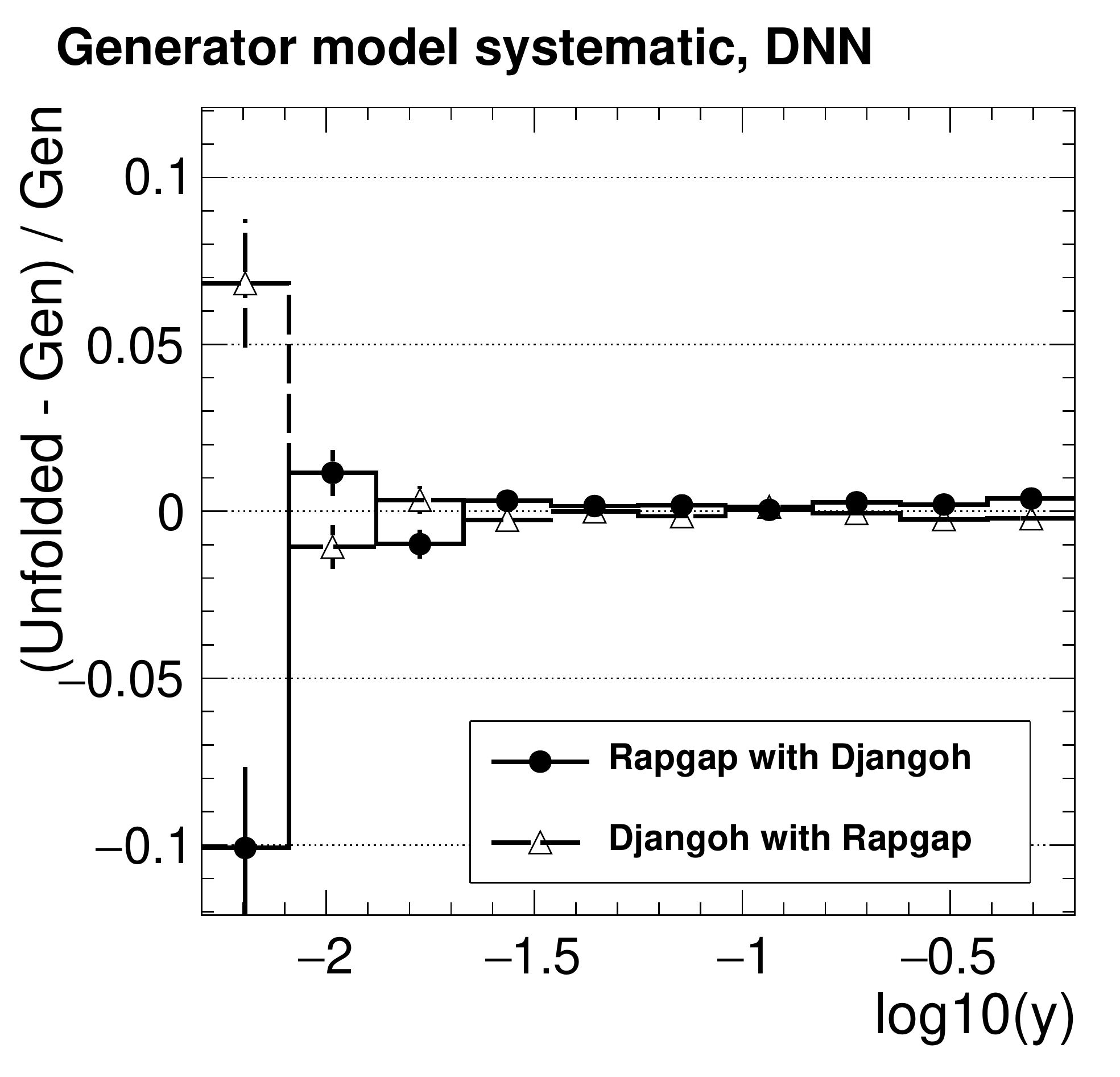} 
\caption{
  Generator model systematic uncertainties for the unfolding of $x$ (top) and $y$ (bottom) for the electron (left), Sigma (middle) and DNN (right) method.
}
\label{fig:dis-unfold-generator-syst2}
   \end{center}
\end{figure}
Figure~\ref{fig:dis-unfold-generator-syst2}  shows the results of unfolding a \textsc{Rapgap} event sample with a response matrix obtained from the \textsc{Djangoh} event generator, and vice versa, for the electron, Sigma and DNN reconstruction methods.
We observe, that the electron method results show very large fluctuations in areas where the resolution is poor, but without any systematic trends.
Also the results with the Sigma reconstruction method show large significant deviations.
In contrast, the DNN method results in an insignificant model dependence in large parts of the $x$ and $y$ distributions. Only at highest $x$ and lowest $y$ some model dependence is observed, albeit smaller than those of the classical reconstruction methods.
Altogether, the DNN reconstruction method results in reduced generator model systematic uncertainty than the classical reconstruction methods, and such would yield a significantly less biased physics result when used in analysis of real data.

\FloatBarrier
\subsection{Event Shapes}

As another example of our method, we explore the global event shape observable 1-jettiness~\cite{Antonelli:1999kx,Stewart:2010tn}, $\tau_1^b$, which is defined as
\begin{align}
    \tau_1^b=\frac{2}{Q^2}\sum_{i\in\text{HFS}}\min\{xP\cdot p_i,(q+xP)\cdot p_i\}\,,
\end{align}
where $x$ is the DIS kinematic quantity from the last section, $P$ is the proton beam four-vector, and $p_i$ are the four-vectors of the HFS objects.
The measurement of $\tau_1^b$ is clearly limited by the acceptance and resolution of the measurements of the HFS, while these measurements are important for the DIS kinematic observables only at lower $y$.
The DNN-based reconstruction extends the case shown in the previous section, and the regression DNN takes into account further input quantities from HFS 4-vector in the current hemisphere (negative $\eta$ in the Breit frame). The direct measurement of $\tau_1^b$ is not used as an input quantity to the DNN.
The outputs of the regression DNN are $Q^2$, $x$, $y$ and $\tau_1^b$.

\begin{figure}
   \begin{center}
\includegraphics[width=0.32\linewidth]{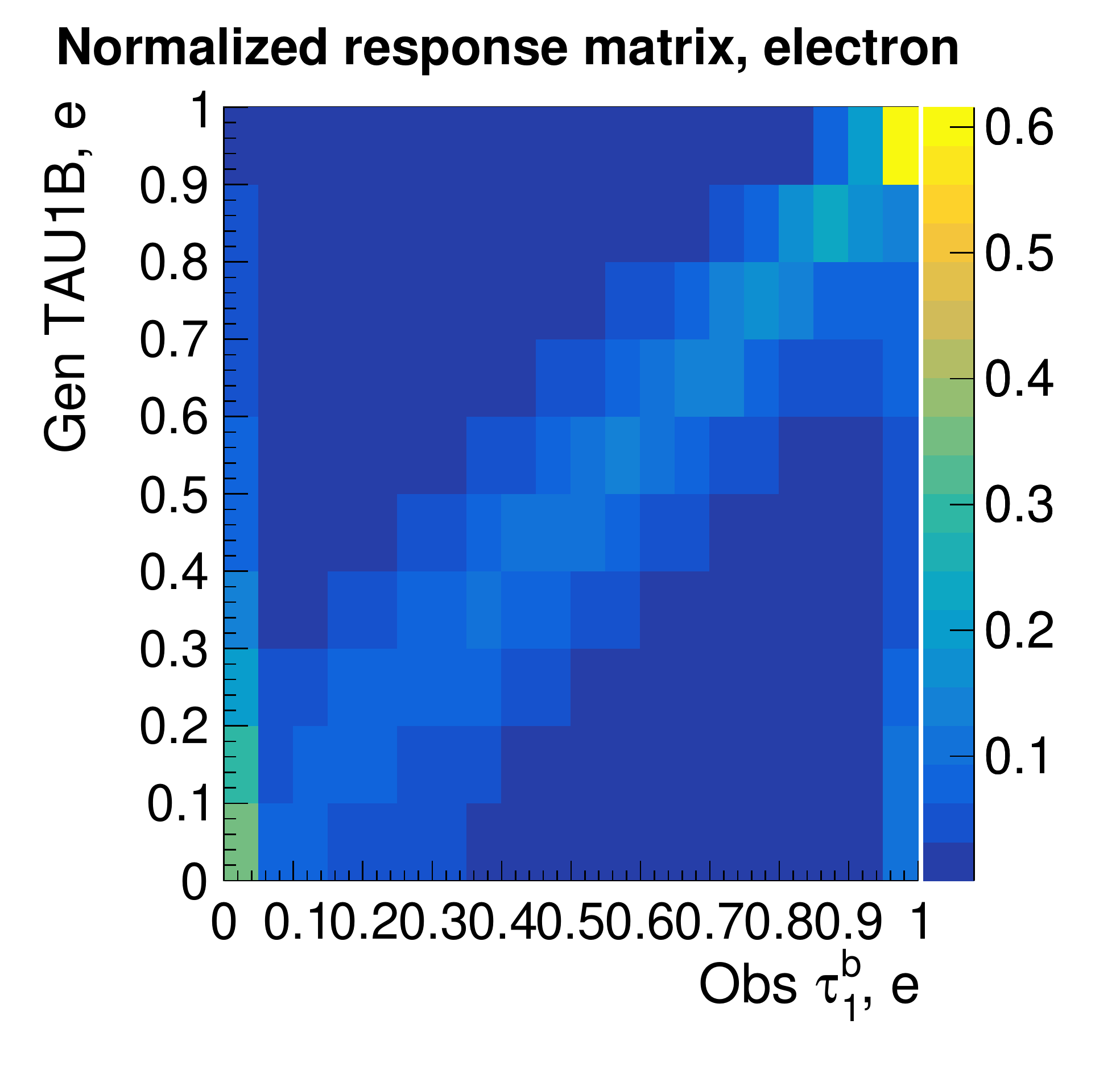}
\includegraphics[width=0.32\linewidth]{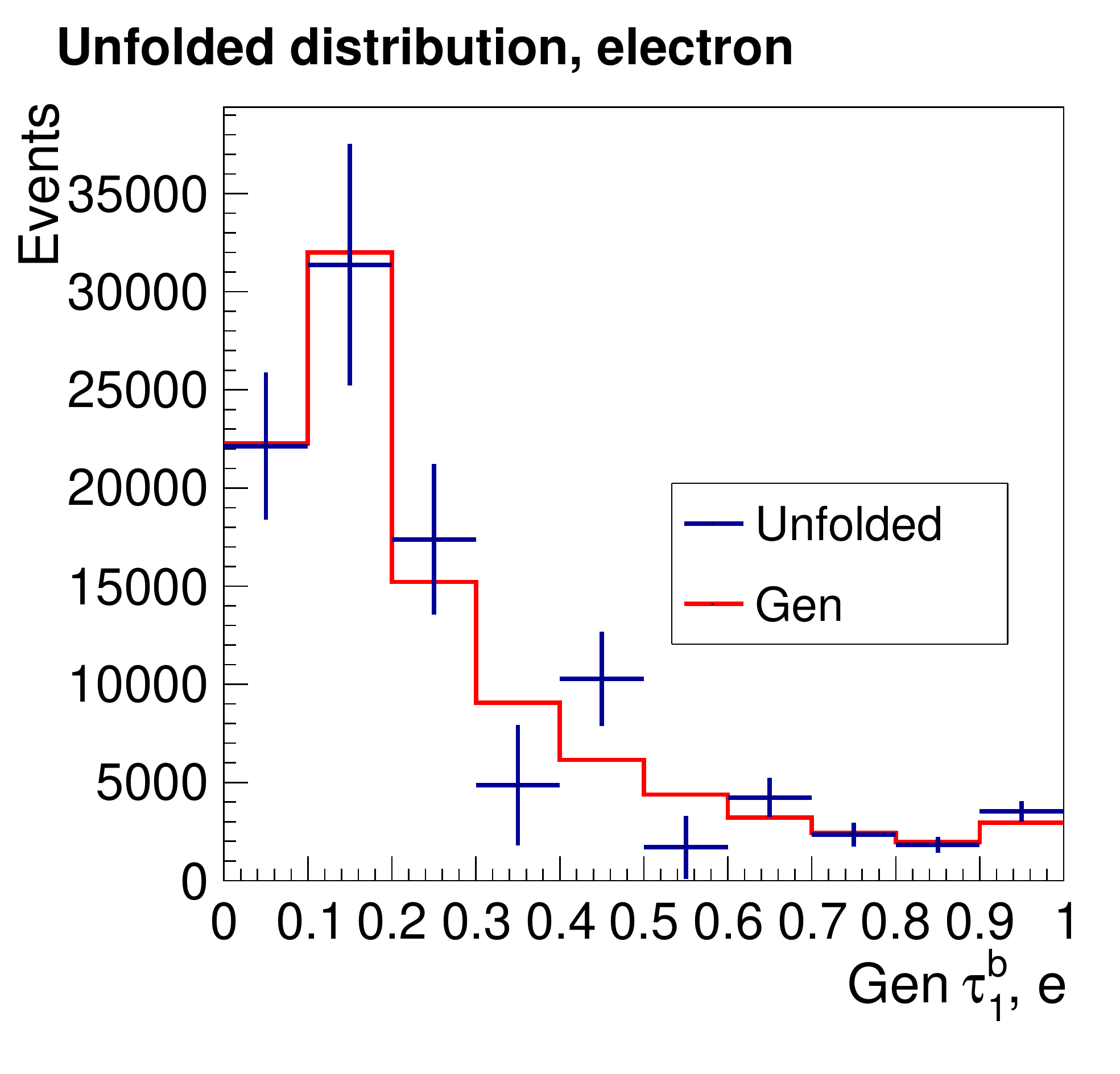}
\includegraphics[width=0.32\linewidth]{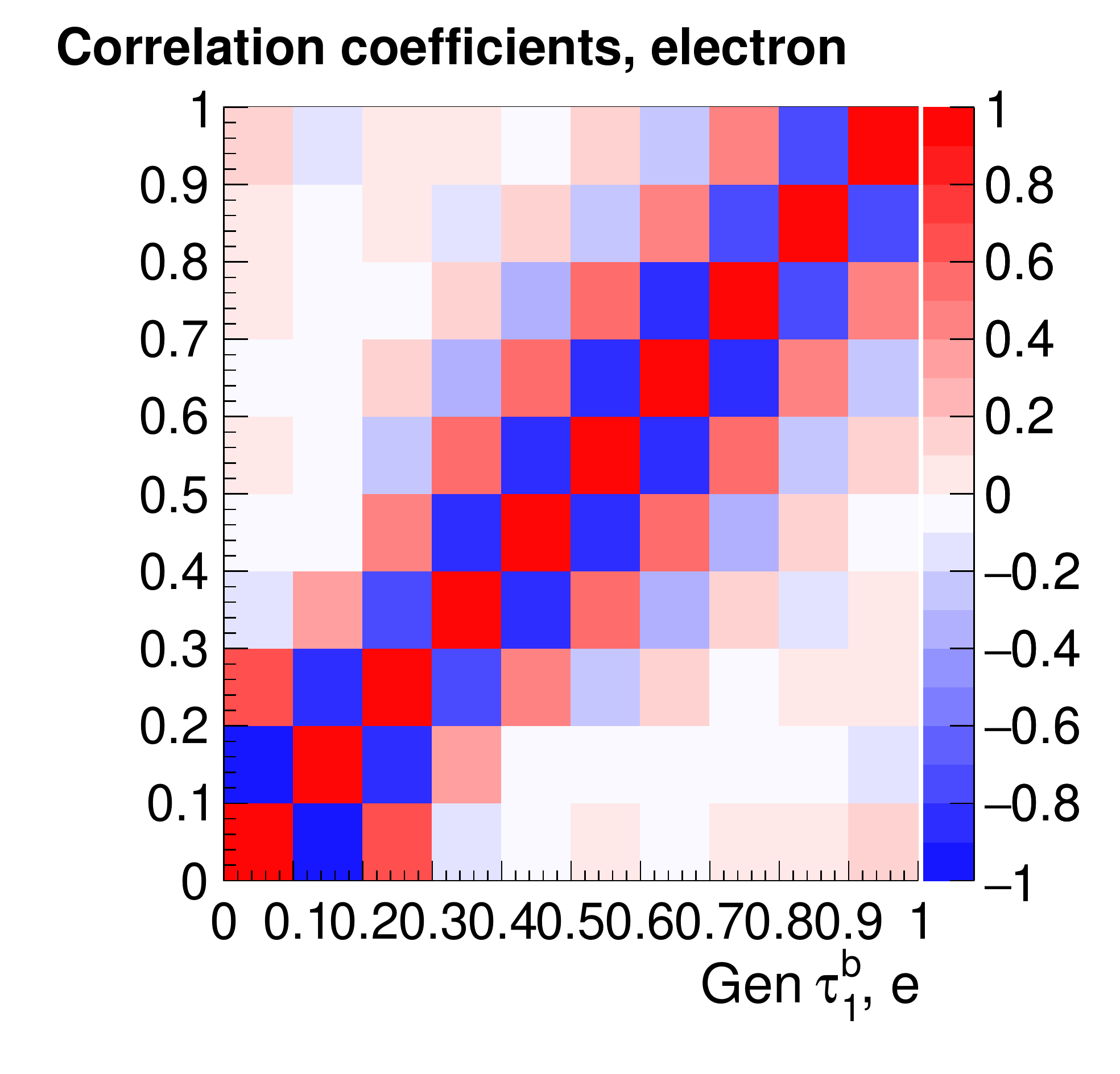} \\
\includegraphics[width=0.32\linewidth]{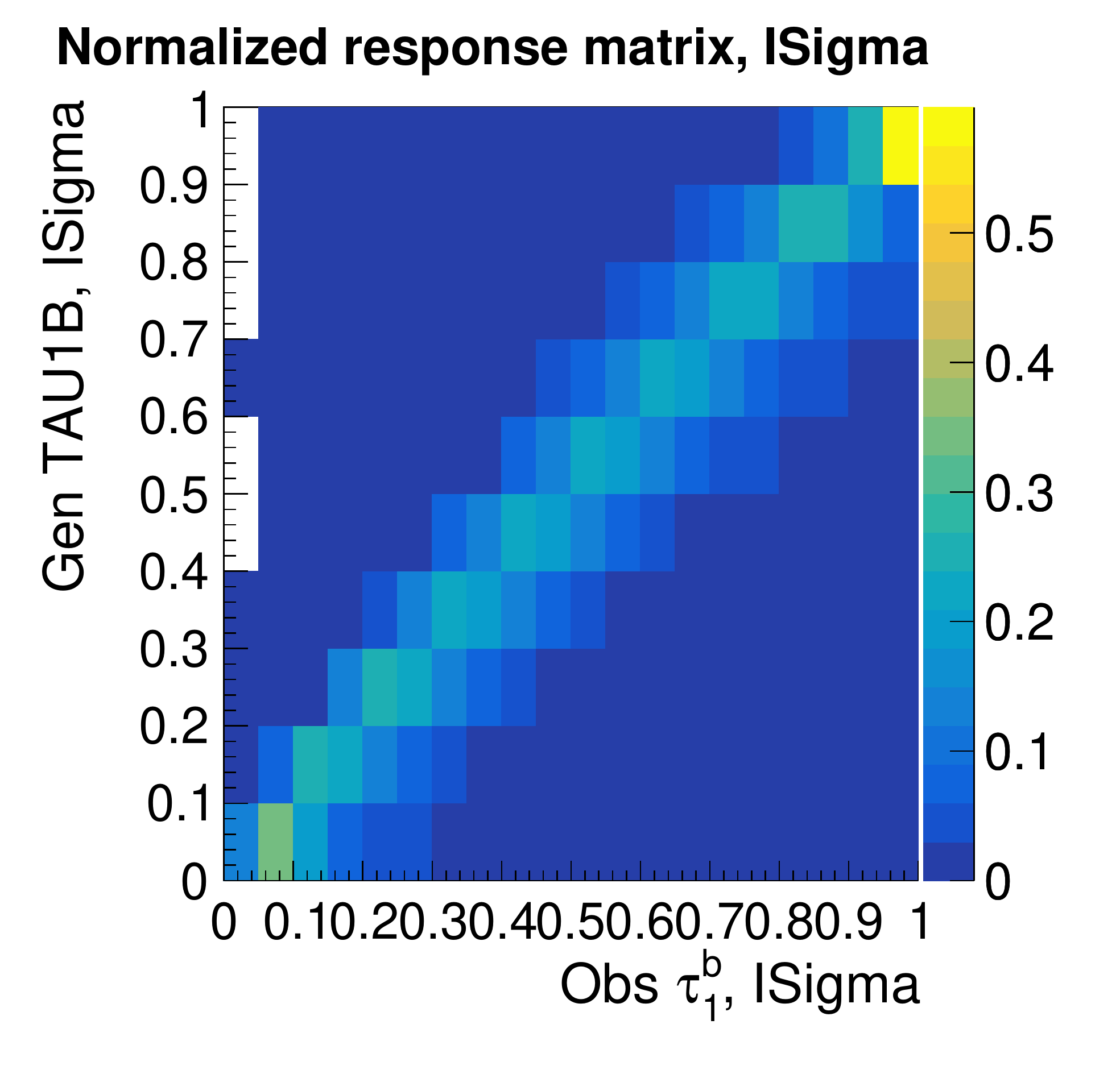}
\includegraphics[width=0.32\linewidth]{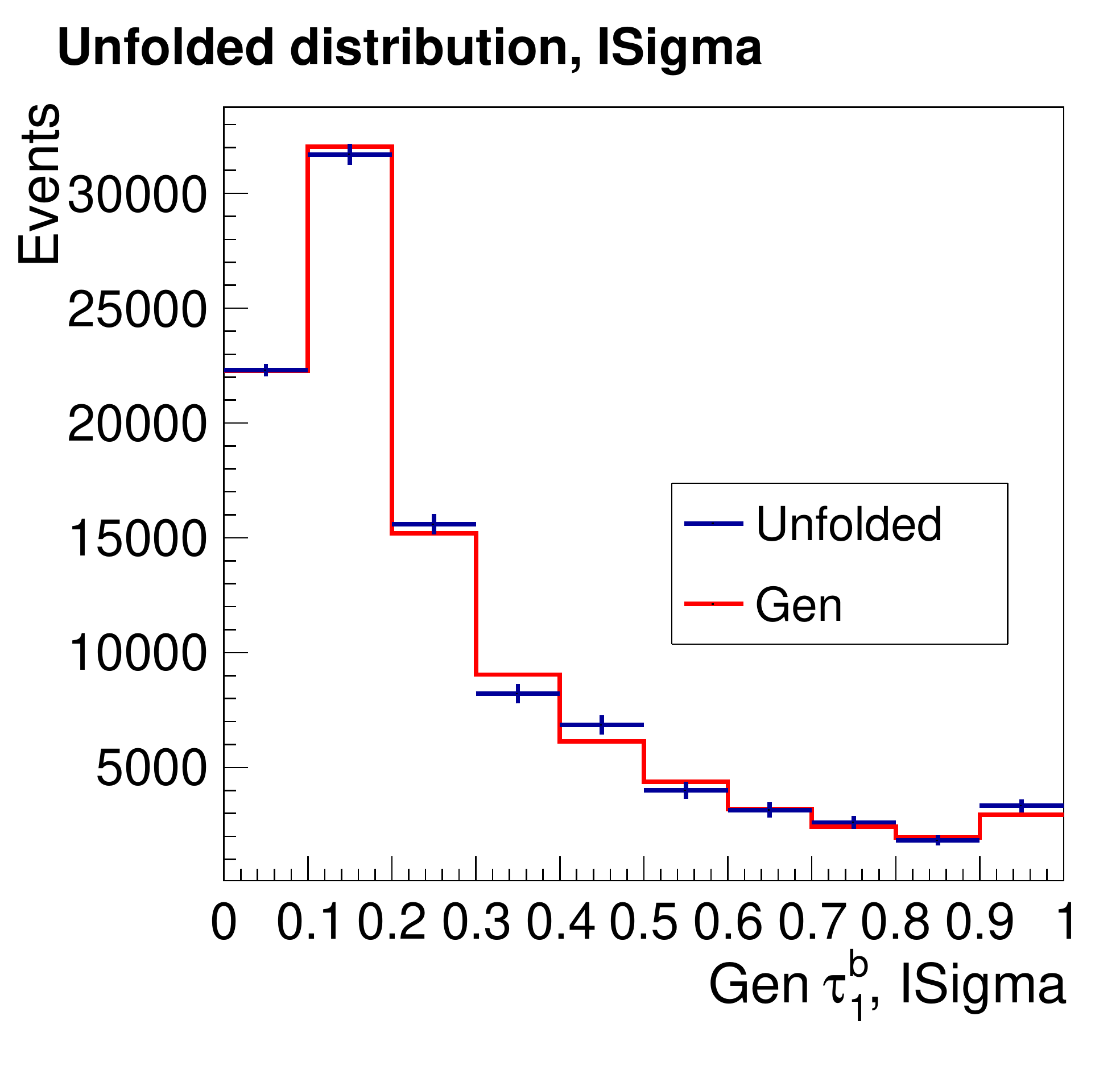}
\includegraphics[width=0.32\linewidth]{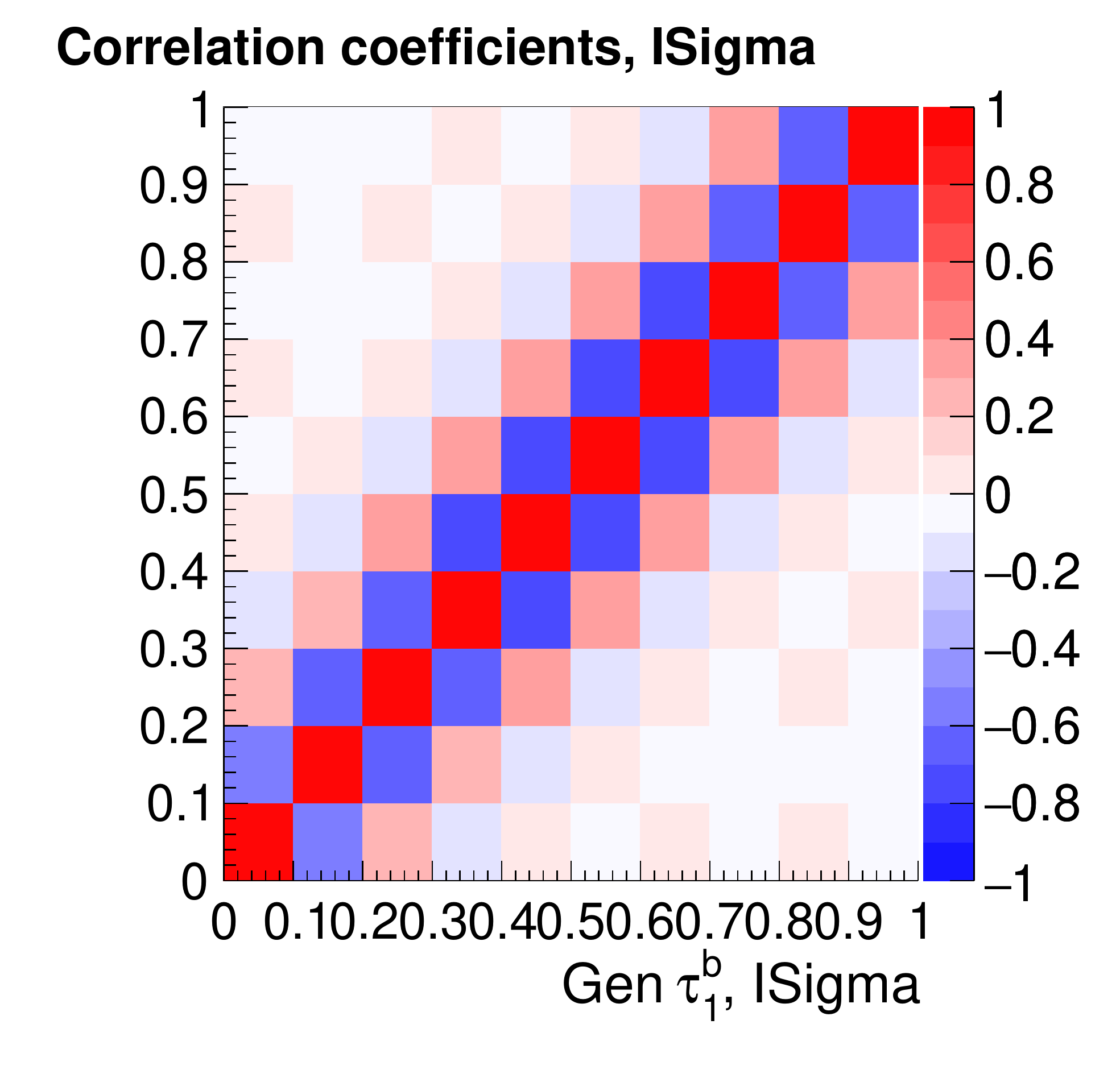} \\
\includegraphics[width=0.32\linewidth]{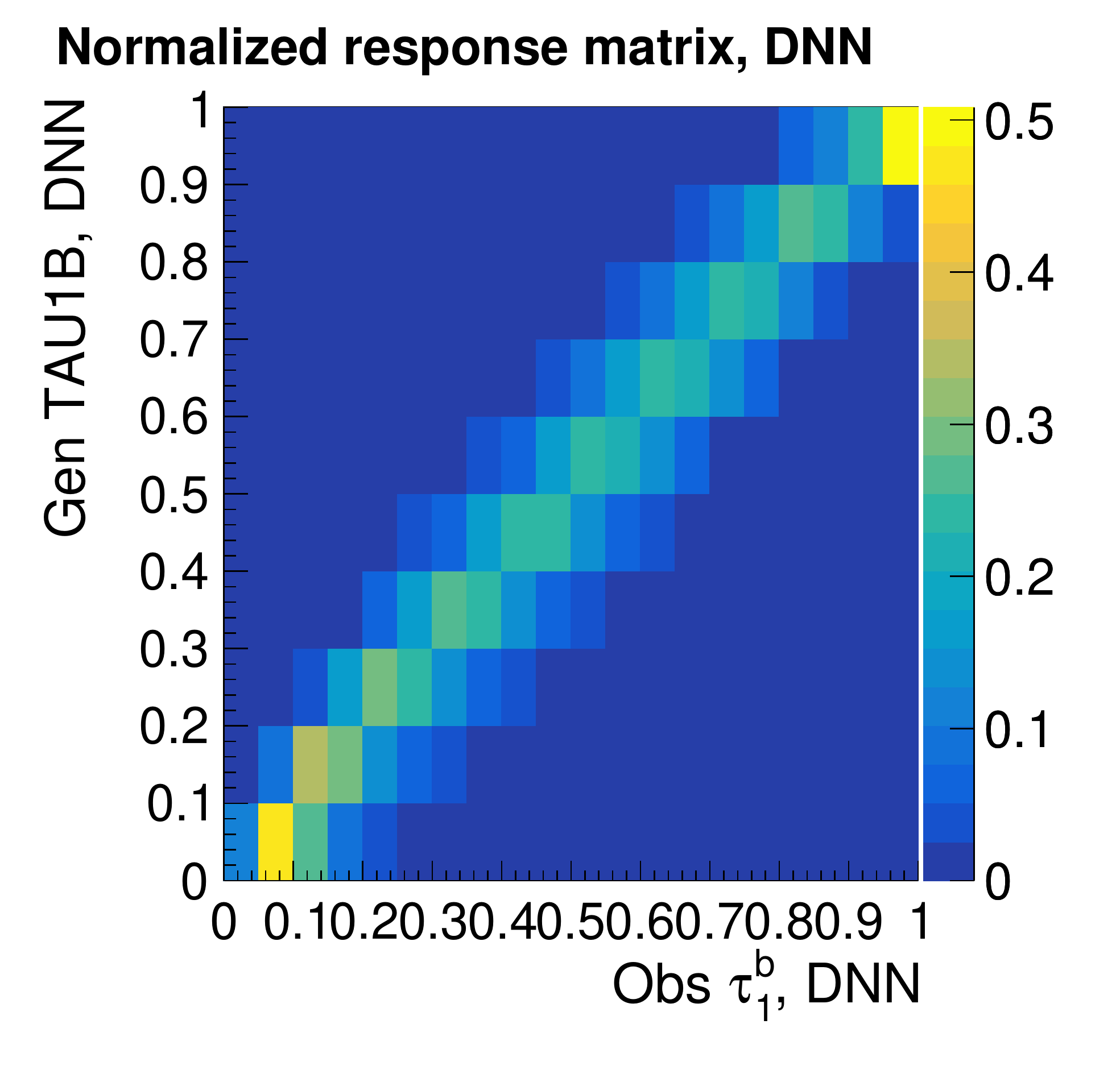}
\includegraphics[width=0.32\linewidth]{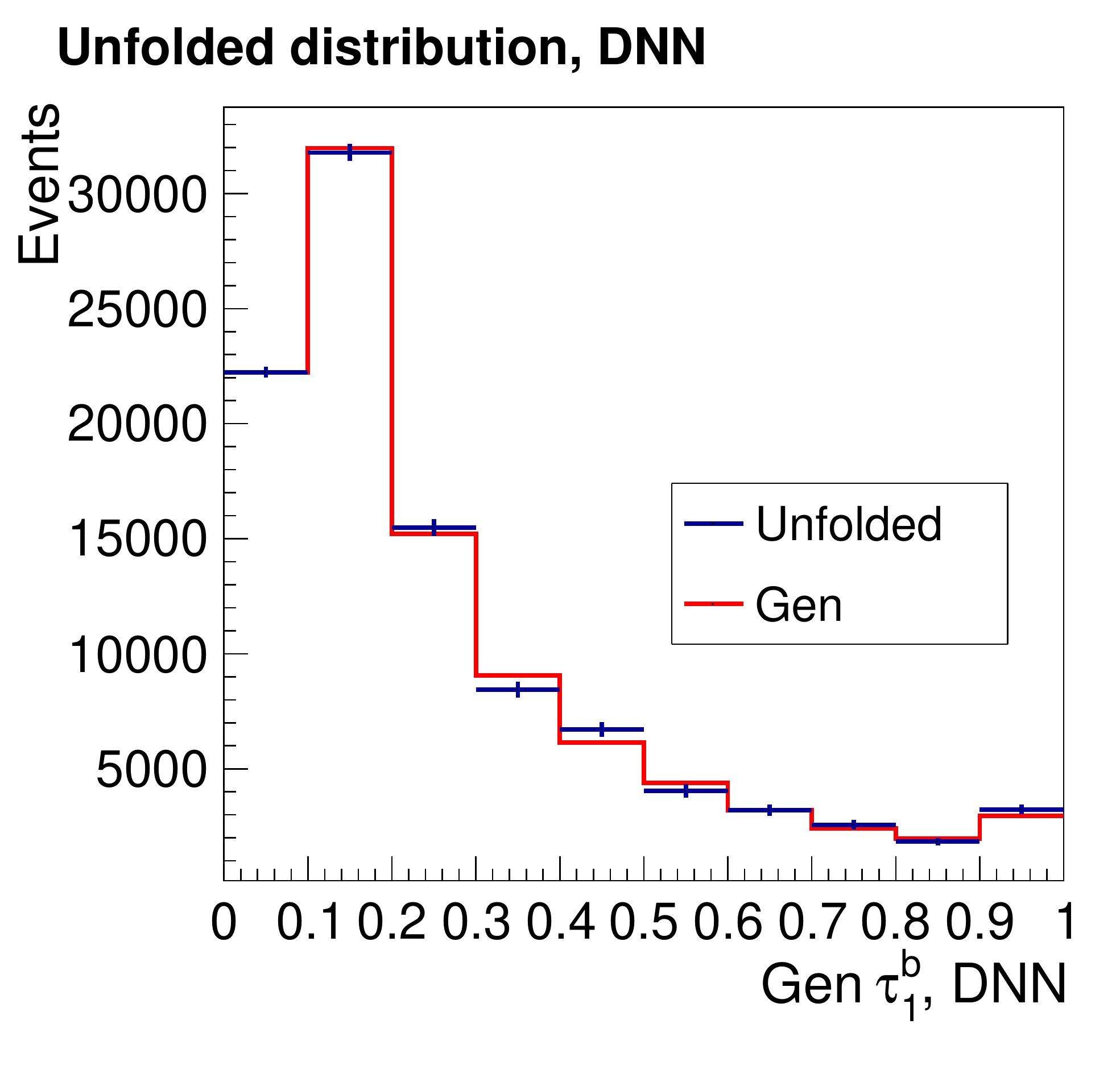}
\includegraphics[width=0.32\linewidth]{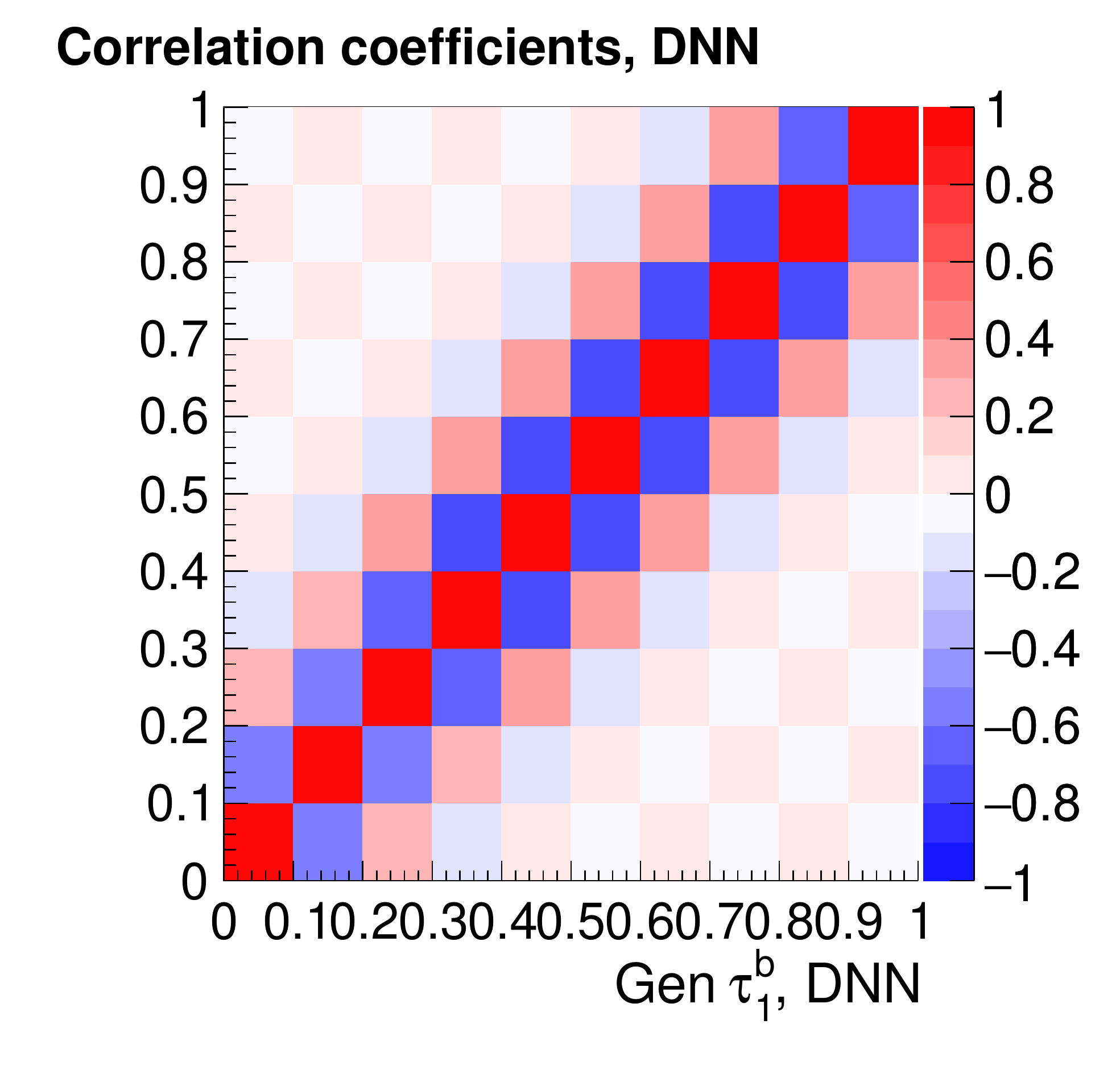} \\
   \caption{
   Examples of unfolding $\tau_1^b$ in DIS in one dimension for samples of $10^5$ events.
      The response matrix (left), unfolded and gen distributions (middle), and unfolding
      correlation matrix (right) are shown for the
      electron (top), ISigma (middle), and DNN (bottom) methods.
   }
         \label{fig:dis-unfold-tau}
   \end{center}
\end{figure}
\begin{figure}
   \begin{center}
\includegraphics[width=0.34\linewidth]{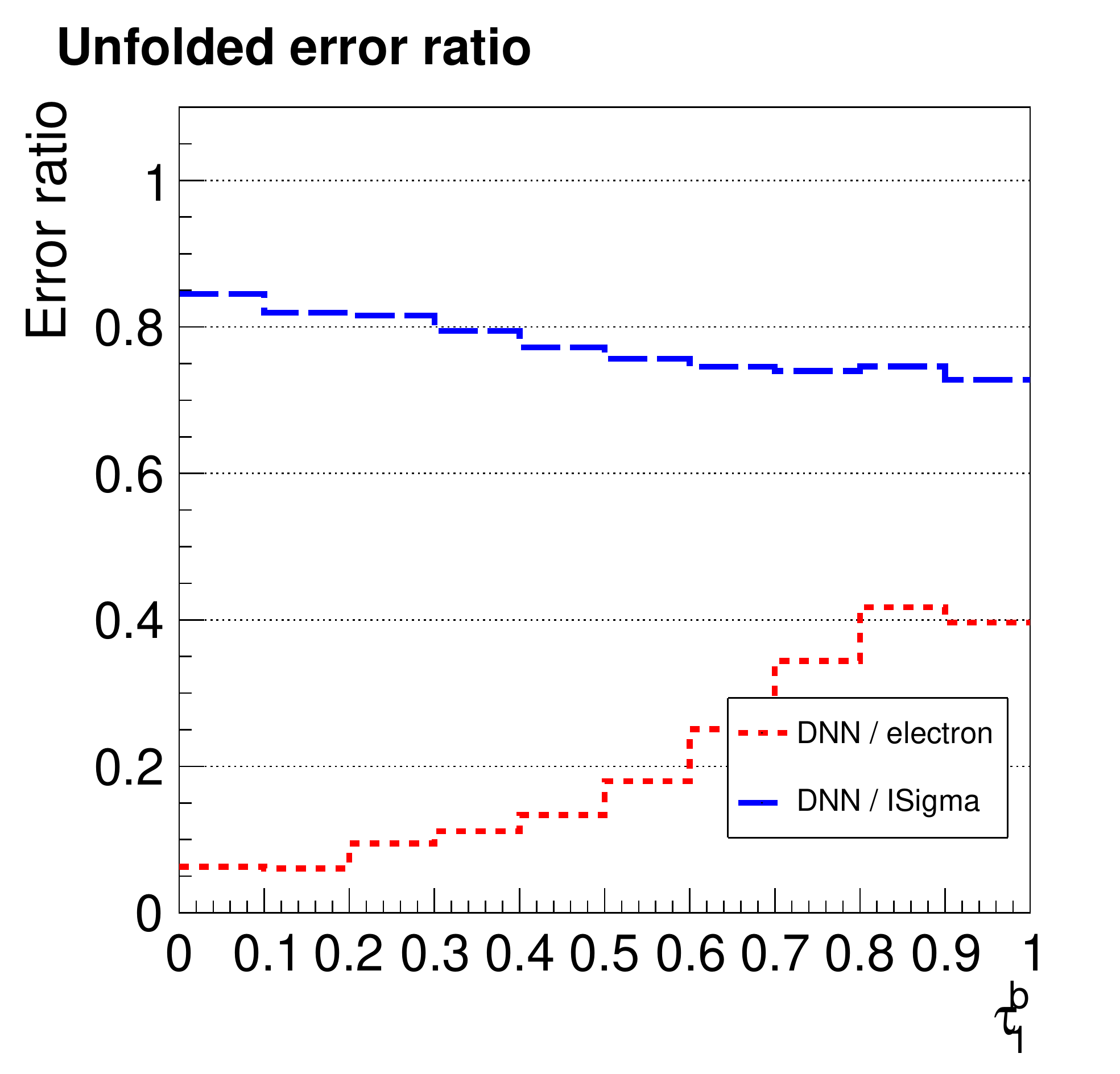} 
\includegraphics[width=0.34\linewidth]{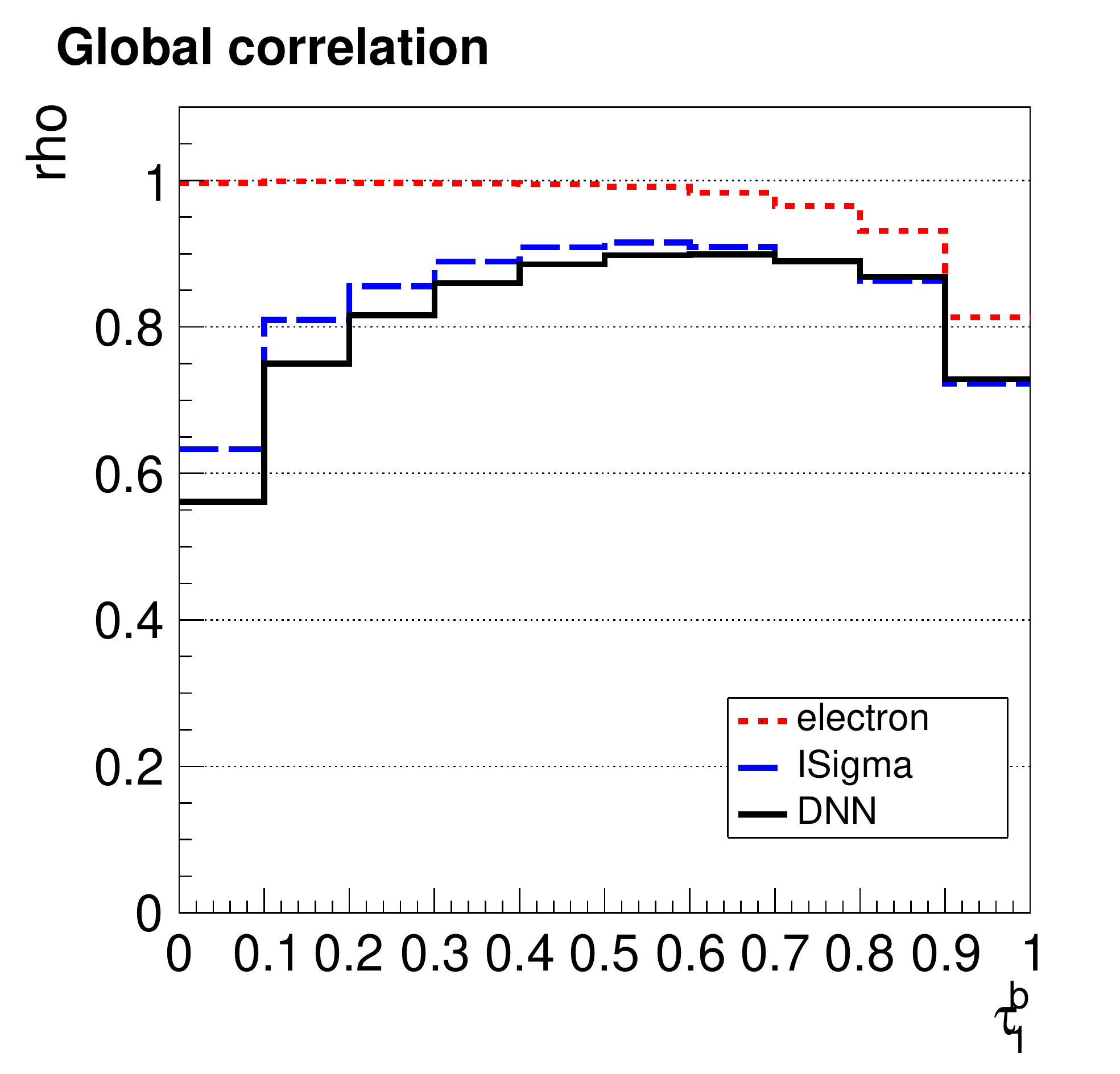}
   \caption{
     Comparison of unfolding results for  $\tau_1^b$ using the electron, ISigma and DNN reconstruction method.
     Distributions for the ratio of the statistical errors of different distributions (left) and the correlation coefficient of the error (right) are shown.
   }
         \label{fig:dis-unfold-tau-2}
   \end{center}
\end{figure}

Figure~\ref{fig:dis-unfold-tau} shows the response matrices and the results of the unfolding for $\tau_1^b$ when using the electron, ISigma~\cite{Bassler:1994uq}, and DNN methods.
The reconstruction method impacts $\tau_1^b$ through $Q^2$ and $x$ in its definition.
The influence matrix is again displayed in Appendix~\ref{sec:PostMigMa}.
Note, differently than for DIS kinematic observables, we select to study the ISigma method instead of the Sigma method, 
since it was observed to have superior resolution for
$\tau_1^b$~\cite{Hessler:2021ubp}.
The migration matrices of the ISigma and DNN methods are significantly more diagonal than that of the electron method, whereas differences between ISigma and DNN methods are small.
The closure of the three unfoldings is found to be unbiased, but the electron method results in large fluctuations due to its poor resolution for $\tau_1^b$.
The correlation matrix has sizable off-diagonal elements for the electron method, while ISigma and DNN are significantly more diagonal.
Altogether, the results from the ISigma and DNN reconstruction appear to be quite similar.
However, the global average correlation coefficient $\rho_\text{avg}$ clearly improves for the DNN method, and it is 0.978, 0.869 and 0.842 for the electron, ISigma and DNN reconstruction method respectively.
A more detailed study is presented in Figure~\ref{fig:dis-unfold-tau-2}, where the ratio of the errors and the correlation coefficients are shown.
The DNN method yields significantly smaller errors in all bins, by about 20\,\%, and also reduced correlations coefficients.

While the results are not as dramatic as for inclusive DIS, the DNN still outperforms the traditional methods.
In general one can expect that the observables that have a very inhomogeneous detector response will benefit the most from a DNN-aided unfolding.  This is why the benefits for inclusive DIS are more prevalent than for $\tau_1^b$.

\FloatBarrier

\section{Summary and outlook}
\label{sec:conclusion}

We have argued that the common practice of requiring analogous definition of observables at particle and detector level is not necessary.
In fact, abandoning the usual constraint grants us freedom to define the observable at detector level using a large number of inputs to better account for detector effects and nuisance physical phenomena like QED radiation. We have proposed to define observables at detector level with deep learning and we have shown that this improves traditional unfolding methods by providing a more diagonal response matrix. Consequently, this results in smaller correlations between bins and smaller uncertainties of the unfolded results. Furthermore, the DNN reconstructed observables exhibit a smaller model dependence, which would reduce the related systematic uncertainties and provide less biased results.  While our examples have focused on inclusive observables in DIS with full detector simulations of the H1 experiment at HERA, this approach can be generalized to any unfolding analysis.  The benefits will be largest when there is a highly inhomogeneous detector response and there are many observables that govern this response which are not explicitly part of the unfolding.  

Our approach deconstructs unfolding into two steps: the construction of observables and then the correction for detector effects.  The first step can be viewed as an event-by-event correction for detector effects, but the second step is still necessary to mitigate prior dependence.  Our approach is complementary to proposes for using machine learning to implement detector corrections~\cite{Gagunashvili:2010zw,Glazov:2017vni,Datta:2018mwd,Andreassen:2019cjw,Bellagente:2020piv,Bellagente:2019uyp,Andreassen:2021zzk,bunse2018unification,Ruhe2019MiningFS, Howard:2021pos,Vandegar:2020yvw}.  A combination of both methods would result in an even more precise and flexible analysis. The ultimate approach would be to unfold all observables simultaneously using all the available information at detector-level~\cite{Andreassen:2019cjw}.  The approach discussed in this paper is much simpler and will likely yield immediate benefits while many of the statistical and computational challenges are being addressed for the full phase space methods.

\section*{Code availability}
The code in this work can be found in: \url{https://github.com/owen234/MLAssistedUnfolding}.

\section*{Acknowledgments}

We thank our colleagues from the H1 Collaboration for allowing us to use the
simulated MC event samples. Thanks to DESY-IT and the MPI f{\"u}r Physik for providing some computing infrastructure and supporting the data preservation project of the HERA experiments.
Parts of the work were supported by the ORIGINS Data Science Laboratory (ODSL).
We thank L.~Gouskos, S.~Kluth, J.~Thaler and Z.~Zhang for helpful conversations and feedback on the manuscript.
M.A. was supported through DOE Contract No.\ DE-AC05-06OR23177 under which JSA operates the Thomas Jefferson National Accelerator Facility, and by the University of California, Office of the President award number 00010100. B.N. was supported by the Department of Energy, Office of Science under contract number DE-AC02-05CH11231.

\clearpage
\appendix
\section{The Influence Matrix}
\label{sec:PostMigMa}
\begin{figure}[hbp!]
   \begin{center}
\includegraphics[width=0.30\linewidth]{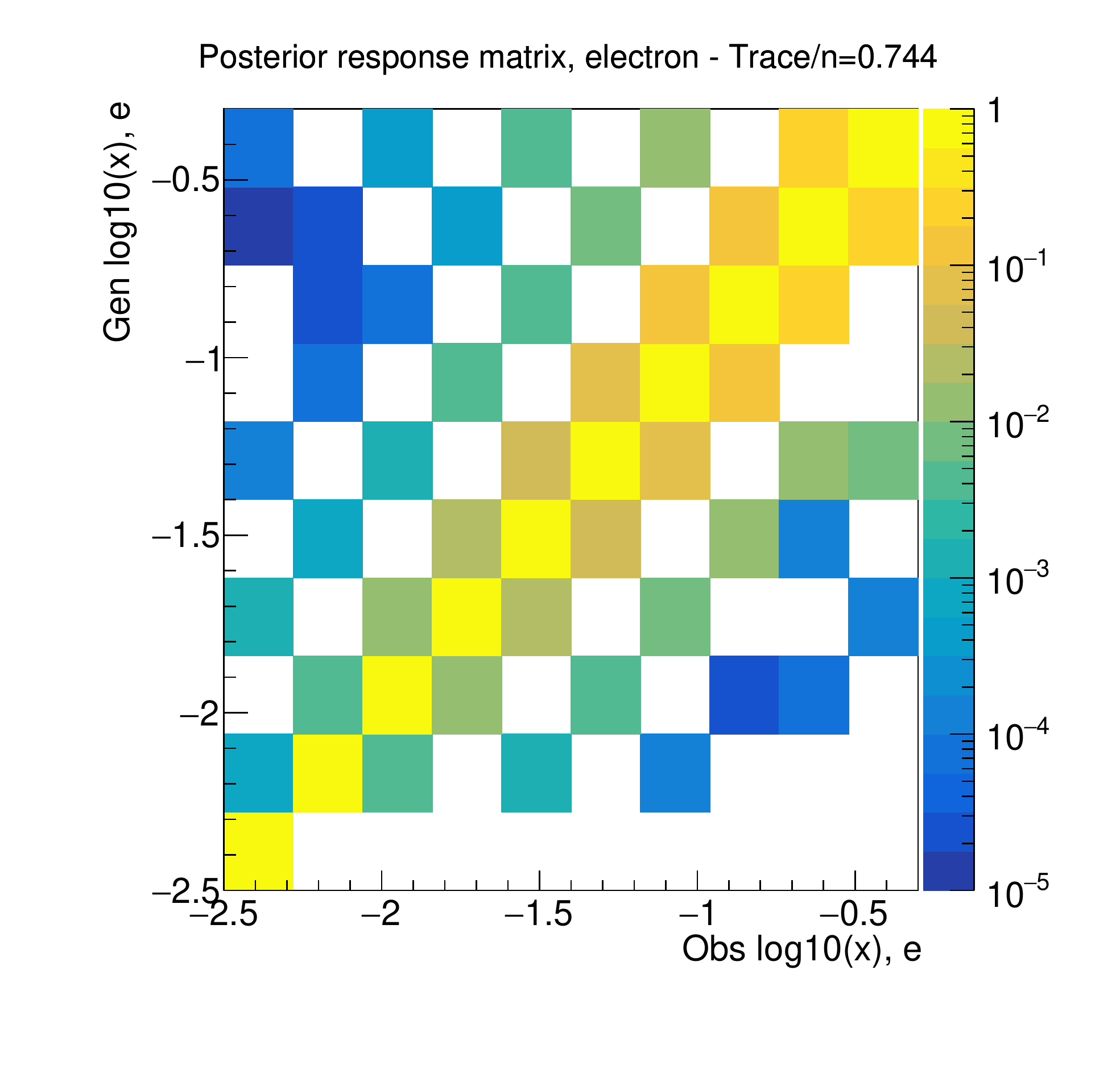}
\includegraphics[width=0.30\linewidth]{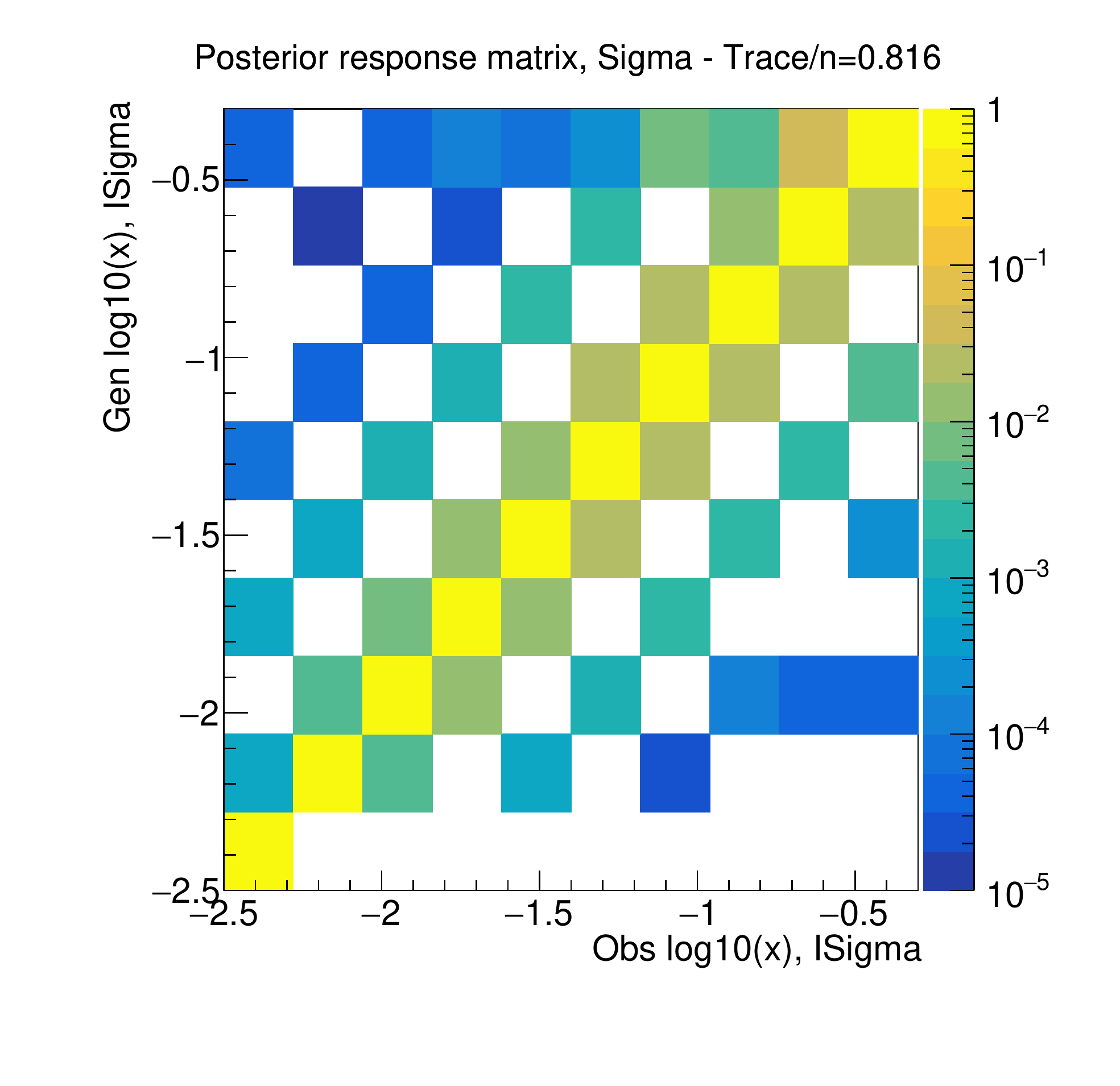}
\includegraphics[width=0.30\linewidth]{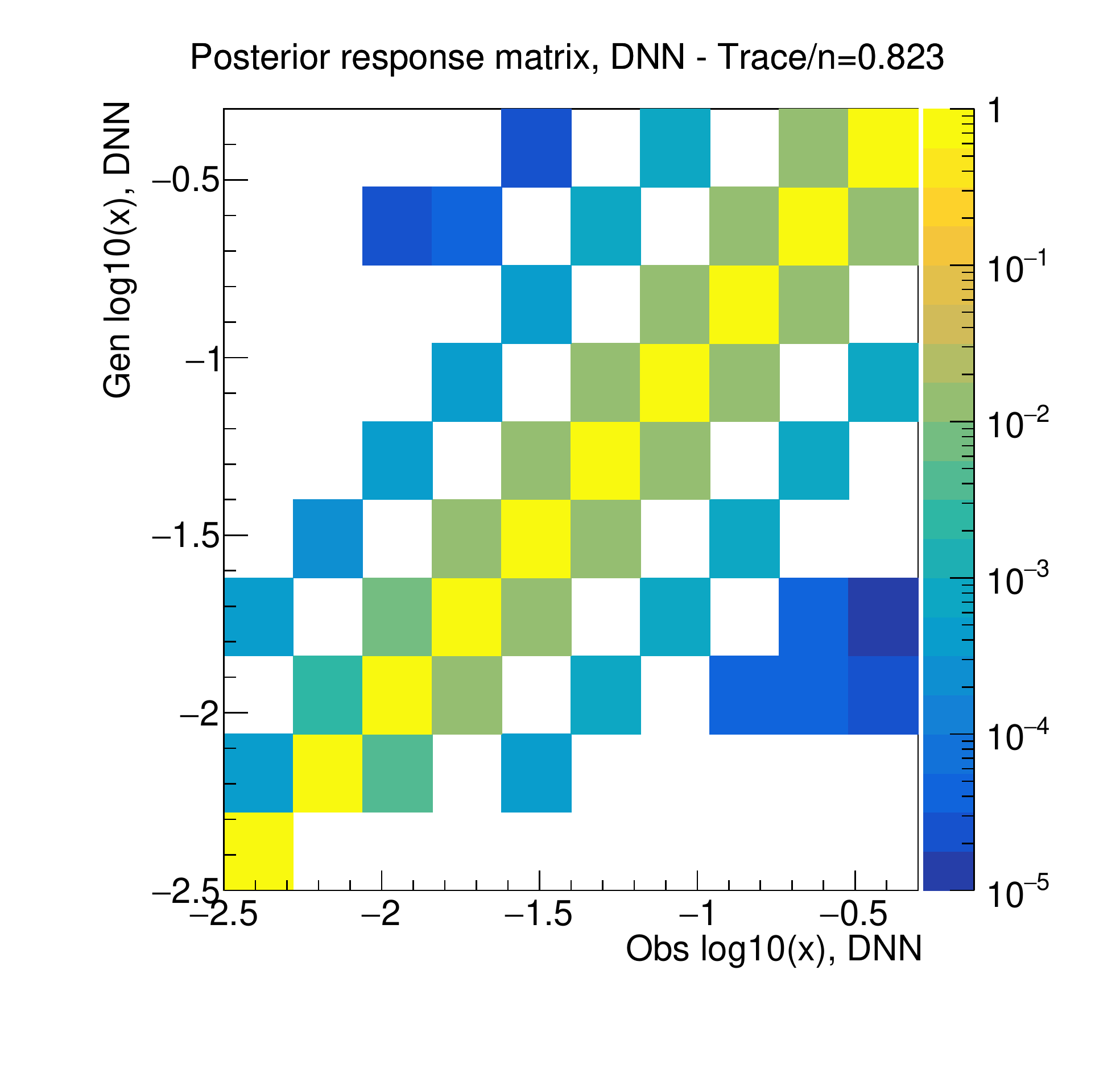}
\\
\includegraphics[width=0.30\linewidth]{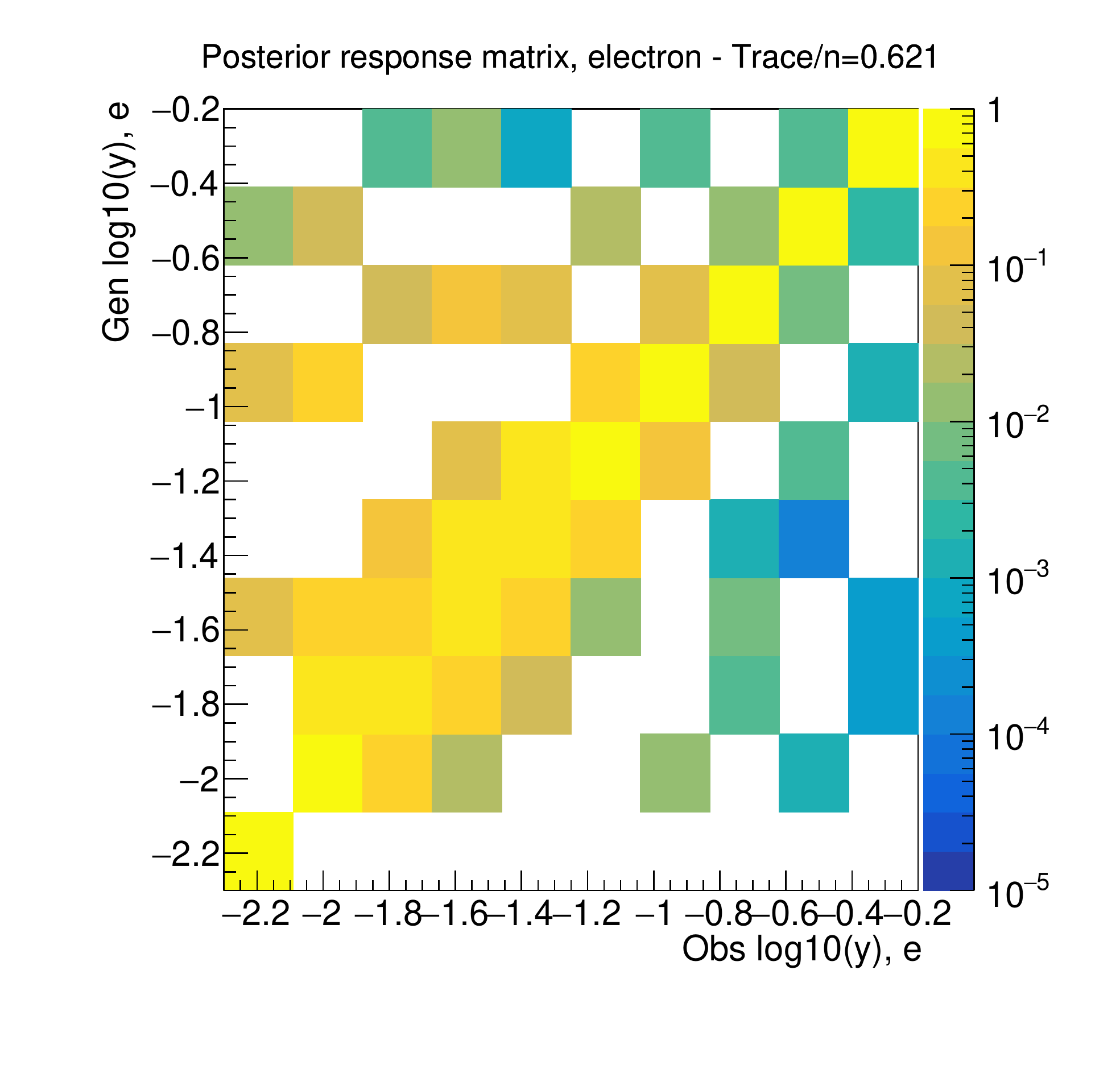}
\includegraphics[width=0.30\linewidth]{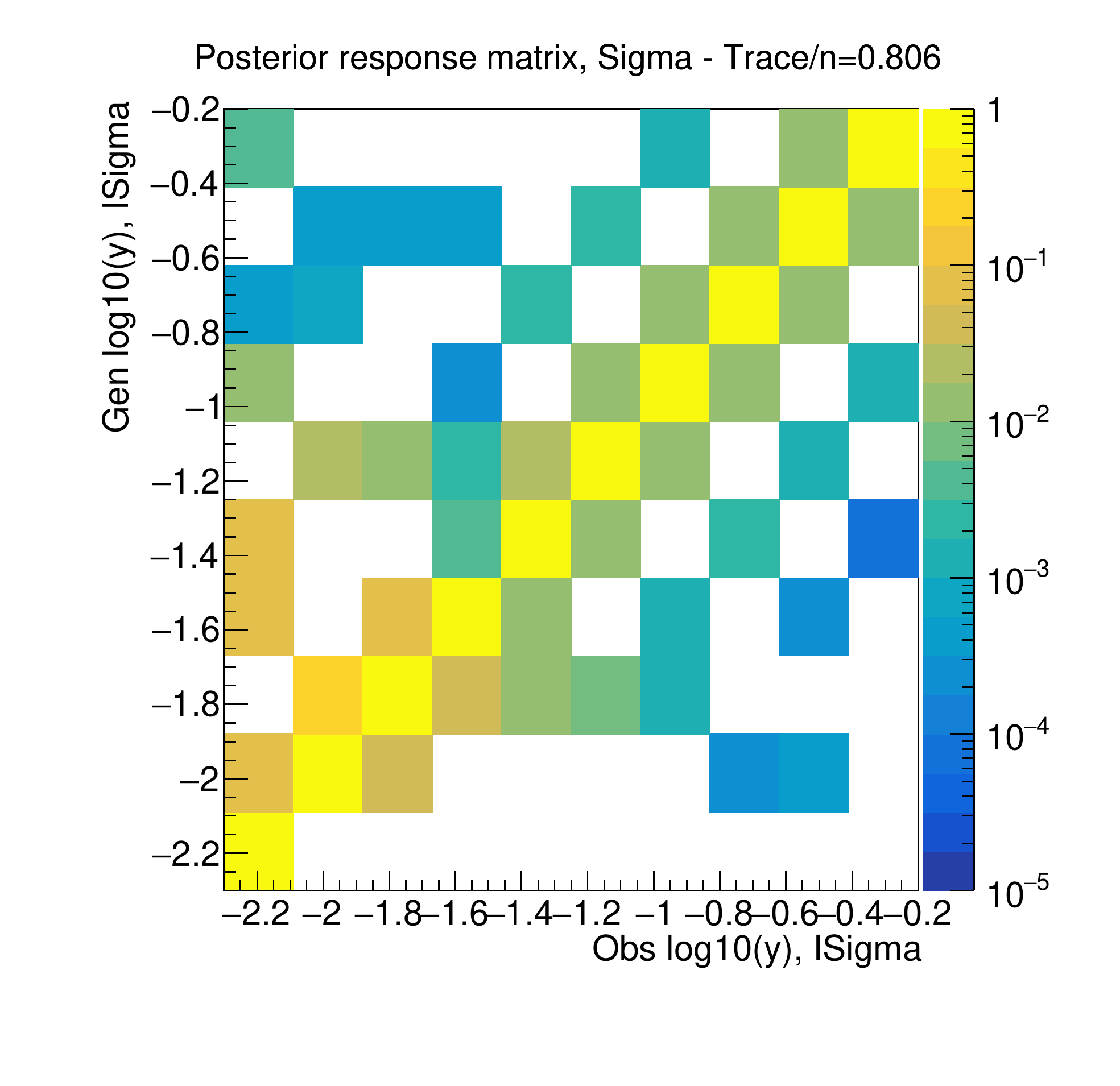}
\includegraphics[width=0.30\linewidth]{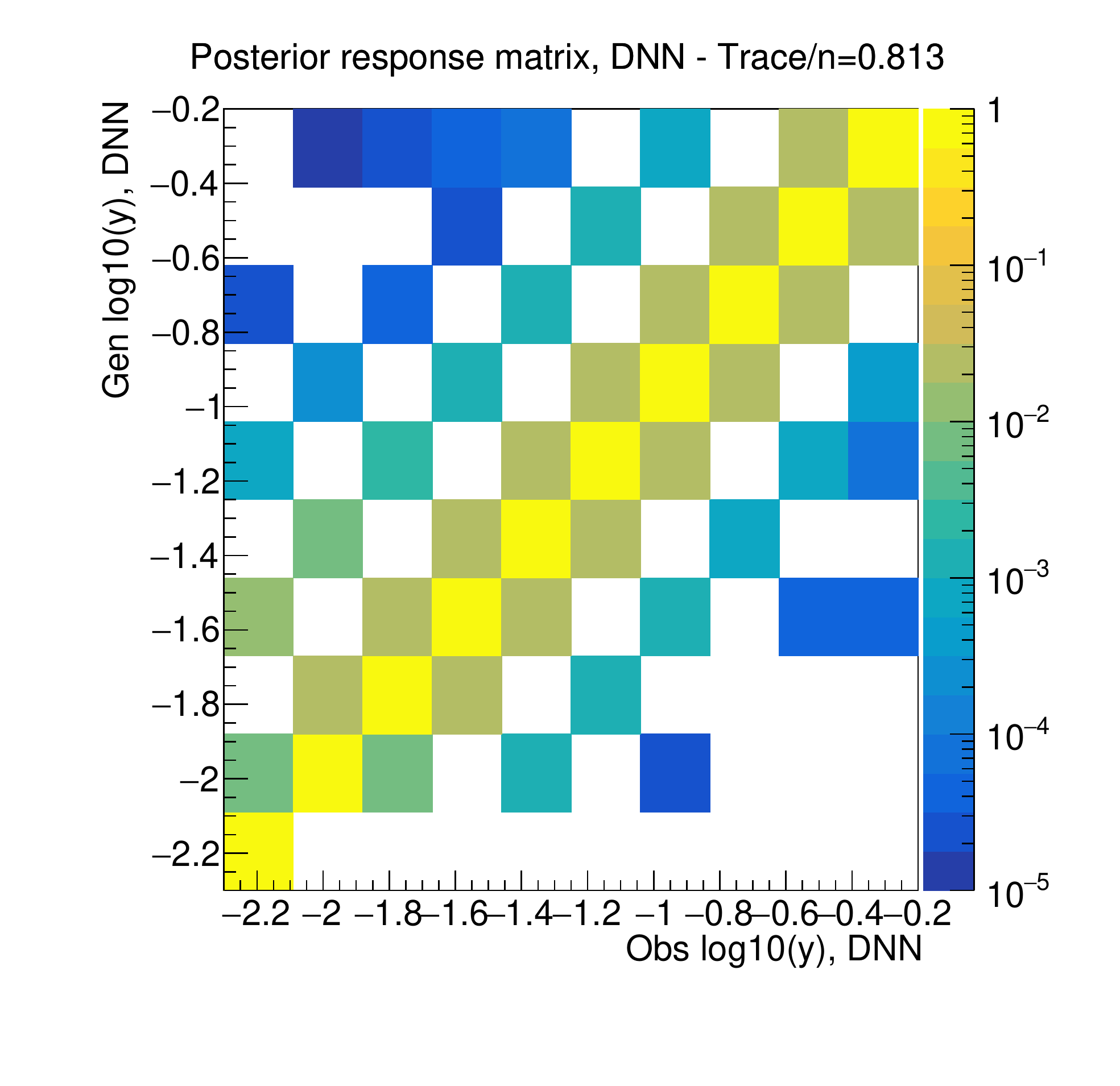}
\caption{
  Influence matrices for the unfolding of $x$ (top) and $y$ (bottom) are shown for the electron (left), Sigma (middle) and DNN (right) reconstruction methods.
}
\label{fig:posterior-matrices}
   \end{center}
\end{figure}

\begin{figure}[hpb!]
   \begin{center}
\includegraphics[width=0.30\linewidth]{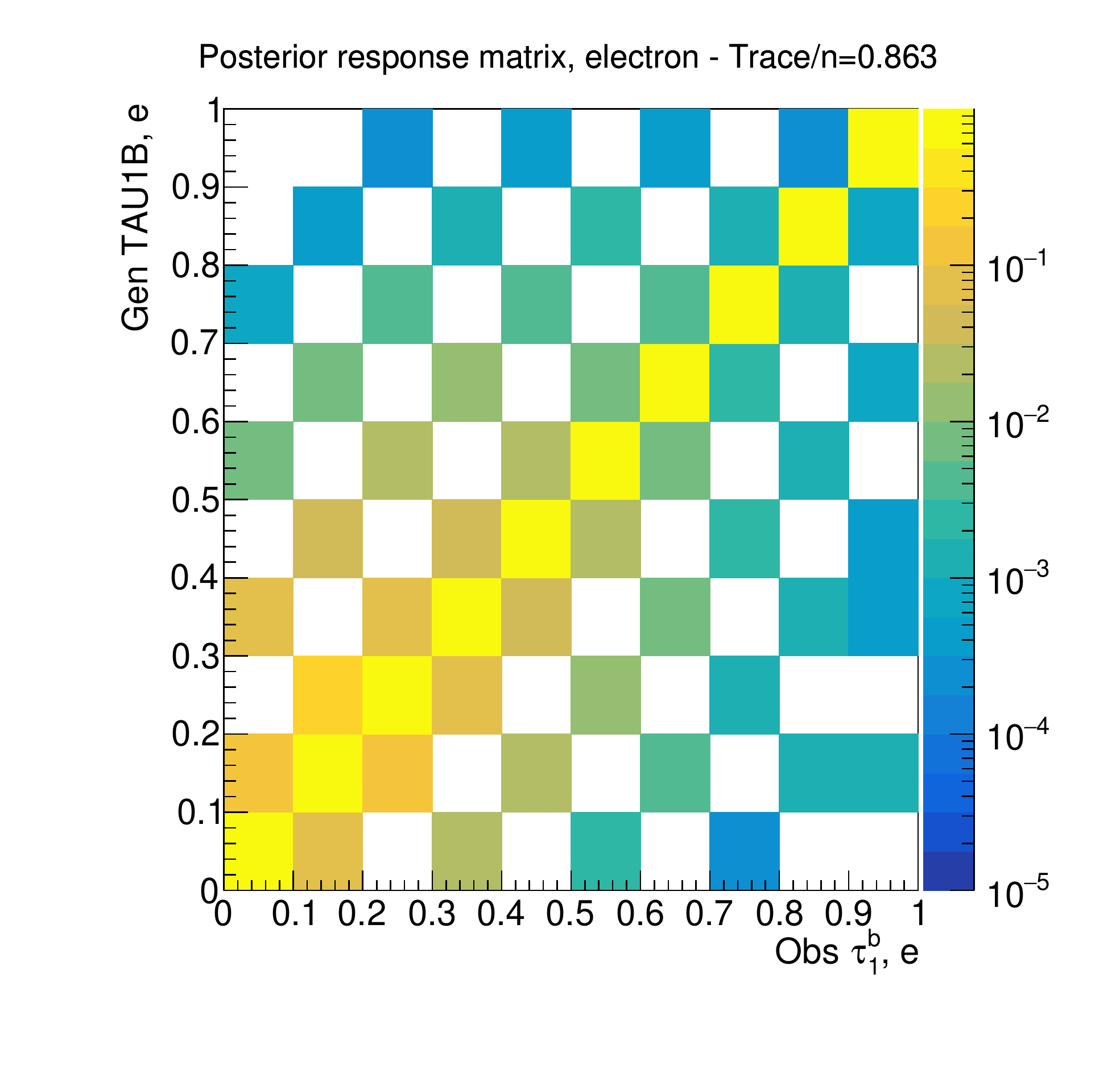}
\includegraphics[width=0.30\linewidth]{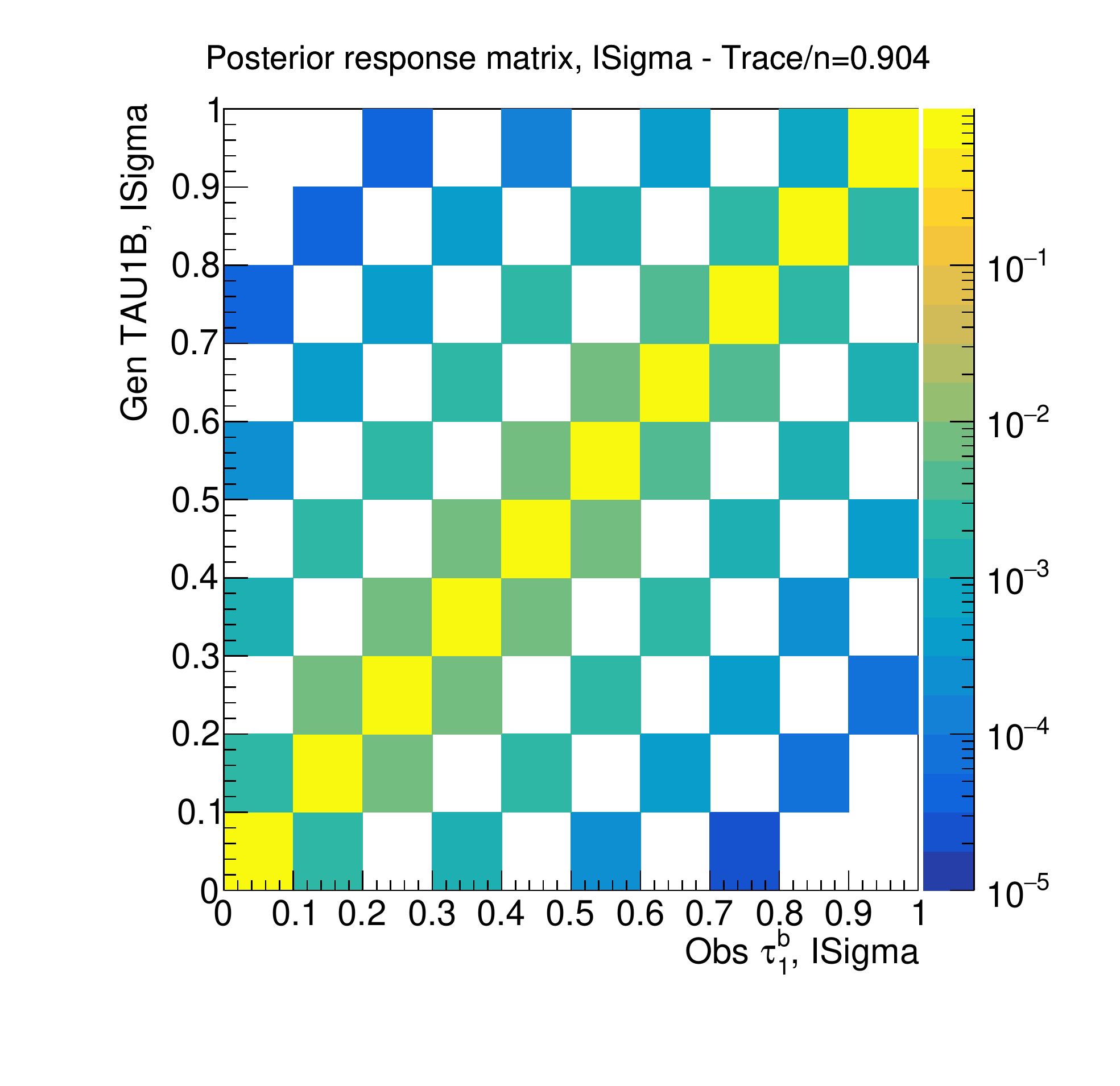}
\includegraphics[width=0.30\linewidth]{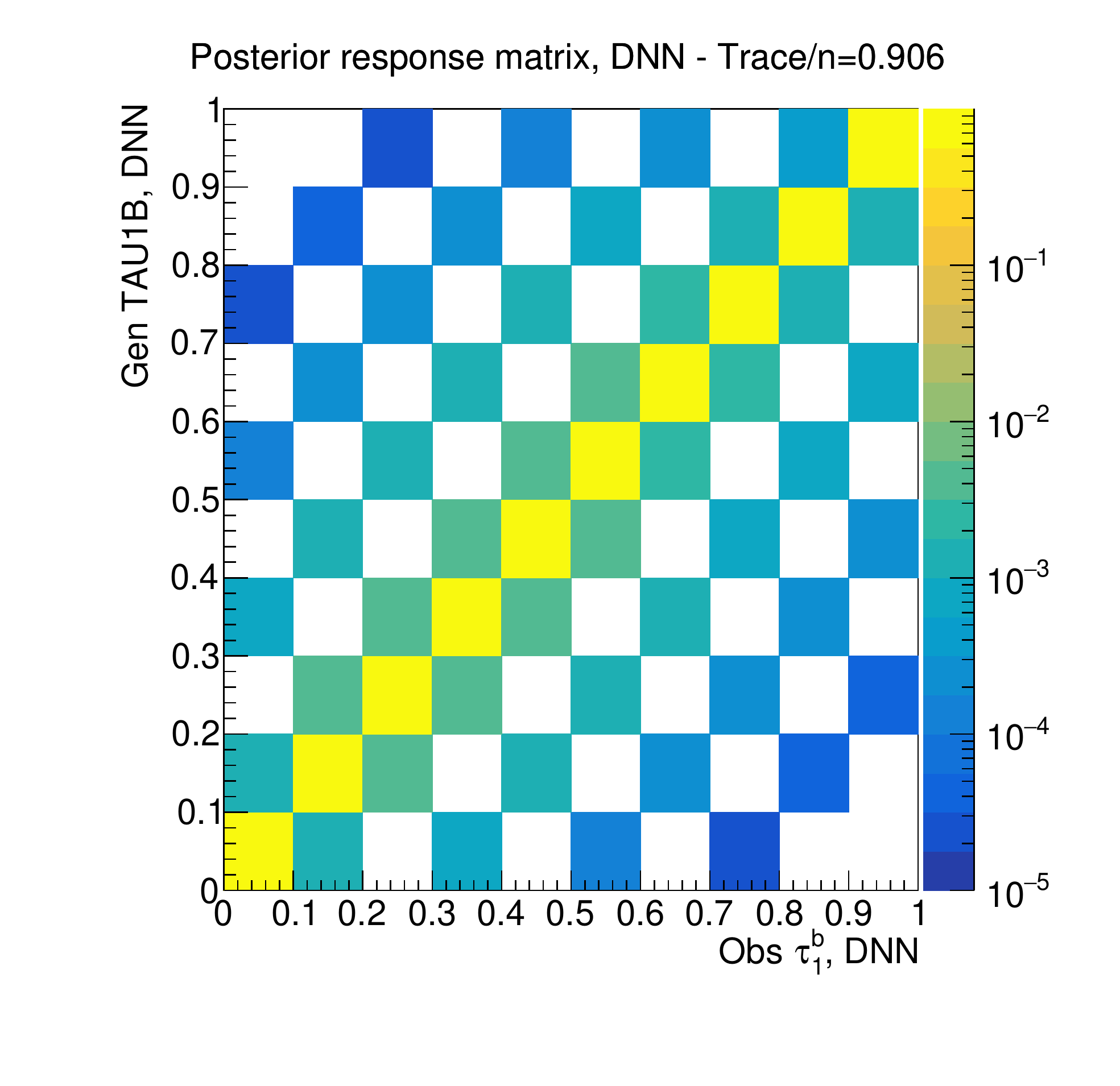}
\caption{
  Influence matrices for the unfolding of $\tau_1^b$ for the electron (left), ISigma (middle) and DNN (right) reconstruction methods.
}
\label{fig:posterior-matrices-tau1b}
   \end{center}
\end{figure}

The properties of the \emph{influence matrix} $P$~\cite{Wahba,doi:10.1002/9783527653416.ch6}, which is also called \emph{posterior response matrix}, were recently studied as an indicator of an unfolding problem in Ref.~\cite{Schmelling:2022rci}.
The matrix $P$ describes how the unfolding result is distorted with respect to the truth, and it can be regarded as the quadratic response matrix including the effect of the regularization.
The influence matrix $P$ for the electron, Sigma and DNN reconstruction methods when unfolding the observables $x$ and $y$ are displayed in Figure~\ref{fig:posterior-matrices}, and for the observable $\tau_1^b$ in Figure~\ref{fig:posterior-matrices-tau1b}.

The influence matrices of the  DNN-based reconstructed observables are more diagonal than those of the classical reconstruction methods. The `chessboard' like structure, which is introduced by the regularization, is more regular for the DNN observables than for the classical reconstruction methods (note the logarithmic scaling of the color coding).
The traces of these matrices normalized to the number of particle-level bins, $\text{tr}(P)/n$, are displayed in Table~\ref{tab:traces}, and are seen to increase for the DNN method.
The Pearson's correlation coefficients of these matrices are found to similar and close to one.
\begin{table}[hbt!]
  \centering
  \small
  \begin{tabular}{cccc}
    \toprule
    Observable & \multicolumn{3}{c}{Reconstruction method} \\
               & electron & (I)Sigma & DNN \\
    \midrule
    $x$        & 0.744 & 0.816 & 0.823 \\
    $y$        & 0.621 & 0.806 & 0.813 \\
    $\tau_1^b$ & 0.863 & 0.904 & 0.906 \\
    \bottomrule
  \end{tabular}
  \caption{Value of the traces of the influence matrices normalized to the number of bins ($n=10$) for the unfolding of $x$, $y$ and $\tau_1^b$ when using the electron, (I)Sigma or DNN reconstruction method.}
  \label{tab:traces}
\end{table}

\FloatBarrier
\section{Closure Tests for Rapgap}
\label{sec:closureRapgap}
\begin{figure}
   \begin{center}
\includegraphics[width=0.34\linewidth]{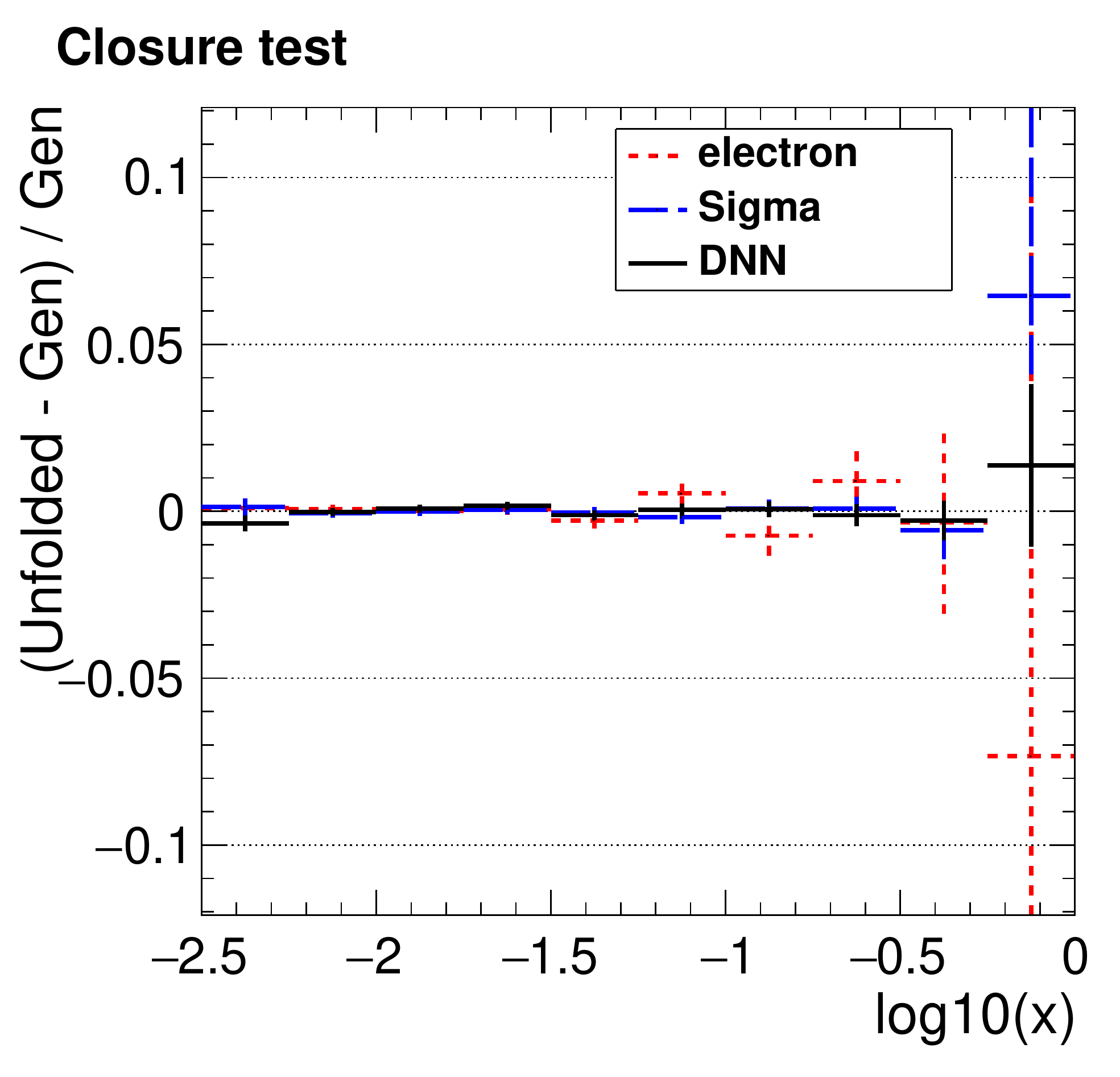}
\includegraphics[width=0.34\linewidth]{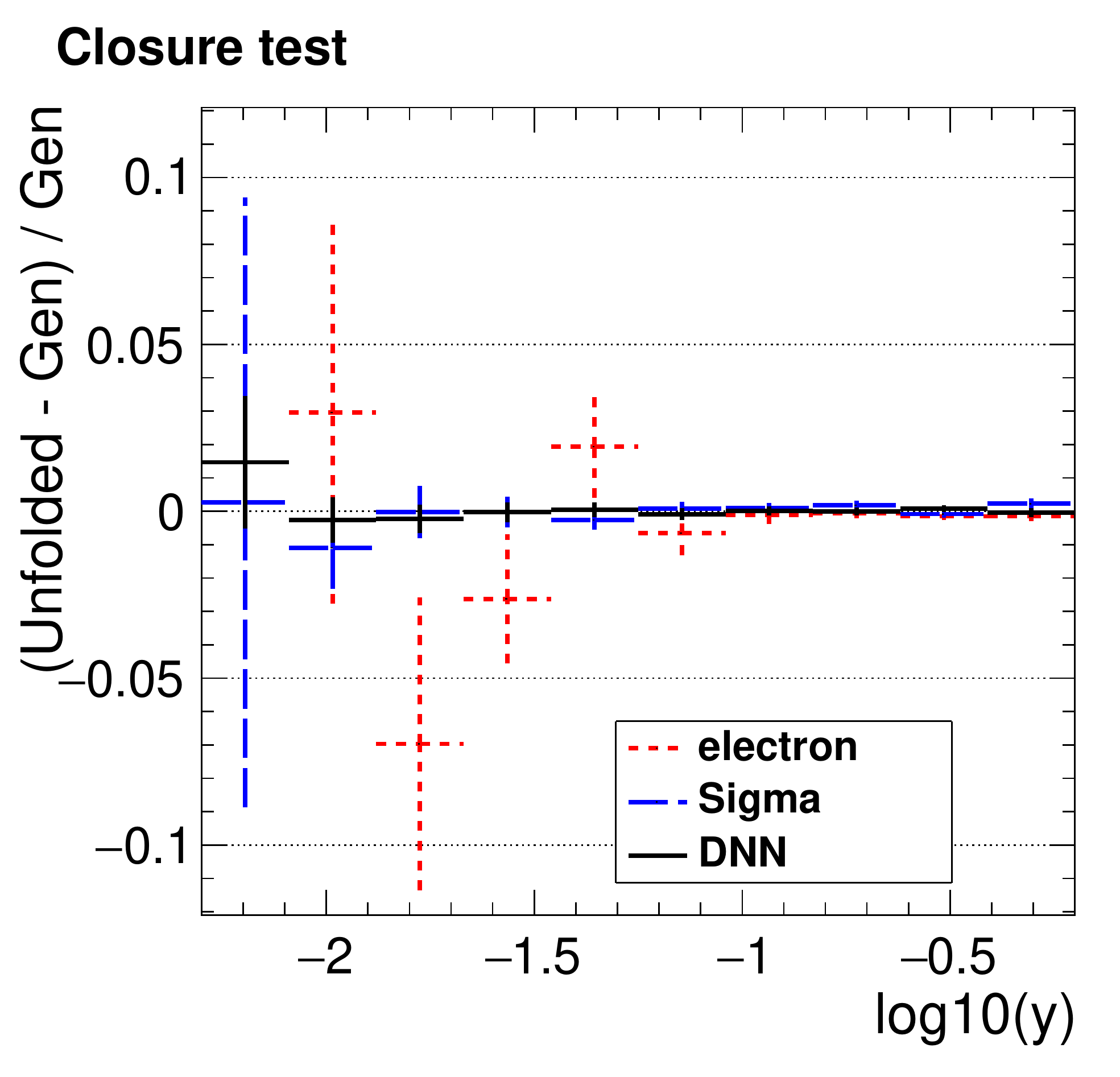} \\
\caption{
  Closure tests for the unfolding of $x$ (left) and $y$ (right) when using the \textsc{Rapgap} event sample.
}
\label{fig:dis-unfold-generator-syst1}
   \end{center}
\end{figure}

Figure~\ref{fig:dis-unfold-generator-syst1} shows the technical closure tests of the unfolding for the \textsc{Rapgap} sample, where the matrix and pseudo-data are obtained from statistically independent \textsc{Rapgap} samples. 
A sample of $10^7$ events generated from the \textsc{Rapgap} sample observed distribution is unfolded with the \textsc{Rapgap} response matrix.
The closure test is successful with the difference between the unfolding result and the generated distribution consistent with zero for all methods and all bins.

\FloatBarrier
 \bibliographystyle{elsarticle-num} 
 \bibliography{cas-refs,HEPML,other-refs}

\end{document}